%% file: sample631.tex
\documentclass[twocolumn]{aastex631}
\input{macros.tex}

\usepackage{amsmath}
\usepackage{soul}

\newcounter{qnumber}

\begin{document}

\title{Two Earth-size Planets and an Earth-size Candidate Transiting the Nearby Star HD~101581\footnote{This paper includes data gathered with the 6.5 meter Magellan Telescopes located at Las Campanas Observatory, Chile.}}

\correspondingauthor{Michelle Kunimoto}
\email{michelle.kunimoto@gmail.com}

\author[0000-0001-9269-8060]{Michelle Kunimoto}
\altaffiliation{Juan Carlos Torres Fellow}
\affiliation{Department of Physics and Kavli Institute for Astrophysics and Space Research, Massachusetts Institute of Technology,\\77 Massachusetts Avenue, Cambridge, MA 02139, USA}
\affiliation{Department of Physics and Astronomy, University of British Columbia, 6224 Agricultural Road, Vancouver, BC V6T 1Z1, Canada}

\author[0000-0003-0525-9647]{Zifan Lin}
\affiliation{Department of Earth, Atmospheric and Planetary Sciences, Massachusetts Institute of Technology,\\77 Massachusetts Avenue, Cambridge, MA 02139, USA}

\author[0000-0003-3130-2282]{Sarah Millholland}
\affiliation{Department of Physics, Massachusetts Institute of Technology,\\77 Massachusetts Avenue, Cambridge, MA 02139, USA}

\author[0000-0002-8400-1646]{Alexander Venner}
\affiliation{University of Southern Queensland, Centre for Astrophysics, USQ Toowoomba, West Street, QLD 4350 Australia}

\author[0000-0003-0595-5132]{Natalie R.~Hinkel}
\affiliation{Physics \& Astronomy Department, Louisiana State University, Baton Rouge, LA 70803, USA}

\author[0000-0002-1836-3120]{Avi Shporer}
\affiliation{Department of Physics and Kavli Institute for Astrophysics and Space Research, Massachusetts Institute of Technology,\\77 Massachusetts Avenue, Cambridge, MA 02139, USA}

\author[0000-0001-7246-5438]{Andrew Vanderburg}
\affiliation{Department of Physics and Kavli Institute for Astrophysics and Space Research, Massachusetts Institute of Technology,\\77 Massachusetts Avenue, Cambridge, MA 02139, USA}

\author[0000-0002-5726-7000]{Jeremy Bailey}
\affiliation{School of Physics, University of New South Wales, Sydney, NSW 2052, Australia}

\author{Rafael Brahm}
\affil{Facultad de Ingenier\'ia y Ciencias, Universidad Adolfo Ib\'a\~nez, Av. Diagonal las Torres 2640, Pe\~nalol\'en, Santiago, Chile}
\affil{Millennium Institute for Astrophysics, Chile}
\affil{Data Observatory Foundation, Chile}

\author[0000-0002-0040-6815]{Jennifer A. Burt}
\affiliation{Jet Propulsion Laboratory, California Institute of Technology, 4800 Oak Grove Drive, Pasadena, CA 91109, USA}

\author[0000-0003-1305-3761]{R. Paul Butler}
\affil{Earth and Planets Laboratory, Carnegie Institution for Science, 5241 Broad Branch Road, NW, Washington, DC 20015, USA}

\author[0000-0003-0035-8769]{Brad Carter}
\affiliation{University of Southern Queensland, Centre for Astrophysics, USQ Toowoomba, West Street, QLD 4350 Australia}

\author[0000-0002-5741-3047]{David R. Ciardi}
\affiliation{Caltech/IPAC-NExScI, M/S 100-22, 770 S Wilson Avenue, Pasadena, CA 91106 USA}

\author[0000-0001-6588-9574]{Karen A.\ Collins}
\affiliation{Center for Astrophysics \textbar \ Harvard \& Smithsonian, 60 Garden Street, Cambridge, MA 02138, USA}

\author[0000-0003-2781-3207]{Kevin I.\ Collins}
\affiliation{George Mason University, 4400 University Drive, Fairfax, VA, 22030 USA}

\author[0000-0001-8020-7121]{Knicole D. Colón}
\affiliation{NASA Goddard Space Flight Center, 8800 Greenbelt Road, Greenbelt, MD 20771, USA}

\author[0000-0002-5226-787X]{Jeffrey D. Crane}
\affiliation{The Observatories of the Carnegie Institution for Science, 813 Santa Barbara Street, Pasadena, CA 91101, USA}

\author[0000-0002-6939-9211]{Tansu Daylan}
\affiliation{Department of Physics and McDonnell Center for the Space Sciences, Washington University, St. Louis, MO 63130, USA}

\author[0000-0002-2100-3257]{Mat\'ias R. D\'iaz}
\affiliation{Las Campanas Observatory, Carnegie Institution of Washington, Colina El Pino, Casilla 601 La Serena, Chile}

\author{John P. Doty}
\affiliation{Noqsi Aerospace Ltd., 15 Blanchard Avenue, Billerica, MA 01821, USA}

\author[0000-0001-6039-0555]{Fabo Feng}
\affiliation{Tsung-Dao Lee Institute \& School of Physics and Astronomy, Shanghai Jiao Tong University, Shanghai 201210, China}

\author[0000-0002-9130-6747]{Eike W. Guenther}
\affil{Thueringer Landessternwarte Tautenburg, Sternwarte 5, 07778 Tautenburg, Germany}

\author[0000-0002-1160-7970]{Jonathan Horner}
\affiliation{University of Southern Queensland, Centre for Astrophysics, USQ Toowoomba, West Street, QLD 4350 Australia}

\author[0000-0002-2532-2853]{Steve~B.~Howell}
\affil{NASA Ames Research Center, Moffett Field, CA 94035, USA}

\author[0000-0002-6384-0184]{Jan Janik}
\affil{Department of Theoretical Physics and Astrophysics, Faculty of Science, Masaryk University, Kotla\v{r}sk\'{a} 2, CZ-611 37, Brno, Czech Republic}

\author{Hugh R. A. Jones}
\affil{Centre for Astrophysics Research, University of Hertfordshire, Hatfield AL10 9AB, UK}

\author[0000-0002-1623-5352]{Petr Kab\'{a}th}
\affil{Astronomical Institute of the Czech Academy of Sciences, Fri\v{c}ova 298, 25165, Ond\v{r}ejov, Czech Republic}

\author[0000-0001-8401-4300]{Shubham Kanodia}
\affil{Earth and Planets Laboratory, Carnegie Institution for Science, 5241 Broad Branch Road, NW, Washington, DC 20015, USA}

\author[0000-0001-7746-5795]{Colin Littlefield}
\affiliation{Bay Area Environmental Research Institute, Moffett Field, CA 94035, USA}
\affiliation{NASA Ames Research Center, Moffett Field, CA 94035, USA}

\author[0000-0002-4047-4724]{Hugh P. Osborn}
\affiliation{Physikalisches Institut, University of Bern, Gesellsschaftstrasse 6, 3012 Bern, Switzerland}
\affiliation{Department of Physics, ETH Zurich, Wolfgang-Pauli-Strasse 2, CH-8093 Zurich, Switzerland}

\author[0000-0003-2839-8527]{Simon O'Toole}
\affiliation{Australian Astronomical Optics, Macquarie University, North Ryde, NSW 1670, Australia}

\author[0000-0001-8120-7457]{Martin Paegert}
\affiliation{Center for Astrophysics \textbar \ Harvard \& Smithsonian, 60 Garden Street, Cambridge, MA 02138, USA}

\author{Pavel Pintr}
\affiliation{Institute of Plasma Physics of Czech Academy of Sciences, U Slovanky 2525/1a, 182 00 Praha 8, Czech Republic}

\author[0000-0001-8227-1020]{Richard P. Schwarz}
\affiliation{Center for Astrophysics \textbar \ Harvard \& Smithsonian, 60 Garden Street, Cambridge, MA 02138, USA}

\author[0000-0002-8681-6136]{Steve Shectman}
\affiliation{The Observatories of the Carnegie Institution for Science, 813 Santa Barbara Street, Pasadena, CA 91101, USA}

\author{Gregor Srdoc}
\affil{Kotizarovci Observatory, Sarsoni 90, 51216 Viskovo, Croatia}

\author[0000-0002-3481-9052]{Keivan G.\ Stassun}
\affiliation{Department of Physics and Astronomy, Vanderbilt University, Nashville, TN 37235, USA}

\author[0009-0008-2801-5040]{Johanna K. Teske}
\affil{Earth and Planets Laboratory, Carnegie Institution for Science, 5241 Broad Branch Road, NW, Washington, DC 20015, USA}
\affiliation{Carnegie Institution for Science \& The Observatories of the Carnegie Institution for Science, 813 Santa Barbara Street, Pasadena, CA 91101, USA}

\author[0000-0002-6778-7552]{Joseph D. Twicken}
\affiliation{SETI Institute, Mountain View, CA  94043, USA}
\affiliation{NASA Ames Research Center, Moffett Field, CA 94035, USA}

\author{Leonardo~Vanzi}
\affiliation{Department of Electrical Engineering and Centre of Astro-Engineering, Pontificia Universidad Catolica de Chile, Av. Vicuña Mackenna 4860, Santiago, Chile}

\author[0000-0002-6937-9034]{Sharon X.~Wang}
\affiliation{Department of Astronomy, Tsinghua University, Beijing 100084, People's Republic of China}

\author[0000-0001-9957-9304]{Robert A. Wittenmyer}
\affiliation{University of Southern Queensland, Centre for Astrophysics, USQ Toowoomba, West Street, QLD 4350 Australia}

\author[0000-0002-4715-9460]{Jon M. Jenkins}
\affiliation{NASA Ames Research Center, Moffett Field, CA 94035, USA}

\author[0000-0003-2058-6662]{George R. Ricker}
\affiliation{Department of Physics and Kavli Institute for Astrophysics and Space Research, Massachusetts Institute of Technology,\\77 Massachusetts Avenue, Cambridge, MA 02139, USA}

\author[0000-0002-6892-6948]{Sara Seager}
\affiliation{Department of Earth, Atmospheric and Planetary Sciences, Massachusetts Institute of Technology,\\77 Massachusetts Avenue, Cambridge, MA 02139, USA}
\affiliation{Department of Physics and Kavli Institute for Astrophysics and Space Research, Massachusetts Institute of Technology,\\77 Massachusetts Avenue, Cambridge, MA 02139, USA}
\affiliation{Department of Aeronautics and Astronautics, Massachusetts Institute of Technology, Cambridge, MA 02139, USA}

\author[0000-0002-4265-047X]{Joshua Winn}
\affiliation{Department of Astrophysical Sciences, Princeton University, 4 Ivy Lane, Princeton, NJ 08540, USA}

\begin{abstract}

We report the validation of multiple planets transiting the nearby ($d = 12.8$ pc) K5V dwarf HD~101581 (GJ 435, TOI-6276, TIC 397362481). The system consists of at least two Earth-size planets whose orbits are near a mutual 4:3 mean-motion resonance, HD~101581 b ($R_{p} = 0.956_{-0.061}^{+0.063}~R_{\oplus}$, $P = 4.47$ days) and HD~101581 c ($R_{p} = 0.990_{-0.070}^{+0.070}~R_{\oplus}$, $P = 6.21$ days). Both planets were discovered in Sectors 63 and 64 TESS observations and statistically validated with supporting ground-based follow-up. We also identify a signal that probably originates from a third transiting planet, TOI-6276.03 ($R_{p} = 0.982_{-0.098}^{+0.114}~R_{\oplus}$, $P = 7.87$ days). These planets are remarkably uniform in size and their orbits are evenly spaced, representing a prime example of the ``peas-in-a-pod'' architecture seen in other compact multi-planet systems. At $V = 7.77$, HD~101581 is the brightest star known to host multiple transiting planets smaller than $1.5~R_{\oplus}$. HD~101581 is a promising system for atmospheric characterization and comparative planetology of small planets.
\end{abstract}

\keywords{Exoplanet Systems (484) --- Exoplanet Dynamics (490) --- Exoplanets (498) --- Transit Photometry (1709)}

\section{Introduction} \label{sec:introduction}

Multi-planet systems (``multis'') represent invaluable laboratories for advancing our understanding of planetary formation, dynamics, and evolution. The presence of multiple planets within a system enables direct comparative planetology because the planets formed within the same protoplanetary disk and evolved around the same host star \cite[e.g.,][]{Millholland2017, Otegi2022}. These systems also enable studies of planet-planet interactions, and the dynamical processes that govern the migration and stability of planetary architectures \cite[e.g.,][]{Lissauer2011, Delisle2017, Petit2020}.

The Kepler mission \citep{Borucki2010} revolutionized our study of multi-planet systems by finding nearly 800 stars that host at least two exoplanets \citep{Thompson2018}. Kepler unveiled a diverse array of system architectures, the most common of which appears to be a ``peas-in-a-pod'' configuration where planets tend to be similarly sized and uniformly spaced \citep{Weiss2018}. The similarity in size appears to be a common outcome of planet formation, further supported by the finding that planets in the same system also tend to have similar masses \citep{Millholland2017}. Correlations between planet size and planet spacing also suggest that dynamics play a key role in shaping final system architectures \cite[e.g.,][]{Huang2022, Luque2023}.

More recently, the Transiting Exoplanet Survey Satellite \citep[TESS;][]{Ricker2015} has identified almost 200 new multis consisting of transiting planets and planet candidates predominantly in short-period orbits ($P \lesssim 20$ days).\footnote{Based on the TESS Object of Interest catalogue on ExoFOP \citep{ExoFOP}, accessed 2024 August 1.} Because TESS searches nearby, bright stars, TESS multis tend to be significantly more amenable than Kepler multis to follow-up observations such as mass measurements with radial velocity observations \cite[e.g.,][]{Gandolfi2018, Trifonov2019, Dragomir2019} and atmospheric characterization with transmission and emission spectroscopy \citep{Hord2024}. Characterizing planets across multiple dimensions is important to take full advantage of multi-planet systems as testbeds for theories of planetary formation, evolution, and dynamics.

We present the detection and statistical validation of a new multi-planetary system with at least two Earth-size planets transiting the bright ($V = 7.77$) K dwarf HD~101581. TESS observations also reveal a potential third Earth-size planet. At 12.8 pc, HD~101581 is the fourth closest system hosting multiple Earth-size planets after LTT-1445 A \citep{Lavie2023}, L 98-59 \citep{Cloutier2019}, and TRAPPIST-1 \citep{Gillon2016}, and is the brightest of all such systems in the optical band (by $\Delta V > 2.82$) discovered so far. 

Our paper is organized as follows: Section \ref{sec:observations} presents the space-based photometry from TESS that led to the detection of the planets and planet candidate. We also describe ground-based observations that supported the validation of the planets, including seeing-limited photometry, high-resolution imaging, and high-resolution spectroscopy. Section \ref{sec:star} provides information about the host star, while Section \ref{sec:planets} details the planetary system parameters. In Section \ref{sec:validation}, we consider various false positive scenarios and present the statistical validation of the system. Section \ref{sec:discussion} discusses several aspects of our results, including dynamical and stability analysis and the potential for follow-up confirmation via radial velocity mass measurements and atmospheric characterization via transmission and emission spectroscopy. Section \ref{sec:conclusion} summarizes our results and presents our conclusions.

\section{Observations} \label{sec:observations}

\subsection{Transiting Exoplanet Survey Satellite}

The Transiting Exoplanet Survey Satellite \cite[TESS;][]{Ricker2015} observed HD~101581 (TIC 397362481) in Sectors 63 (2023 March 10 - April 06, Camera 2 CCD 3) and 64 (2023 April 06 - May 04, Camera 2 CCD 4) at a cadence of 2 minutes, totalling 55 days of observations. The data were processed in the TESS Science Processing Operations Center \cite[SPOC;][]{SPOC} at NASA Ames Research Center. The SPOC conducted a transit search with an adaptive, noise-compensating matched filter \citep{Jenkins2002, Jenkins2010, Jenkins2020} and detected two transiting signals at $P = 6.207$ and $4.465$ days in both sectors, as presented in SPOC Data Validation (DV) reports \citep{Twicken2018, Li2019}. These signals were reported as TESS Objects of Interest \cite[TOIs;][]{TOI} TOI-6276.01 and TOI-6276.02, respectively, on 2023 April 27. A third candidate with a period of $7.871$ days was identified by SPOC and alerted as TOI-6276.03 on 2024 February 01.

\begin{figure*}
\centering
    \includegraphics[width=0.9\linewidth]{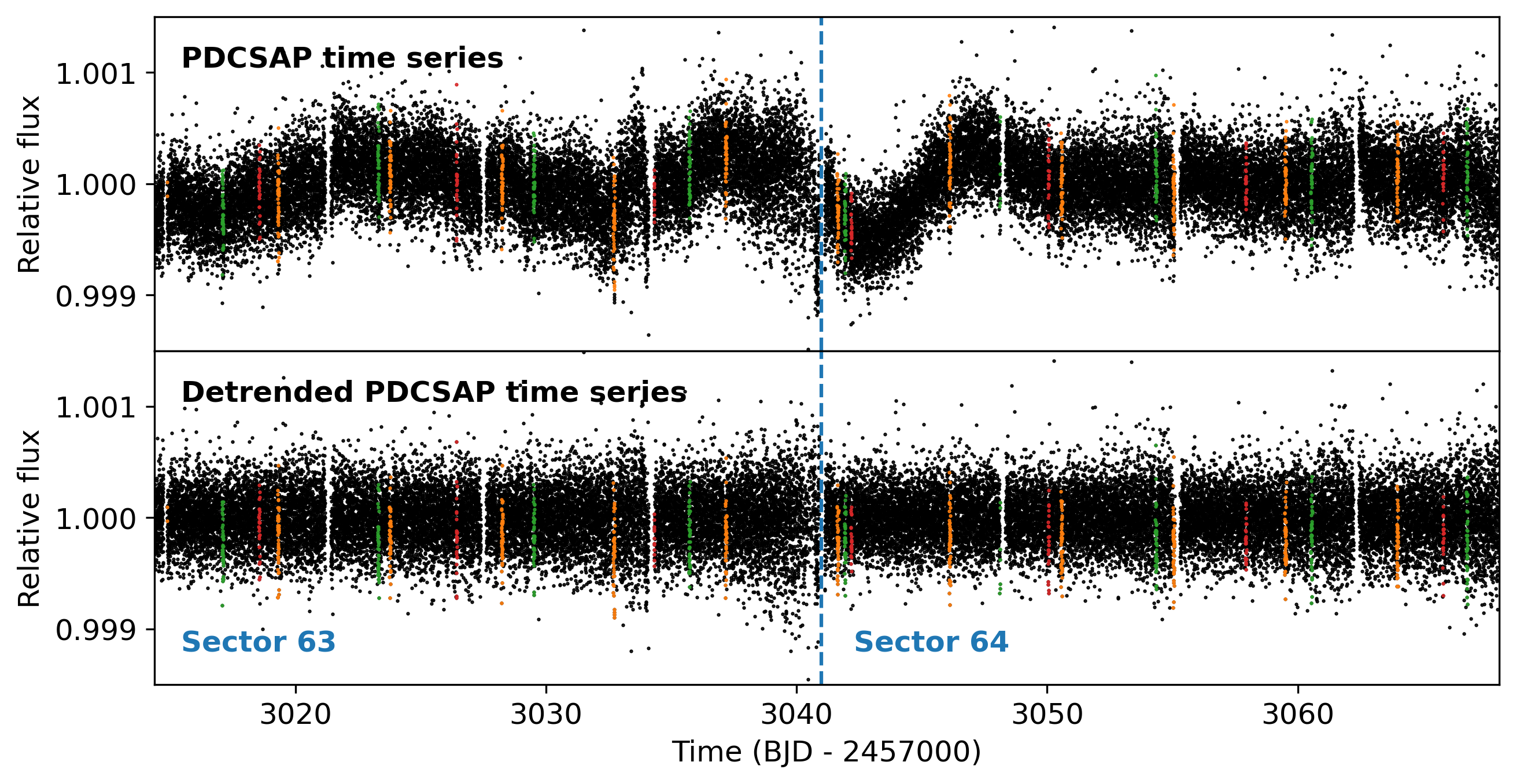}
    \caption{SPOC PDCSAP light curve for TOI-6276 from Sectors 63 and 64 before detrending (top) and after detrending (bottom) with \texttt{wotan}. The transits of TOI-6276.01, 6276.02, and 6276.03 are indicated by green, orange, and red dots, respectively, based on our TLS search.}\label{fig:lc}
\end{figure*}

For our analysis, we downloaded the SPOC Sector 63 and 64 light curves from the Mikulski Archive for Space Telescopes (MAST).\footnote{https://mast.stsci.edu} We used the Presearch Data Conditioning Simple Aperture Photometry \cite[PDCSAP;][]{Smith2012,Stumpe2012,Stumpe2014} light curves with all data points with a nonzero quality flag removed, and further removed low-frequency trends using the biweight time-windowed slider implemented in the \texttt{wotan} Python package \citep{Wotan} with a window of 0.5 days. Because detrending can distort transit shapes, we set the detrending algorithm to ignore cadences within one transit duration of the transit mid-points when determining the parameters of the trend functions. We ran a blind search for periods between 1 and 27 days (half the timespan of the data) on the combined multi-sector light curve using Transit Least Squares \cite[TLS;][]{TLS} and recovered all three transit signals with properties consistent with the DV reports (Table~\ref{tab:initial}), shown in Figure \ref{fig:lc}. We did not uncover any new candidates in further searches of the data.

\begin{table*}[]
    \centering
    \begin{tabular}{ccccc}
    \hline\hline
        TOI & Orbital period & Transit epoch & Transit depth & SNR \\
        & (days) & (BJD - 2457900) & (ppm) & \\
        \hline
        TOI-6276.01 & 6.207 & 3017.103 & 192 & 12.4 \\
        TOI-6276.02 & 4.466 & 3014.849 & 197 & 15.6 \\
        TOI-6276.03 & 7.874 & 3018.560 & 155 & 7.9 \\
        \hline
    \end{tabular}
    \caption{Initial ephemerides and signal-to-noise ratios (SNR) of the three candidates transiting HD~101581, based on a TLS search of the combined Sector 63 and 64 TESS light curves. Transit epochs are given as the Barycentric Julian Date (BJD) offset by 2457000 days, also known as the Barycentric TESS Julian Date.}
    \label{tab:initial}
\end{table*}

\subsection{Ground-Based Photometry}\label{sec:sg1}

We obtained ground-based photometric observations for TOI-6276.01 and 6276.02 through the TESS Follow-up Observing Program (TFOP) Subgroup 1, which specializes in seeing-limited photometry to aid in the validation of TESS planet candidates. Observations were taken using telescopes in the Las Cumbres Observatory global telescope network \cite[LCOGT;][]{LCO}. Images were calibrated using the standard LCOGT \texttt{BANZAI} pipeline \citep{BANZAI}, and photometric data were extracted with \texttt{AstroImageJ} \citep{AstroImageJ}. All time series are available on the Exoplanet Follow-up Observing Program website\footnote{\url{https://exofop.ipac.caltech.edu}} \cite[ExoFOP;][]{ExoFOP}. While the shallow transit depths of the candidates preclude on-target detection from the ground, these observations were used to look for deep eclipses on nearby eclipsing binaries (NEBs) that could result in the shallow transits seen in TESS data due to blending on the large TESS pixels.

Observations of a full transit of TOI-6276.01 were attempted on 2023 May 27 by the 1m telescope at the Cerro Tololo Inter-American Observatory, which uses a $4096\times4096$ pixel SINISTRO camera with a pixel scale of 0.389$^{\prime\prime}$px$^{-1}$. 164 images in the $z^{\prime}$ band were taken over 243 minutes. The field out to 2.5$^{\prime}$ was cleared of NEBs using aperture radii between 5 and 9 px (1.9 - 3.5$^{\prime\prime}$). The inner and outer radii of the sky annuli were 20 and 35 px, respectively.

Observations of full transits of TOI-6276.02 were attempted on 2023 May 03 and 2024 March 15 by the 1m telescope at the South African Astronomical Observatory, which also uses a SINISTRO camera. On the first night, 260 images in the $z^{\prime}$ band were taken over 230 minutes. The field out to 2.5$^{\prime}$ was likewise cleared of NEBs using aperture radii between 5 and 13 px (1.9 - 5.1$^{\prime\prime}$), with the inner and outer radii of the sky annulus chosen to be 22 and 33 px. On the second night, 232 images in the $z^{\prime}$ band were taken over 359 minutes, clearing the field of NEBs using 5 px aperture radii, 24 px inner sky annulus radii, and 36 px outer sky annulus radii.
 
\subsection{High-Resolution Imaging}\label{sec:sg3}

We obtained high-resolution observations of TOI-6276 using both adaptive optics (AO) and speckle imaging through TFOP Subgroup 3. We searched for nearby stars (either bound or chance-aligned) which could be false positive sources for any of the candidates. These companions can also dilute the TESS photometry, resulting in under-estimated planet radii. All reduced data are available on ExoFOP.

\begin{figure}
    \centering
    \includegraphics[width=\linewidth]{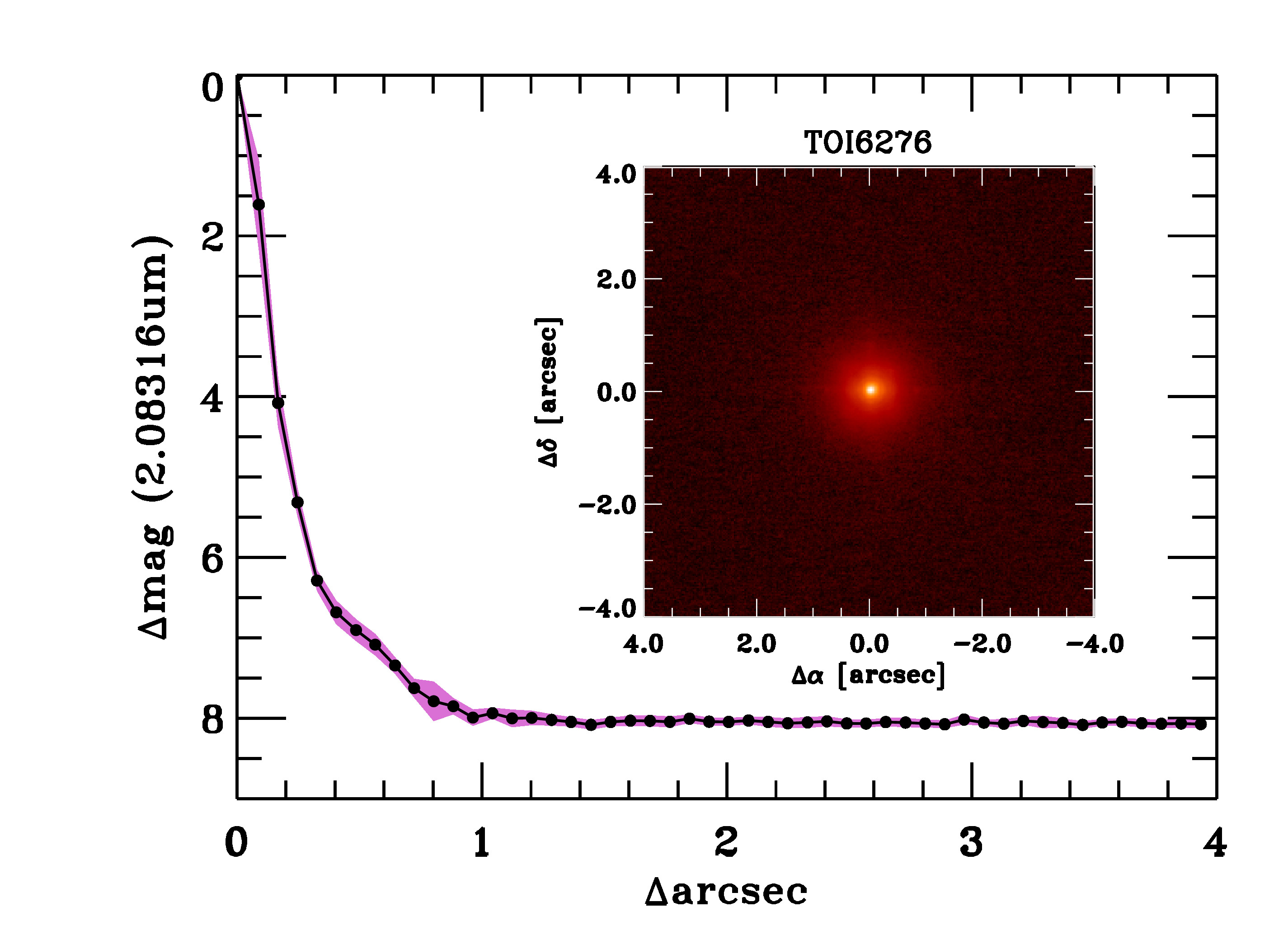}
    \includegraphics[width=0.8\linewidth]{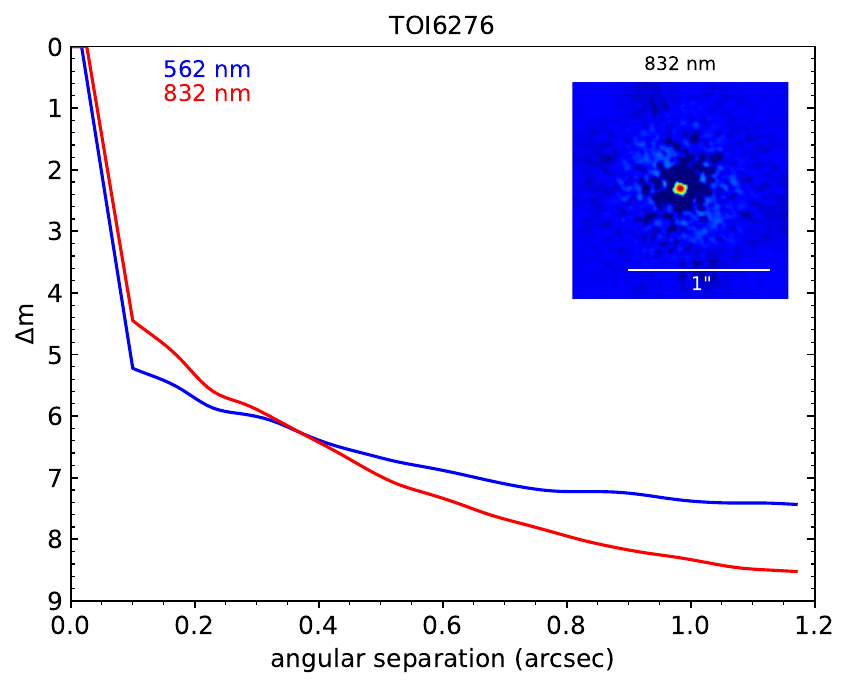}
    \caption{Contrast curves and high-resolution images from archival VLT/NACO AO observations (top) and Gemini-S/Zorro speckle observations (bottom). No stellar companions were detected in any observations.}
    \label{fig:HRI}
\end{figure}

TOI-6276 was observed on 2005 May 07 as part of a VLT/NACO AO survey to search for stellar companions of planet-hosting stars \citep{NACO}, with the star included as part of a control sample of stars without known planets. The high-resolution images were taken at 2.0831 $\mu$m with a pixel scale of 0.027$^{\prime\prime}$px$^{-1}$ and estimated PSF of 0.08$^{\prime\prime}$. We also obtained speckle images taken with the Zorro instrument on the 8m Gemini-South telescope \citep{Scott2021} on 2023 June 30. The high-resolution images were collected at 562 and 832 nm with a pixel scale of 0.01$^{\prime\prime}$px$^{-1}$ and an inner working angle of 20 mas. The data were reduced using the standard Fourier techniques as outlined in \citet{Howell2011}. To within the angular and magnitude limits achieved by the observations, no stellar companions were detected within the sensitivity limits of VLT/NACO ($\Delta m = 7.0$ at 0.5$^{\prime\prime}$) or Gemini-S/Zorro ($\Delta m = 6.9$ at 0.5$^{\prime\prime}$), as shown in Figure \ref{fig:HRI}.

\subsection{High-Resolution Spectroscopy}\label{sec:sg24}

We obtained high-resolution spectra and radial velocity (RV) measurements through TFOP Subgroup 2 to search for evidence of the target star being a spectroscopic binary or having a stellar-mass companion based on large RV variations. The full list of derived RVs are provided in Table \ref{tab:rvs}.

\begin{table*}[]
    \centering
    \begin{tabular}{c|c|c|c|c}
    \hline\hline
        Instrument & Time & RV & RV error & $S$-index \\
        & (BJD - 2450000) & (m s$^{-1}$) & (m s$^{-1}$) & \\
        \hline
        PFS & 5663.62928 & -8.65 & 1.04 & 0.3316 \\
& 6088.55647 & 1.68 & 1.01 & 0.2804 \\
& 6092.55211 & 3.51 & 1.21 & 0.2911 \\
& 6282.85282 & 2.23 & 0.93 & 0.3672 \\
& 6347.77346 & -1.0 & 1.11 & 0.3935 \\
& 6355.73528 & -3.09 & 1.08 & 0.3394 \\
& ... & ... & ... & ... \\
\hline
UCLES & 2389.99163 & -2.08 & 1.81 & - \\
& 2424.97284 & 1.43 & 2.3 & - \\
& 2452.92427 & 3.75 & 2.24 & - \\
& 2452.93193 & 0.23 & 2.0 & - \\
& 2454.89403 & 3.38 & 2.04 & - \\
& 2454.90169 & -5.5 & 2.17 & - \\
& ... & ... & ... & ... \\
\hline
HARPS & 3015.86777 & 0.529 & 0.983 & - \\
& 3016.86072 & -0.086 & 0.908 & - \\
& 3759.82147 & 4.517 & 0.909 & - \\
& 4118.85185 & -3.266 & 0.964 & - \\
& 4137.85350 & -0.933 & 0.585 & - \\
& 4171.77809 & 0.21 & 0.809 & - \\
\hline
PUCHEROS+ & 10075.63965 & 105 & 46 & - \\
& 10075.66055 & -93 & 48 & - \\
& 10075.63965 & 105 & 46 & - \\
& 10075.66055 & -93 & 48 & - \\
& 10079.66104 & -24 & 43 & - \\
& 10080.62353 & -167 & 45 & - \\
& ... & ... & ... & ... \\
\hline
    \end{tabular}
    \caption{Radial velocity measurements from the Magellan II/PFS, AAT/UCLES, La Silla/HARPS, and La Silla/PUCHEROS+ instruments. The PUCHEROS+ RVs have had a mean of 14.358 km s$^{-1}$ subtracted from the provided values. A portion of this table is shown for demonstration. The full table is available in a machine readable format.}
    \label{tab:rvs}
\end{table*}

\subsubsection{La Silla/PUCHEROS+}

We obtained 14 spectra from the upgraded version of the Pontificia Universidad Catolica High Echelle
Resolution Optical Spectrograph \cite[PUCHEROS+, based on][]{Vanzi2012}, which is currently installed at the 1.52-m telescope at La Silla observatory, Chile, within the PLATOSPec project.\footnote{\url{https://stel.asu.cas.cz/plato/}} PUCHEROS+ is an $R \sim 18000$ spectrograph with a spectral coverage of $400 - 700$ nm. Observations were taken between 2023 May and 2024 March (see \ref{tab:rvs}). Spectra were processed by the CERES+ pipeline \citep{CERES}, which extracts 1D, order-by-order spectra from the raw images, generates the corresponding wavelength solutions, corrects for instrumental drifts, and then computes both the RV shift and the bisector span of the cross correlation function used to measure the RV. The radial velocities were computed with the cross-correlation technique by using a binary mask \citep{Baranne1996,Fellgett1955,Griffin1967}.

\subsubsection{Magellan II/PFS}

\begin{figure*}
    \includegraphics[width=\linewidth]{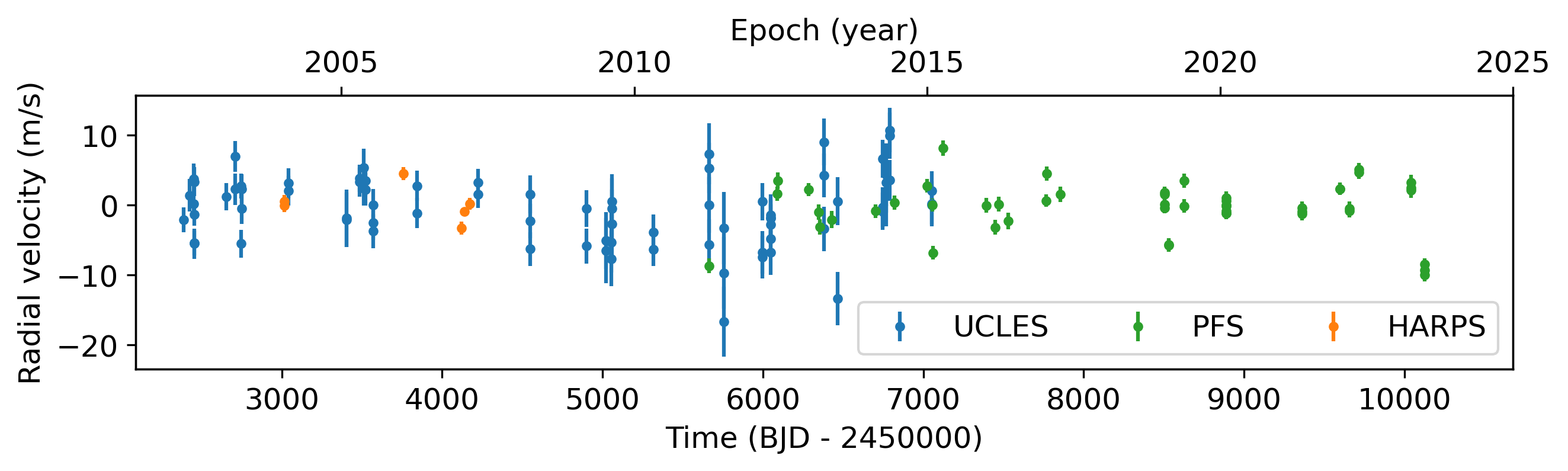}
    \caption{RV observations from PFS (green), UCLES (blue), and HARPS (orange). PUCHEROS+ RVs are not included due to the large uncertainties ($\sim$50 m s$^{-1}$). While the expected RV semi-amplitudes of the three planets are too small to be resolved ($<0.4$ m s$^{-1}$), the observations do not feature large RV variations corresponding to stellar-mass companions.}\label{fig:RVs}
\end{figure*}

HD~101581 was observed between 2011 April and 2023 June as part of the Magellan Exoplanet Long Term Survey (LTS), which is carried out on the 6.5-m Magellan II Telescope at Las Campanas Observatory in Chile using the Carnegie Planet Finder Spectrograph \cite[PFS;][]{Crane2006, Crane2008, Crane2010}. PFS is a high-resolution optical echelle spectrograph with a total wavelength coverage of $391 - 734$ nm. The use of an iodine cell to measure RVs results in a wavelength range used for RV derivation of $500 - 620$ nm. We obtained 20 RVs with a mean uncertainty of 1.05 m s$^{-1}$ from before PFS underwent an upgrade in 2018 February, and 30 RVs with a mean uncertainty of 0.93 m s$^{-1}$ afterwards. RVs were extracted following the technique described by \citet{Butler1996}. Given that the planets have expected RV semi-amplitudes of $K\sim0.4$ m s$^{-1}$ (\S\ref{sec:followup}), these RVs are too imprecise to allow the determination of the mass of any of the three planets. However, the lack of large RV variations (top panel of Figure \ref{fig:RVs}) is an important indicator that HD~101581 lacks stellar-mass and high-mass planetary companions (\S\ref{sec:eclipsing}).

The PFS spectral wavelength range covers the Ca II H \& K lines, enabling measurements of $S$-indices to monitor stellar activity. $S$-indices are correlated with spot activity on the stellar surface \cite[e.g.,][]{Wilson1978}, and serve as a proxy for chromospheric activity that could cause RV shifts that mimic those induced by planets. $S$-indices were derived using the algorithm outlined by \citet{Santos2000}.

\subsubsection{AAT/UCLES}

HD~101581 was observed between 2002 April and 2015 January as part of the Anglo-Australian Planet Search \citep{Tinney2001}, which was carried out on the 3.9-m Anglo-Australian Telescope (AAT) using the University College of London Echelle Spectrograph \cite[UCLES;][]{Diego1990}. UCLES is a high-resolution echelle spectrograph covering $482 - 855$ nm, limited to $500 - 620$ nm for RV derivation using an iodine cell. We obtained 79 RVs with a mean uncertainty of 2.95 m s$^{-1}$, finding a similar lack of large RV variation as in the PFS observations (Figure \ref{fig:RVs}).

\subsubsection{La Silla/HARPS}

HD~101581 was observed between 2004 January and 2007 March by the High Accuracy Radial velocity Planet Searcher \cite[HARPS;][]{Mayor2003} on the ESO 3.6-m telescope at La Silla Observatory as part of the HARPS GTO planet search program \citep{Sousa2008}. HARPS is a high-resolution echelle spectrograph covering 380 -- 690 nm. We obtained 6 RVs with a mean uncertainty of 0.93 m s$^{-1}$ through the \texttt{RVBank} archive \citep{Trifonov2020}. The adopted values correspond to RVs extracted from spectra by the SpEctrum Radial Velocity AnaLyser (SERVAL) pipeline \citep{Zechmeister2018} and corrected for systematic effects including barycentric Earth radial velocity, secular acceleration of the RV \citep{Kurster2003}, Fabry-Perot drift, and nightly zero-points.

\subsection{Astrometry}

Precise astrometric observations of HD~101581 have been obtained in the course of the \textit{Hipparcos} and \textit{Gaia} missions \citep{Hipparcos, HipparcosNew, Gaia}. We use the cross-calibrated \textit{Hipparcos-Gaia} proper motion data from \citet{Brandt2018, Brandt2021} to place limits on long-term tangential motion of the star. No evidence of astrometric acceleration is detected above the level of precision (sky-projected velocity $\sigma_{Gaia}=$~0.02~mas~yr$^{-1}\approx1.2$~m~s$^{-1}$), suggesting that any companions must be very long period or comparatively light \cite[e.g.][]{CastroGinard2024}.

\subsection{Archival Images}

\begin{figure*}[t!]
    \centering
    \includegraphics[width=\linewidth]{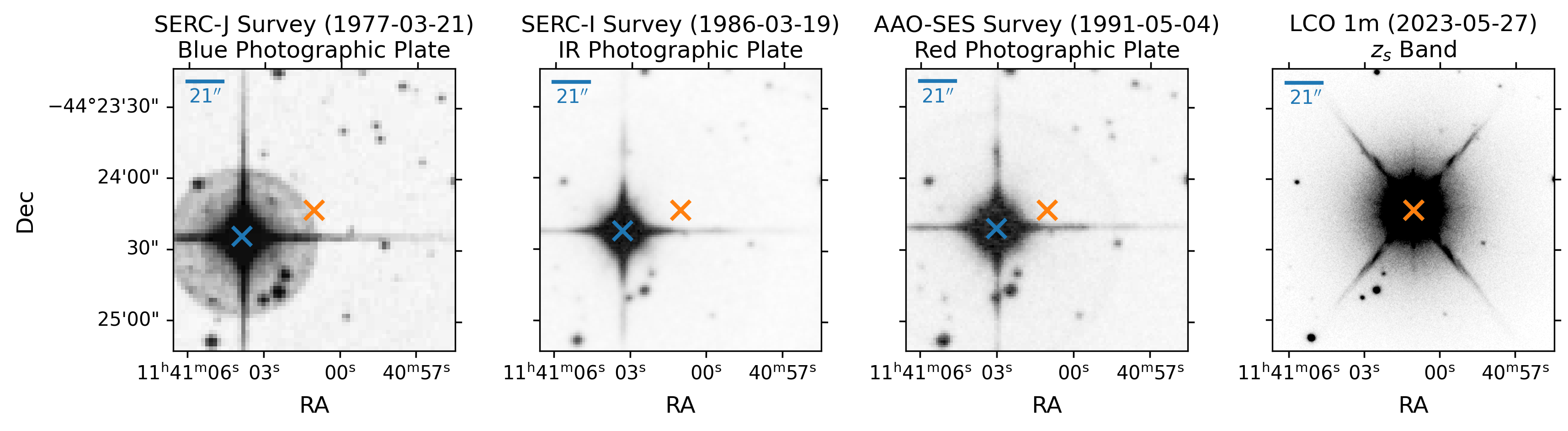}
    \caption{The field within $60^{\prime\prime}$ of HD~101581 in blue (far left), infrared (center left), and red (center right) filters from the SERC and AAO-SES Surveys, and from LCO/CTIO observations taken shortly after the end of the TESS observations in 2023 (right). The location of HD~101581 at the time of each image is marked with a blue cross, while the location at the start of TESS Sector 63 in 2023 March is marked in orange. The width of a TESS pixel ($21^{\prime\prime}$) is marked in the top left of each panel. HD~101581 has moved $\sim32^{\prime\prime}$ between 1977 and 2023.}
    \label{fig:finders}
\end{figure*}

HD~101581 was observed in blue and infrared bands as part of the SERC Southern Sky Survey in 1977 and 1986, as well as in the red band as part of the AAO-SES Survey in 1991. We obtained the corresponding images from the NASA/IPAC Infrared Science Archive,\footnote{\url{https://irsa.ipac.caltech.edu/applications/finderchart}} as shown in Figure \ref{fig:finders}. Due to its high proper motion, the star has moved $32^{\prime\prime}$ between the oldest image in 1977 and the start of the TESS observations in 2023 March. There are no other stars visible at its current position down to the limiting magnitude of the SERC-J survey ($B \approx 23$).

\section{Stellar Parameters}\label{sec:star}

HD~101581 is a K4.5V dwarf \citep{Gray2006} and is listed in the TICv8.2 \citep{Paegert2021} as TIC 397362481 with radius $R_{\star} = 0.631\pm0.054~R_{\odot}$, mass $M_{\star} = 0.740\pm0.087~M_{\odot}$, effective temperature $T_{\rm eff} = 4634\pm143$ K, and surface gravity $\log g~\text{(cgs)}= 4.71\pm0.11$. A summary of stellar properties and their sources is given in Table \ref{tab:star}.

\begin{table*}[h]
    \centering
    \caption{Stellar Parameters for HD~101581 (TOI-6276). The parameters from our isochrones analysis (\S\ref{sec:isochrones}) were adopted for deriving planetary parameters (\S\ref{sec:planets}).}
    \begin{tabular}{l|l|l|l}
        \hline\hline
        Parameter & Value & Description & Source \\ 
        \hline\hline
        \multicolumn{4}{c}{\textit{TIC Parameters}} \\
        \hline
        ID & 397362481 & TESS Input Catalog ID & TICv8.2 \citep{Paegert2021} \\
        $T_{\rm eff}$ & $4634\pm143$ & Effective temperature (K) & TICv8.2\\
        $\log g$ & $4.71\pm0.11$ & Surface gravity (cgs) & TICv8.2\\
        $R_{\star}$ & $0.631\pm0.054$ & Stellar radius ($R_{\odot}$) & TICv8.2 \\
        $M_{\star}$ & $0.740\pm0.087$ & Stellar mass ($M_{\odot}$) & TICv8.2 \\
        $[$Fe/H$]$ & $-0.505\pm0.027$ & Metallicity (dex) & TICv8.2 \\
        \hline
        \multicolumn{4}{c}{\textit{Astrometric Parameters}} \\
        \hline
        $\alpha$ & 11:41:01.482 & Right ascension (J2000, epoch 2016) & \gaia\ DR3 \citep{Gaia,GaiaDR3} \\
        $\delta$ & -44:24:14.81 & Declination (J2000, epoch 2016) & \gaia\ DR3 \\
        $\bar{\omega}$ & $78.2268\pm0.0182$ & Parallax (mas) & \gaia\ DR3\ \\
        $\mu_{\alpha}$ & $-660.634\pm0.016$& Proper motion right ascension (mas~yr$^{-1}$) & \gaia\ DR3 \\
        $\mu_{\delta}$ & $242.096\pm0.013$ & Proper motion declination (mas~yr$^{-1}$) &\gaia\ DR3 \\
        \hline
        \multicolumn{4}{c}{\textit{Photometric Parameters}} \\
        \hline
        $T$ & $6.721\pm0.006$ & TESS band magnitude (mag) & TICv8.2 \\
        $B$ & $8.865\pm0.020$ & $B$ band magnitude (mag) & UCAC4 \citep{UCAC4} \\
        $V$ & $7.770\pm0.030$ & $V$ band magnitude (mag) & UCAC4\\
        $G$ & $7.394\pm0.003$ & $G$ band magnitude (mag) & \gaia\ DR3 \\
        $G_{\mathrm{BP}}$ & $7.979\pm0.003$ & $G_{\mathrm{BP}}$ band magnitude (mag) & \gaia\ DR3 \\
        $G_{\mathrm{RP}}$ & $6.666\pm0.004$ & $G_{\mathrm{RP}}$ band magnitude (mag) & \gaia\ DR3 \\
        $J$ & $5.792\pm0.021$ & $J$ band magnitude (mag) & \twomass\ \citep{2MASS} \\
        $H$ & $5.273\pm0.075$ & $H$ band magnitude (mag) & \twomass \\
        $K_{s}$ & $5.101\pm0.016$ & $K$ band magnitude (mag) & \twomass \\
        $W1$ & $5.033\pm0.206$ & $W1$ band magnitude (mag) & \wise\ \citep{WISE}\\
        $W2$ & $4.880\pm0.099$ & $W2$ band magnitude (mag) & \wise \\
        $W3$ & $5.075\pm0.014$ & $W3$ band magnitude (mag) & \wise \\
        $W4$ & $5.027\pm0.028$ & $W4$ band magnitude (mag) & \wise \\  
        \hline
        \multicolumn{4}{c}{\textit{Derived Parameters}} \\
        \hline
        $\mathrm{[C/H]}$ & $-0.40\pm0.09$ &  C abundance & Hypatia Catalog \citep{Hinkel14} (\S\ref{sec:hypatia})\\
        $\mathrm{[Na/H]}$ & $-0.32\pm0.06$ &  Na abundance & Hypatia Catalog \\
        $\mathrm{[Mg/H]}$ & $-0.41\pm0.05$ &  Mg  abundance & Hypatia Catalog \\
        $\mathrm{[Al/H]}$ & $-0.31\pm0.07$ & Al  abundance & Hypatia Catalog \\
        $\mathrm{[Si/H]}$ & $-0.43\pm0.11$ & Si  abundance & Hypatia Catalog \\
        $\mathrm{[Ca/H]}$ & $-0.37\pm0.20$ & Ca  abundance & Hypatia Catalog \\
        $\mathrm{[Sc/H]}$ & $0.13\pm0.14$ & Sc  abundance & Hypatia Catalog \\
        $\mathrm{[Ti/H]}$ & $-0.18\pm0.17$ &  Ti abundance & Hypatia Catalog \\
        $\mathrm{[V/H]}$ & $0.13\pm0.11$ & V  abundance & Hypatia Catalog \\
        $\mathrm{[Fe/H]}$ & $-0.51\pm0.12$ & Fe  abundance & Hypatia Catalog \\
        $\mathrm{[Ni/H]}$ & $-0.48\pm0.08$ & Ni  abundance & Hypatia Catalog \\
        $\mathrm{[YII/H]}$ & $-0.54\pm0.08$ & YII  abundance & Hypatia Catalog \\
        $\mathrm{[BaII/H]}$ & $-0.43\pm0.15$ & BaII  abundance & Hypatia Catalog \\
        $\mathrm{[EuII/H]}$ & $-0.08\pm0.10$ &  EuII  abundance & Hypatia Catalog \\
        $v\sin{i}$ & $2.47\pm0.30$ & Projected rotational velocity ($\mathrm{km~s}^{-1}$) & HARPS spectral classification (\S\ref{sec:spec}) \\
        $F_{\rm bol}$ & $3.599 \pm 0.012 \times 10^{-8}$ & Bolometric flux (erg~s$^{-1}$~cm$^{-2}$) & SED analysis (\S\ref{sec:sed}) \\
        $T_{\rm eff}$ & $4675\pm53$ & Effective temperature (K) & Isochrones analysis (\S\ref{sec:isochrones}) \\
        $\log g$ & $4.654\pm0.057$ & Surface gravity (cgs) & Isochrones analysis \\
        $[$Fe/H$]$ & $-0.343\pm0.059$ & Metallicity (dex) & Isochrones analysis \\   
        $R_{\star}$ & $0.630\pm0.027$ & Stellar radius ($R_{\odot}$) & Isochrones analysis \\
        $M_{\star}$ & $0.653\pm0.028$ & Stellar mass ($M_{\odot}$) & Isochrones analysis \\
        $\tau_\star$ & $6.88\pm4.27$ & Stellar age (Gyr) & Isochrones analysis \\
        \hline
        \end{tabular}
    \label{tab:star}
\end{table*}

\subsection{Stellar Abundances}\label{sec:hypatia}
A total of 24 elements have been measured within the photosphere of HD~101581 per the Hypatia Catalog\footnote{All abundance measurements can be found online at \url{ www.hypatiacatalog.com}.} \citep{Hinkel14}. The data were compiled from four individual literature sources, where their values were renormalized to the same Solar scale so that the results were on the same baseline. When multiple sources determined the same elements within HD~101581, the median value was used and half of the range or spread in measurements is adopted as the error. A subselection of the elemental abundances for HD~101581 is shown in Table \ref{tab:star}, where the overwhelming conclusion is that HD~101581 is significantly deficient in all elements with respect to the Sun -- including $\alpha$-elements, odd-Z, iron-peak, beyond the iron-peak, and neutron-capture. Given that the planets are all Earth-size, we convert to molar ratios and find that Fe/Mg = 0.66 and Si/Mg = 1.10, which lies higher than at least 1$\sigma$ with respect to other Hypatia stars \citep[e.g.,][]{Hinkel18}.

In addition, stellar parameters were compiled via literature sources found within the Hypatia Catalog, since the determination of stellar abundances requires modeling the stellar atmosphere. The compiled $T_{\rm eff} = 4646\pm82$ K, $\log{g} = 4.46\pm0.29$, $R_{\star} = 0.64\pm0.01~R_{\odot}$, $M_{\star} = 0.71\pm0.04~M_{\odot}$ agree with the TICv8.2 determined values to within the reported uncertainties. 

\subsection{Spectral Characterization}\label{sec:spec}

We used multiple methods to measure stellar atmospheric parameters from the PUCHEROS+ spectra. First, the CERES pipeline performs an estimation of the stellar atmospheric parameters by cross-correlating each spectrum with a grid of synthetic models adapted from \citet{Coelho2005}. Averaging over the 14 PUCHEROS+ spectra gives $T_{\mathrm{eff}} = 4590\pm150$ K, $\log{g} = 4.1\pm0.5$, and metallicity of [Fe/H] $= -0.5$ dex. We further analyzed the co-added spectra using the ZASPE code \citep{ZASPE}, which compares the spectra to a grid of synthetic models generated from the ATLAS9 model atmospheres \citep{ATLAS9}, to find $T_{\mathrm{eff}} = 4743\pm100$ K, $\log{g} = 4.19\pm0.30$, and [Fe/H] = $-0.47\pm0.05$ dex. The PUCHEROS+ spectra also constrain the stellar projected rotation velocity to $v\sin{i} < 10$ km s$^{-1}$, with measurements of lower $v\sin{i}$ limited by the resolution of the spectrograph.

\citet{Perdelwitz2024} used the SPECIES codebase \citep{Soto2018} to extract parameters from the archival HARPS spectra for 3612 stars including HD 101581, finding $T_{\mathrm{eff}} = 4709\pm62$ K, $\log{g} = 4.11\pm0.13$, [Fe/H] = $-0.58\pm0.05$ dex, and $v\sin{i} = 2.47\pm0.30$ km s$^{-1}$. The $T_{\mathrm{eff}}$ and $\log{g}$ values from the Hypatia, PUCHEROS+, and HARPS results all agree within $1\sigma$, while the HARPS-based measurement of metallicity is slightly lower ($1.6\sigma$) than that from PUCHEROS+. Both analyses conclude that HD 101581 is a metal-poor star.

\subsection{SED Analysis}\label{sec:sed}

\begin{figure}
\centering
\includegraphics[width=\linewidth,trim=80 70 50 50,clip]{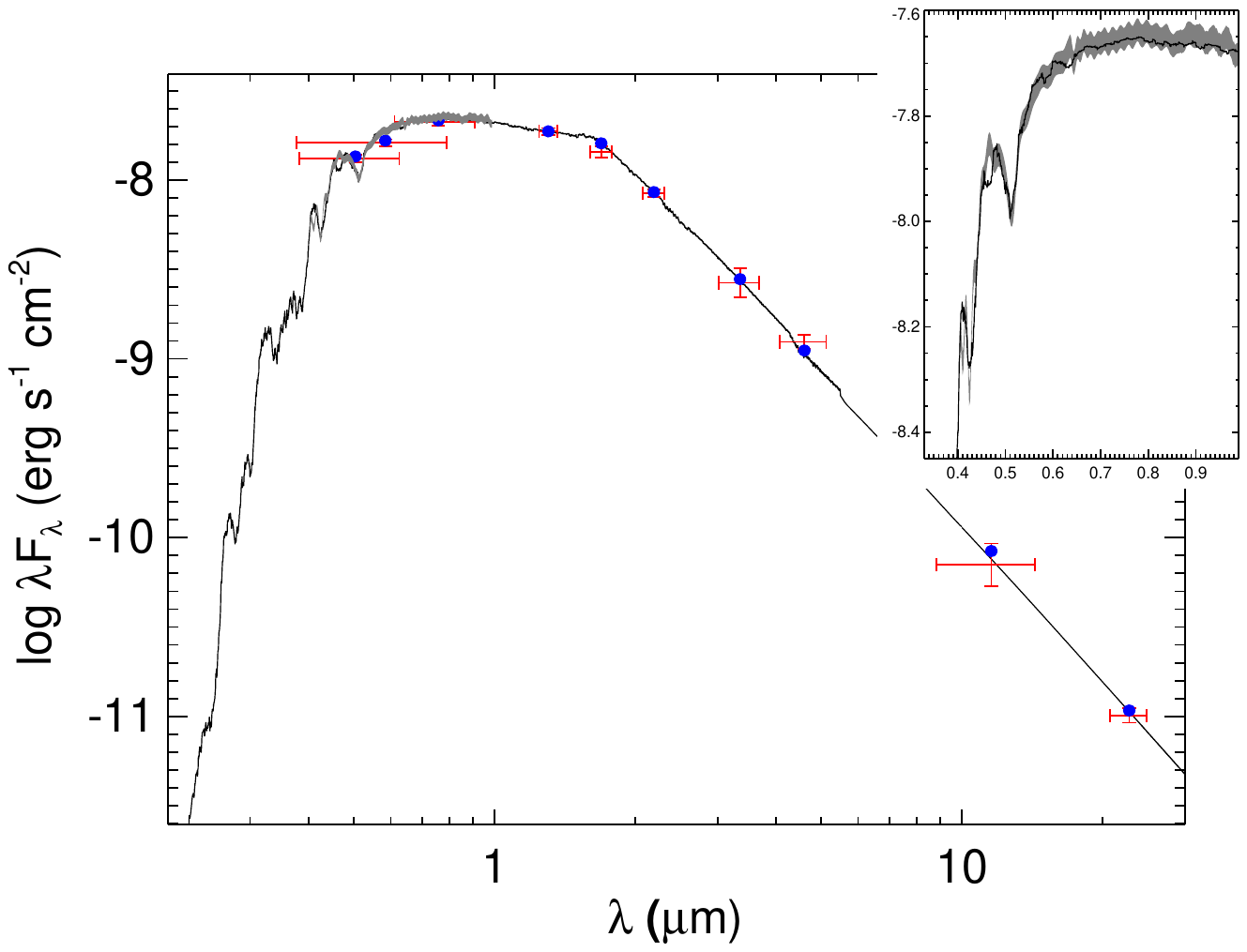}
\caption{Spectral energy distribution of HD~101581. Red symbols represent the observed photometric measurements, where the horizontal bars represent the effective width of the passband. Blue symbols are the model fluxes from the best-fit PHOENIX atmosphere model (black). The absolute flux-calibrated {\it Gaia\/} spectrum is shown as a grey swathe in the inset figure. \label{fig:sed}}
\end{figure}

We performed an analysis of the broadband spectral energy distribution (SED) of the star together with the {\it Gaia\/} DR3 parallax, in order to determine an empirical measurement of the stellar radius \citep{Stassun:2016,Stassun:2017,Stassun:2018}. We extracted the $JHK_S$ magnitudes from \twomass, the W1--W4 magnitudes from \wise, and the $G_{\rm BP} G_{\rm RP}$ magnitudes from \gaia. We also utilized the absolute flux-calibrated \gaia\ spectrum. Together, the available photometry spans the full stellar SED over the wavelength range 0.4--20~$\mu$m (see Figure~\ref{fig:sed}). 

We performed a fit using PHOENIX stellar atmosphere models \citep{Husser:2013}, adopting from the Hypatia analysis the effective temperature ($T_{\rm eff}$), metallicity ([Fe/H]), and surface gravity ($\log g$). We fitted for the extinction $A_V$, limited to the maximum line-of-sight value from the Galactic dust maps of \citet{Schlegel:1998}. The resulting fit (Figure~\ref{fig:sed}) has $A_V = 0.06 \pm 0.03$, with a reduced $\chi^2$ of 1.0. Integrating the (unreddened) model SED gives the bolometric flux at Earth, $F_{\rm bol} = 3.599 \pm 0.012 \times 10^{-8}$ erg~s$^{-1}$~cm$^{-2}$. Taking the $F_{\rm bol}$ together with the {\it Gaia\/} parallax directly gives the bolometric luminosity, $L_{\rm bol} = 0.18332 \pm 0.00059$~L$_\odot$. The Stefan-Boltzmann relation then gives the stellar radius, $R_\star = 0.662 \pm 0.023$~R$_\odot$. In addition, we estimated the stellar mass using the empirical relations of \citet{Torres:2010}, giving $M_\star = 0.64 \pm 0.04$~M$_\odot$. 


\subsection{Isochrones Analysis}\label{sec:isochrones}

To derive a self-consistent set of physical parameters for the host star, we fit the observed properties of HD~101581 to MIST evolutionary models \citep{Dotter2016, Choi2016} using the stellar model grid package \texttt{isochrones} \citep{Morton2015}. We defined a single-star model using parallax from Gaia DR3, observed magnitudes ($JHK_{s}$, $GG_{\mathrm{BP}}G_{\mathrm{RP}}$, W1--W3), and $T_{\mathrm{eff}}$ and [Fe/H] from the Hypatia literature analysis. Following the convention of \citet{Eastman2019}, we inflated the Gaia magnitude uncertainties to 0.02.

The fit parameters were equivalent evolutionary phase, age, metallicity, distance, and extinction. We used the default priors from \texttt{isochrones}, except for a broad flat prior for metallicity. The parameter space was explored with the \texttt{emcee} ensemble sampler \citep{ForemanMackey2013, Goodman2010} with 100 walkers for 30,000 steps, at which point the fit converged based on the chain being more than 50 times longer than the estimated autocorrelation time. The first 1,000 steps were discarded as burn-in. Based on the \texttt{isochrones} fit, HD~101581 is an old, metal-poor K-dwarf (age = $6.88\pm3.74$ Gyr, [Fe/H] = $-0.344\pm0.059$, $T_{\mathrm{eff}} = 4675\pm53$ K, log$g$ (cgs) = $4.654\pm0.012$) with mass $M_{\star} = 0.653\pm0.015~M_{\odot}$ and radius $R_{\star} = 0.630\pm0.005~R_{\odot}$.

The above uncertainties are under-estimated because they do not take into account differences in assumptions between different stellar grid models. We used the \texttt{kiauhoku} package \citep{Claytor2020} to estimate the properties of HD~101581 based on interpolating effective temperature, luminosity, and metallicity to stellar grid models from YREC \citep{Demarque2008}, MIST, DSEP \citep{Dotter2008}, and GARSTEC \citep{WeissSchlattl2008} codes and databases. We find maximum differences between models of $\approx4\%$, $4\%$, $1\%$, and $30\%$ in mass, radius, log$g$, and age, respectively. These systematic errors have been added in quadrature to the fit uncertainties and are reflected in Table \ref{tab:star}.

We defer to the \texttt{isochrones}-fitted parameters when requiring stellar parameters in the remainder of this paper.

\subsection{Stellar Activity and Rotation}

HD~101581 has a chromospheric emission parameter $\log{R^{\prime}_{HK}}$ of  $-4.759$ from \citet{Gray2006} and $-4.70$ from \citet{Jenkins2006}, consistent with a low-activity star. The SPOC light curve (\S\ref{sec:observations}) also does not show any indications of flares. 

We independently estimated $\log{R^{\prime}_{HK}}$ from the $S$-index and $B-V$ color of the star following the method outlined by \citet{Noyes1984}. Based on the mean $S$-index of $0.388$ from the PFS observations and color of $B - V = 1.095$ from UCAC4 \citep{UCAC4}, we found $\log{R^{\prime}_{HK}} = -4.85$. The overall average of all three values, $\log{R^{\prime}_{HK}} = -4.77$, suggests a stellar rotation period of $P_{\text{rot}} \sim 30$ days based on the relationship for K stars provided by \citet{Mascareno2016}, which is much longer than all of the planet periods. A $\sim30$-day rotation signal may be visible in the PDCSAP light curve (Figure \ref{fig:lc}), but measuring rotation periods longer than $\sim13$ days using TESS is challenging due to the telescope orbit \citep[e.g.][]{Hedges2020, Claytor2022}.

\begin{figure}
    \includegraphics[width=\linewidth]{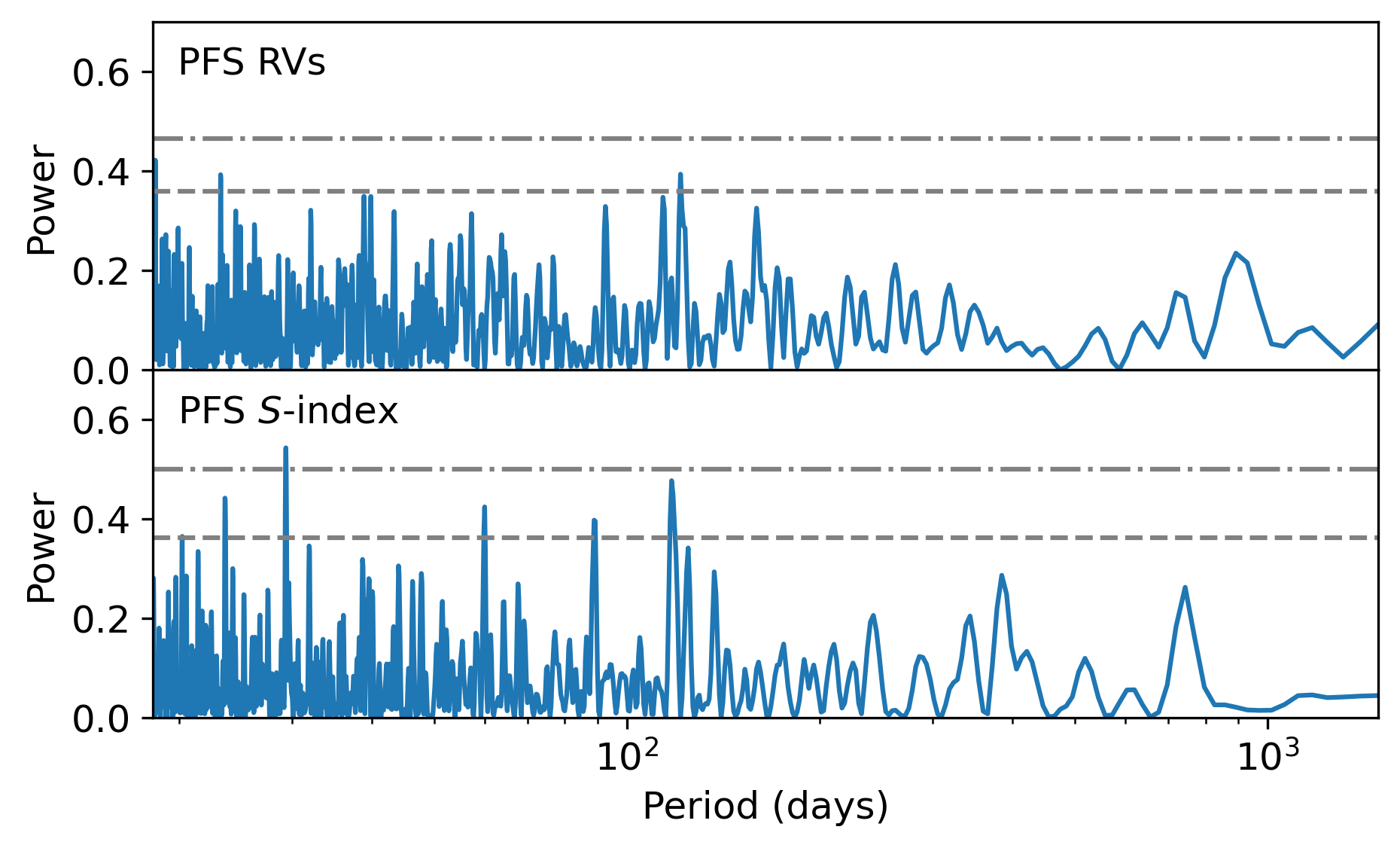}
    \caption{GLS periodograms of the RVs and $S$-indices measured by PFS. The horizontal lines mark 1\% (dotted line) and 0.1\% (dash-dotted line) false alarm probability levels. Only one peak is stronger than the 0.1\% level, at $P = 29.4$ days in the power spectrum for $S$-indices, which likely corresponds to the $P_{\text{rot}} \sim 30$-day rotation period estimated through relation to the chromospheric emission parameter $\log{R^{\prime}_{HK}} = -4.77$.}\label{fig:periodogram}
\end{figure}

We also produced generalized Lomb-Scargle (GLS) periodograms \citep{Zechmeister2009} from the PFS RVs and $S$-indices to search for significant periodicities (Figure \ref{fig:periodogram}). Frequencies were sampled up to 10 times the average Nyquist frequency of the observations. Only one peak in the $S$-index power spectrum is above a $0.1\%$ false alarm probability (FAP) level, at $P = 29.4$ days, which is consistent with the expected stellar rotation period.

\section{Planet Parameters}\label{sec:planets}

\begin{figure}
    \centering
    \includegraphics[width=\linewidth]{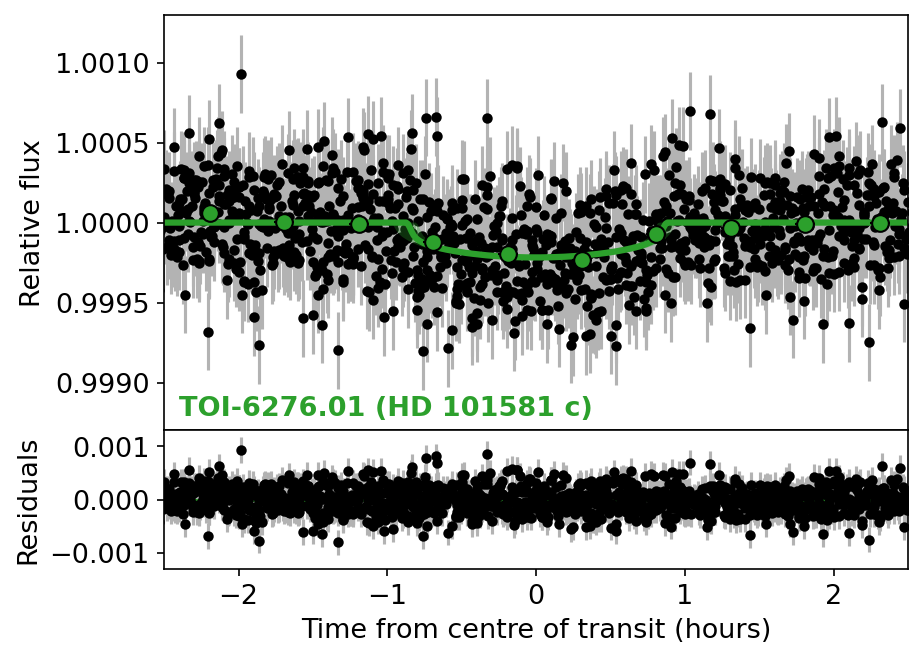}
    \includegraphics[width=\linewidth]{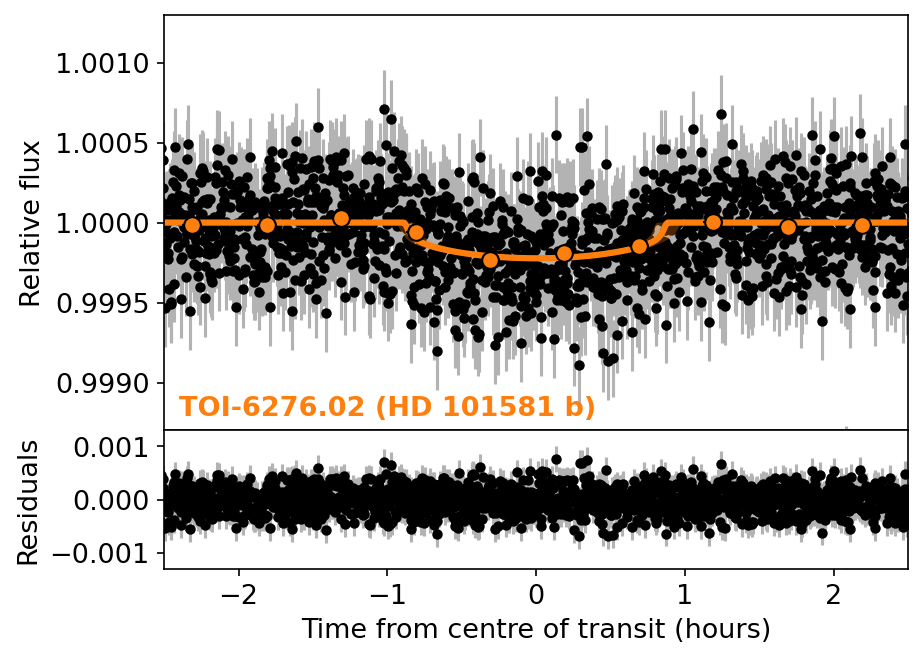}
    \includegraphics[width=\linewidth]{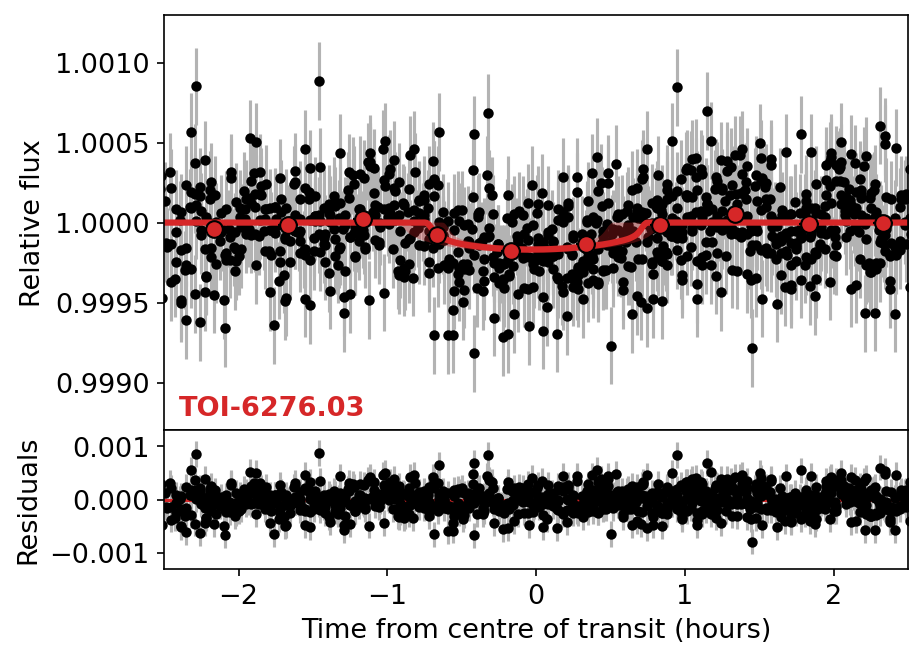}
    \caption{Plots of the transit model fits, with residuals after subtracting the median models provided in the lower panel of each phase diagram. Black points show observations with offsets subtracted and jitter terms added in quadrature with uncertainties, while colored circles represent binned data. Colored lines represent median model values. In each planet's phase diagram, the best-fit models for the other two planets have been subtracted from the data.}
    \label{fig:fits}
\end{figure}

To estimate the physical and orbital parameters of each of the planet candidates, we fit a three-planet transit model to the detrended multi-sector light curve using \texttt{exoplanet} \citep{exoplanet:zenodo}. We assumed circular Keplerian orbits and parameterized the transit model by orbital period ($P$), transit epoch ($T_{0}$), planet-to-star radius ratio ($R_{p}/R_{\star}$), and impact parameter ($b$), with stellar radius and mass fixed to the median \texttt{isochrones}-fitted values ($R_{\star} = 0.630~R_{\odot}$, $M_{\star} = 0.653~M_{\odot}$. We adopted a quadratic limb-darkening law parameterized by $q_{1}$, $q_{2}$ from \citet{exoplanet:kipping13}, and fit for a flux offset as well as jitter term added in quadrature to the uncertainties of the SPOC observations. Uniform or normal priors were used for parameters as shown in Table \ref{tab:priors}. We sampled four chains for 2000 tuning steps and 2000 draw steps each, which converged according to the Gelman-Rubin convergence statistic for each parameter satisfying $\hat{r} < 1.01$ \citep{Gelman1992}. The corresponding posterior parameters are summarised in Table~\ref{tab:parameters}, with stellar parameter uncertainties from the \texttt{isochrones} results propagated through the uncertainties in derived parameters. The best-fit light curves are shown in Figure \ref{fig:fits}. The best fits indicate that HD~101581 hosts three Earth-size exoplanets ($R_{p} \sim 1~R_{\oplus}$) in co-planar orbits ($i \sim 88^{\circ}$).

\begin{table*}[]
    \centering
    \begin{tabular}{c|ccc|l}
        \hline\hline
        Parameter & TOI-6276.01 & TOI-6276.02 & TOI-6276.03 & Description \\
        & HD~101581 c & HD~101581 b & - & \\
\hline\hline
\multicolumn{5}{c}{\textit{Planet Parameters}} \\
\hline
$P$ & $6.20401_{-0.00044}^{+0.00054}$ & $4.46569_{-0.00032}^{+0.00029}$ & $7.8708_{-0.0011}^{+0.0016}$ & Orbital period (days) \\
$T_0$ & $3017.114_{-0.0024}^{+0.0023}$ & $3014.8496_{-0.0023}^{+0.0028}$ & $3018.5696_{-0.0072}^{+0.004}$ & Transit epoch (BJD - 2457000) \\
$R_p/R_\star$ & $0.0144_{-0.0008}^{+0.0007}$ & $0.0139_{-0.0006}^{+0.0006}$ & $0.0143_{-0.0013}^{+0.0015}$ & Planet-to-star radius ratio \\
$b$ & $0.713_{-0.057}^{+0.032}$ & $0.616_{-0.079}^{+0.036}$ & $0.854_{-0.051}^{+0.026}$ & Impact parameter \\
\hline
\multicolumn{5}{c}{\textit{Derived Parameters}} \\
\hline
$R_{p}$ & $0.990_{-0.070}^{+0.070}$ & $0.956_{-0.063}^{+0.061}$ & $0.982_{-0.098}^{+0.114}$ & Planet radius ($R_{\oplus}$)\\
$i$ & $87.93_{-0.15}^{+0.19}$ & $87.78_{-0.2}^{+0.27}$ & $87.88_{-0.14}^{+0.15}$ & Inclination angle ($^{\circ}$) \\
$T_{\text{dur}}$ & $1.76_{-0.12}^{+0.15}$ & $1.77_{-0.12}^{+0.13}$ & $1.45_{-0.14}^{+0.19}$ & Transit duration\footnote{From first to last contact} (hrs)\\
$a$ & $0.0573_{-0.0009}^{+0.0009}$ & $0.046_{-0.0007}^{+0.0007}$ & $0.0671_{-0.001}^{+0.001}$ & Semi-major axis (AU) \\
$S$ & $52_{-5}^{+6}$ & $80_{-8}^{+9}$ & $38_{-4}^{+4}$ & Instellation flux ($S_{\oplus}$)\\
$T_{\text{eq}}$ & $747_{-20}^{+21}$ & $834_{-23}^{+23}$ & $690_{-19}^{+19}$ & Equilibrium temperature\footnote{Assuming albedo = 0} (K)\\
\hline
\multicolumn{5}{c}{\textit{Photometric Parameters}} \\
\hline
$\mu$ & \multicolumn{3}{c|}{ $-4_{-4}^{+4}$} & Flux offset (ppm) \\
$\sigma$ & \multicolumn{3}{c|}{ $143_{-4}^{+4}$} & Flux jitter (ppm)\\
$u_{1}$ & \multicolumn{3}{c|}{ $0.91_{-0.52}^{+0.45}$} & Limb-darkening coefficient 1\\
$u_{2}$ & \multicolumn{3}{c|}{ $-0.14_{-0.4}^{+0.55}$} & Limb-darkening coefficient 2\\
\hline
\end{tabular}
    \caption{Fitted transit model and planet parameters for the three candidates orbiting HD~101581.}
    \label{tab:parameters}
\end{table*}

We also ran an alternative fit letting the orbital eccentricity ($e$) and argument of pericenter ($\omega$) for each planet vary. We used the distribution from \citet{VanEylen2019}, appropriate for multi-planet systems, for the eccentricity prior. The argument of pericenter was constrained to the range $-\pi$ to $\pi$, and the sampling was performed in two-dimensional vector space ($\sqrt{e}\sin{\omega}$, $\sqrt{e}\cos{\omega}$) to avoid the sampler seeing a discontinuity at values of $\pi$. The results were consistent with the circular fit, and the eccentricities were consistent with zero, with 84th percentile (1$\sigma$) upper limits of 0.08, 0.09, and 0.09. Future RV observations could further constrain these eccentricities.

We next perform a simplistic model of the precision RV data (AAT, PFS, HARPS) in order to quantify upper limits on the masses of the three transiting planet candidates. We model the RV signals assuming zero eccentricity and enforcing Gaussian priors on $P$ and $T_0$ based on the values in Table~\ref{tab:parameters} for each companion. In order of increasing orbital period, we place upper limits on the RV semi-amplitudes of $K<(1.9, 2.0, 1.5)$~m~s$^{-1}$ and constraints on the planetary masses of $M_{p}<(3.6, 4.2, 3.6)~M_\oplus$ at 3$\sigma$ confidence. At this level of precision, the limits on planetary densities are not probative ($\rho\lesssim20$~g~cm$^{-3}$). Thus, despite the substantial amount of observations, the RV signals of the transiting planet candidates remain below our current detection limits.

\section{Statistical Validation}\label{sec:validation}

Lacking planet masses, we attempt to statistically validate the TOIs by ruling out possible false positive scenarios where the signals are not due to orbiting planets.

\subsection{TOI-6276 is Not an Eclipsing Binary System} \label{sec:eclipsing}

The first false positive scenario to consider is that the transit signals of HD~101581 are caused by an eclipsing stellar companion. We begin by using the \texttt{MOLUSC} framework \citep{MOLUSC} to constrain the range of unseen stellar companions that could produce the observed transits by simulating a large number of possible companions and eliminating those that should have been detected by \textit{Gaia} in astrometry, high-resolution imaging, and/or RVs. We generated one million companions to TOI-6276 with orbital inclinations forced to be consistent with an eclipsing system and compared each star's detectability with \textit{Gaia} DR3 astrometry, all three contrast curves, and the PFS, UCLES, and HARPS RV observations. Only 3.5\% of the simulated companions survived the comparison, 99\% of which had low masses ($< 0.074~M_{\odot}$). Companions with $P < 200$ days, including those at all the transiting planet periods, were entirely ruled out.

We can significantly improve on these limits by incorporating constraints from the long-term observations with precision RVs and \textit{Hipparcos-Gaia} astrometry \citep{Brandt2018, Brandt2021}. Across $>$20 years of observation both RVs and astrometry are constant at the level of a few m~s$^{-1}$, ruling out companions with orbital periods below $P<8000$~d down to Jupiter-like masses. We further attempt to fit linear acceleration terms to these observations, and measure values consistent with zero at high precision: $\frac{d RV}{dt}=-0.05\pm0.12$ m~s$^{-1}$~yr$^{-1}$, $\frac{d\mu_{\alpha}}{dt}=-0.17\pm0.15$ m~s$^{-1}$~yr$^{-1}$, and $\frac{d\mu_{\delta}}{dt}=-0.06\pm0.14$ m~s$^{-1}$~yr$^{-1}$. The absence of any significant accelerations argues against the presence of stellar-mass companions.

As this contemporaneously constrains the motion of the star in both radial and tangential directions, we may extrapolate to make mass constraints at wider separations. The expression for companion mass $M$ (in $M_\odot$) as a function of velocity change is

\begin{equation}
\label{equation:acceleration}
\begin{split}
M=5.342\times 10^{-6}\times~& d^2 \times \rho^2 \times \left(\frac{d TV}{dt}\right)^{-2} \times \\
&\left[\left(\frac{d RV}{dt}\right)^2 + \left(\frac{d TV}{dt}\right)^2\right]^{\frac{3}{2}}
\:,
\end{split}
\end{equation}
where $d$ is the stellar distance in parsecs, $\rho$ is the projected separation in arcseconds, and $TV$ represents the tangential velocity $=\sqrt{\mu_{\alpha}^2 + \mu_{\delta}^2}$ \citep[][equation 2]{Bowler2021}. As $d$ is known, we can convert the limits on the acceleration terms into a limit on companion masses as a function of projected separation.

\begin{figure*}
    \centering
    \includegraphics[width=\columnwidth]{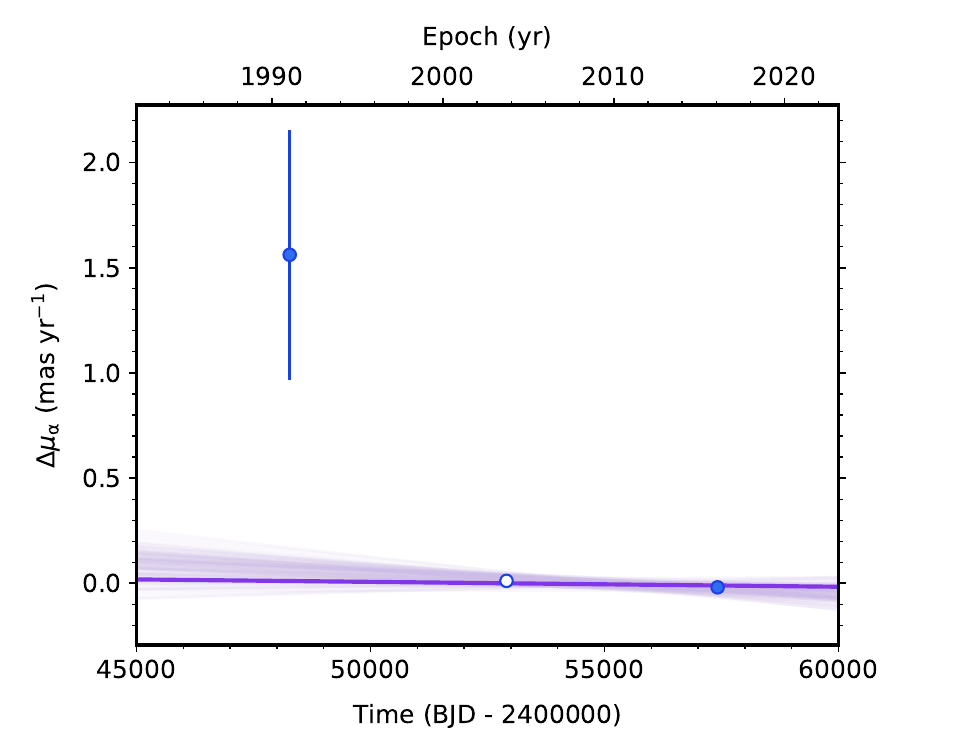}
    \includegraphics[width=\columnwidth]{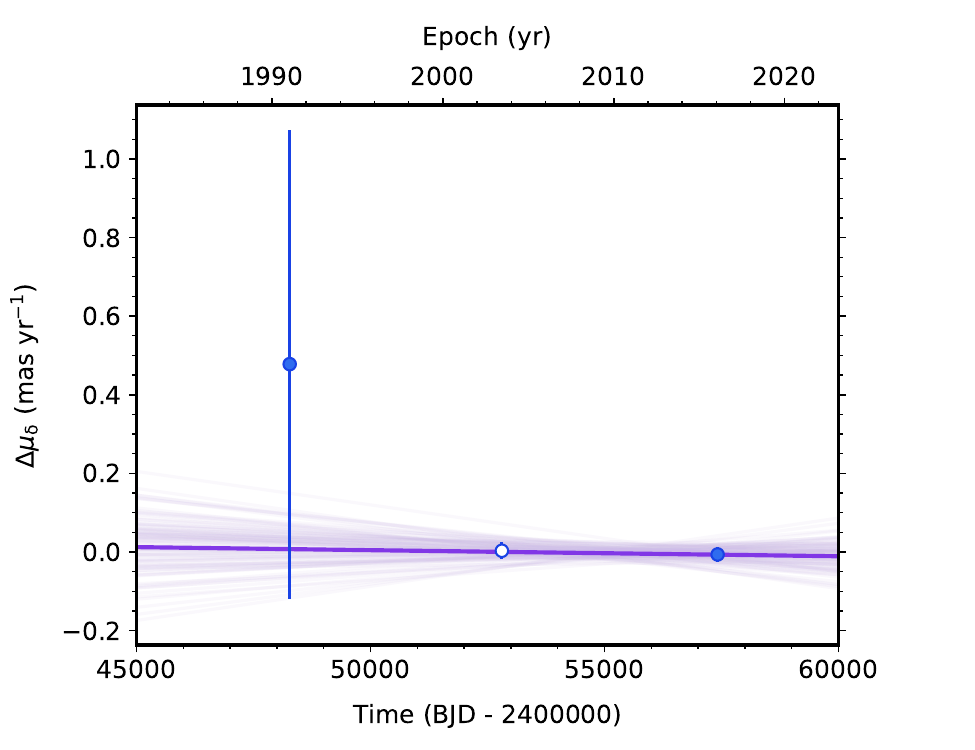}
    \includegraphics[width=\columnwidth]{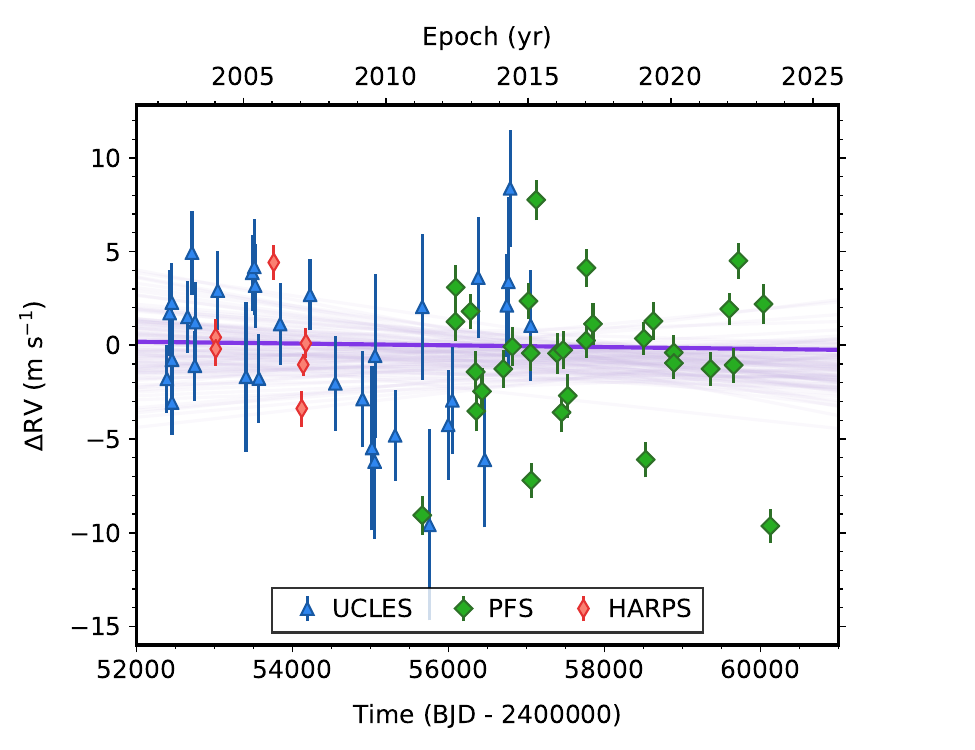}
    \includegraphics[width=\columnwidth]{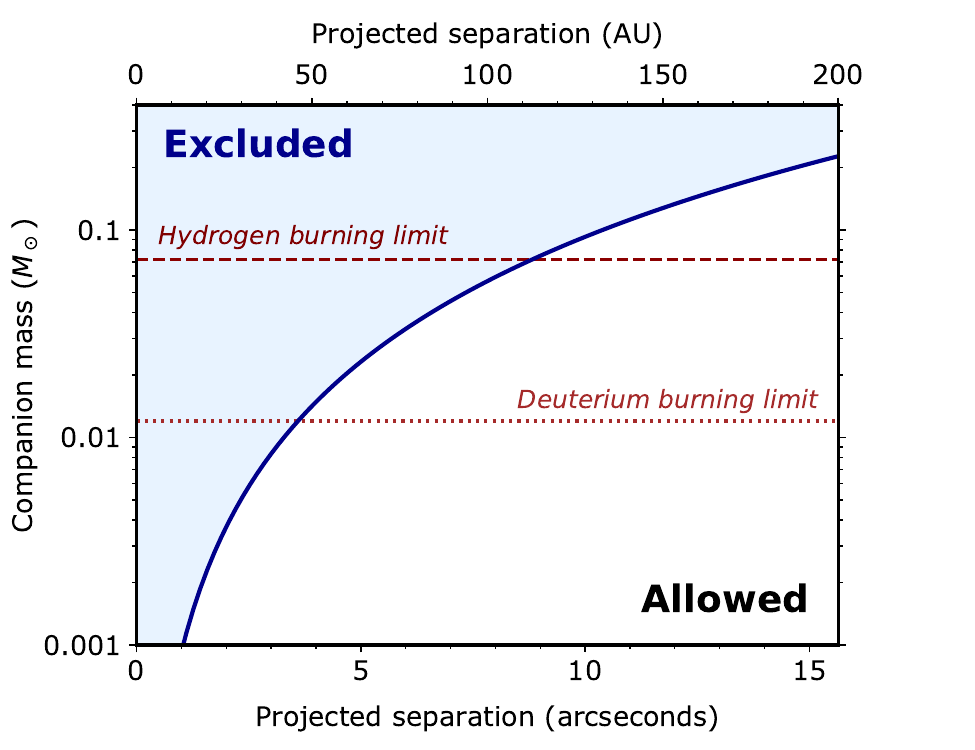}
    \caption{Linear acceleration model for HD~101581 in \textit{Hipparcos-Gaia} astrometry (right ascension, top left; declination, top right), and radial velocity (bottom left) across $>$20 years of observations. In each dimension we find that the acceleration terms are consistent with zero at high precision. We leverage this to achieve strong limits on the existence of bound unresolved companions as a function of projected separation (bottom right). Based on the velocities alone, stellar-mass companions to HD~101581 can be entirely excluded within $<$110~AU.}
    \label{fig:acceleration}
\end{figure*}

For HD~101581 we set 3$\sigma$ upper limits on the absolute accelerations of $|\frac{d RV}{dt}|<0.41$ m~s$^{-1}$~yr$^{-1}$ and $|\frac{d TV}{dt}|<0.63$ m~s$^{-1}$~yr$^{-1}$. We plot the resulting limits on wide companions in Figure~\ref{fig:acceleration}. We exclude the existence of any bound companions with masses above the hydrogen-burning limit ($\gtrsim$0.075~$M_\odot$) within a projected separation of $<$9 arcseconds from HD~101581 ($<$110~AU), wholly excluding any unresolved stellar companions to HD~101581. Within $<$3.5 arcseconds ($<$45~AU) we can even exclude any companions above the deuterium-burning limit ($>$13~$M_J$). On the weight of the rich observing record, the eclipsing binary scenario for HD~101581 can be excluded at high confidence.

\subsection{TOI-6276 is Not a Nearby or Background Eclipsing System}

The signals may not be associated with the target star, but rather due to a nearby or chance-aligned eclipsing binary (or transiting planet) system that contaminated the TESS photometry.

First, we limit the possible transit source locations using the difference image centroid offsets provided in the SPOC DV reports. TOI-6276.01 has a measured centroid offset of $11.4\pm2.7^{\prime\prime}$ ($4.3\sigma$) from the TIC position of HD~101581. While not within $3\sigma$ of the target star, the measured centroids are only visually near TIC~397362481 ($T = 17.7$), a star too faint to explain the transits. TOI-6276.02 and TOI-6276.03 have offsets of $6.6\pm4.1^{\prime\prime}$ ($1.6\sigma$) and $8.9\pm4.9^{\prime\prime}$ ($1.8\sigma$), placing them within $3\sigma$ of HD~101581. We believe that image saturation due to the brightness of HD~101581 ($T = 6.7$) affects the measured offsets, and thus all the signals are consistent with being on-target. Our LCOGT observations also cleared the field out to 2.5$^{\prime}$ at the ephemerides of TOI-6276.01 and TOI-6276.02, limiting possible sources to stars not resolved by seeing-limited photometry.

We further rule out potential NEBs by placing upper limits on the magnitude of a fully blended star that may cause the observed transit signals. In the case of photometric contamination by blended light, the observed TESS transit depth ($\delta_{\mathrm{obs}}$) is given by:

\begin{equation}
    \delta_{\mathrm{obs}} \simeq \delta_{\mathrm{EB}} \frac{f}{1 + f},
\end{equation}

\noindent where $\delta_{\mathrm{EB}}$ is the depth of the eclipse in the absence of a blend, $f$ is the flux ratio $f \equiv F_{\mathrm{EB}}/F_{s}$, $F_{\mathrm{EB}}$ is the flux of the contaminating EB, and $F_{s}$ is the flux of HD~101581. We may also place an upper bound on $\delta_{\mathrm{EB}}$ using Eqn.~21 from \citet{Seager2003}, assuming an equatorial eclipse ($b = 0$):

\begin{equation}
    \delta_{\mathrm{EB}} \leq \frac{(1 - t_{F}/t_{T})^{2}}{(1 + t_{F}/t_{T})^{2}},
\end{equation}

\noindent where $t_{F}$ is the duration of the flat part of the transit (time from end of ingress to start of egress) and $t_{T}$ is the total duration of the transit (time from start of ingress to end of egress). Combining these equations gives a lower bound on $f$ as a function of transit observables $\delta_{\mathrm{obs}}$ and $t_{F}/t_{T}$. We then use the relation between stellar magnitudes and fluxes ($\Delta m = -2.5\log_{10}{f}$) to find the largest difference in magnitude from HD~101581 that a star could have to explain the observed transits.

We estimated $\delta_{\mathrm{obs}}$, $t_{F}$, and $t_{T}$ for each planet by applying the equations described by \citet{Seager2003} to the transit model fit results, finding box-shaped transits with $t_{F}/t_{T} \sim 0.96$, $0.97$, and $0.93$ for TOI-6276.01, .02, and .03, respectively. This rules out blended stars fainter than $\Delta T = 2.2$, $1.6$, and $4.2$ at the 99.9th percentile level. All known stars within $1^{\prime}$ have $\Delta T > 7$, and therefore are too faint to be possible NEB sources. Our high-resolution imaging observations from Gemini-S/Zorro also revealed no bright companions ($\Delta m < 5$) down to a separation of $\sim0.2^{\prime\prime}$ in the field close to the time of the TESS observations. 

Finally, we see that the 2023 location of HD~101581 (corresponding to the last TESS observations) is clear of stars in the 1977 archival SERC-J Survey image down to the limiting magnitude of $B \approx 23$, further ruling out chance-aligned background EBs as false positive sources.

In summary, we rule out contamination from an NEB as the source of the transit signals based on SPOC centroid offsets, seeing-limited and high-resolution imaging follow-up, and archival images.

\subsection{TOI-6276 is Not a Hierarchical Triple}

The final scenario involves an EB gravitationally bound to HD~101581. This type of EB would contaminate the TESS photometry in the same way as the NEB case, but would evade detection due to its close proximity to HD~101581. In Section~\ref{sec:eclipsing} we used the absence of any significant RV or astrometric acceleration to rule out any bound stellar-mass companions within $<9^{\prime\prime}$. This conclusion is also supported by the lack of detectable companions in the VLT/NACO high-resolution images (out to $4^{\prime\prime}$ away) and Gemini-S/Zorro images (out to $1.2 ^{\prime\prime}$ away). Any stars at more distant separations would be easily detectable in imaging, but archival images of HD 101581 from the SERC-I, SERC-J, AAO-SES, and LCO surveys do not suggest the presence of bound EBs. The absence of any bound stellar companions therefore rules out this false positive scenario.


\subsection{False Positive Probabilities}

We used the \texttt{TRICERATOPS} statistical validation tool \citep{TRICERATOPS} with the properties of surviving stellar companions from \texttt{MOLUSC} provided as inputs to quantify each candidate's false positive probabilities (FPPs) and nearby false positive probabilities (NFPPs). \texttt{TRICERATOPS} was run 10 times to find the mean and standard deviation of FPPs and NFPPs. Based on their low FPP and NFPP values ($< 10^{-3}$; Table \ref{tab:fpps}), we consider TOI-6276.01 and TOI-6276.02 to be statistically validated and refer to them as HD~101581 c and HD~101581 b, respectively, based on their distance from the star. 

\begin{table*}[t!]
    \centering
    \begin{tabular}{cccccc}
    \hline\hline
        TOI & Name & Orbital period & SNR & FPP & NFPP \\
        \hline
        TOI-6276.01 & HD~101581 c & 6.204 & 12.4 & $2.4\pm0.6\times10^{-6}$ & $1.1\pm0.1\times10^{-8}$ \\
        TOI-6276.02 & HD~101581 b & 4.466 & 15.6 & $2.5\pm5.0\times10^{-4}$ & $1.4\pm0.3\times10^{-8}$ \\
        TOI-6276.03 & - & 7.871 & 7.9 & $0.010\pm0.003$ & $9.0\pm0.8\times10^{-5}$ \\
        \hline
    \end{tabular}
    \caption{False positive probabilities (FPPs) and nearby false positive probabilities (NFPPs) computed by \texttt{TRICERATOPS} for the transiting planets in the HD~101581 system. While TOI-6276.03 meets the criteria for statistical validation against astrophysical false positive scenarios, we do not yet consider it validated because of its low transit SNR.}
    \label{tab:fpps}
\end{table*}

TOI-6276.03 has a higher FPP value of $\mathrm{FPP} = 0.01$, which does not satisfy criteria for statistical validation (commonly adopted as $< 0.01$). While a multiplicity boost due to its membership in a multi-planet system would sufficiently reduce the FPP below this threshold \citep{Lissauer2012, TOI}, the relatively weak transit signal (SNR $\sim8$) is an additional reason why we do not consider it a statistically validated planet. \texttt{TRICERATOPS} only considers astrophysical false positive scenarios and does not consider the possibility of false alarms, such as those caused by instrumental systematics or intrinsic stellar variability, and so it should not be used to validate low-SNR planets. Re-detection in a future TESS sector or with another telescope would increase confidence in the planet scenario. HD~101581 will be next observed in Sector 90 (2025 March 12 -- April 9) based on the \texttt{tess-point} high precision TESS pointing tool \citep{tesspoint}, which should improve the SNR of all signals by a factor of $\sim\sqrt{3/2} = 1.22$.

\section{Discussion}\label{sec:discussion}

\subsection{Planet Mass, Radius, and Mean-motion Resonance}
HD~101581 is a metal-poor K dwarf hosting two validated terrestrial planets, HD~101581 b and c, and a third Earth-size candidate, TOI-6276.03. Adopting mass-radius relations appropriate for planets with $R_{p} < 1.23~R_{\oplus}$ from \citet{ChenKipping2017},

\begin{equation}
    \frac{M_{p}}{M_{\oplus}} = 0.9718\bigg(\frac{R_{p}}{R_{\oplus}}\bigg)^{3.58},
\end{equation}

\noindent we find that the three planets have predicted masses of $M_{p} \approx 0.83_{-0.18}^{+0.21}~M_{\oplus}$, $0.94_{-0.22}^{+0.26}~M_{\oplus}$, and $0.91_{-0.29}^{+0.44}~M_{\oplus}$, respectively.

Alternatively, we estimate the planet masses assuming that their iron-to-silicate mass fraction ($f_{\rm iron}^{\rm planet}$), which is defined as the mass of Fe divided by the total mass of Fe and Si-bearing species (MgSiO$_3$, Mg$_2$SiO$_4$, and SiO$_2$), equals the stellar value ($f_{\rm iron}^{\rm star}$). Using this definition, \citet{adibekyan_compositional_2021} found a statistically significant correlation between $f_{\rm iron}^{\rm planet}$ and $f_{\rm iron}^{\rm star}$ for rocky planets. Following their methodology, we find that $f_{\rm iron}^{\rm star}$ of metal-poor HD~101581 is $0.29^{+0.10}_{-0.09}$ (using metallicity data from the Hypatia Catalog). Assuming a fully differentiated two-layer structure, where all of the iron in a planet resides in its core and all of the silicates are stored within its mantle in the form of MgSiO$_3$, and a mean iron-to-silicate mass fraction of 0.29, we find that HD~101581 b, HD~101581 c, and TOI-6276.03 have masses of $M_p\approx0.84^{+0.20}_{-0.17}~M_\oplus$, $0.94^{+0.26}_{-0.21}~M_\oplus$, and $0.92^{+0.41}_{-0.29}~M_\oplus$, respectively. These masses are almost identical to those estimated from the \citet{ChenKipping2017} mass-radius relations.

Similar to most Kepler multiplanet systems, the HD~101581 system follows the ``peas-in-a-pod'' trend where planets in the same system tend to be similarly sized and regularly spaced \citep{Weiss2018, Millholland2017}. The planet radii are remarkably uniform. \citet{Weiss2023} defined the fractional dispersion of the planet radii within a given system as $\sigma_{\mathcal{R}} = \mathrm{Var}\{\log_{10}(R_{p,j}/R_{\oplus})\}$, where $j$ indexes over the planets in the system. Using this metric, $\sigma_{\mathcal{R}} = 0.007$, which is smaller than any of the Kepler systems with four or more planets considered in \citet{Weiss2023}. As for orbital spacing, the inner and outer pairs are separated by only $\Delta = 15.7$ and $11.3$ mutual Hill radii, slightly more packed than the typical $\Delta \sim 20$ spacing among Kepler multis \citep{Weiss2018}. This tight spacing implies that the orbits of the planets most likely have low eccentricity, as more eccentric orbits would likely prove dynamically unstable \cite[e.g.,][]{PuWu2015}.

The inner and outer planet pairs also have period ratios of 1.389 and 1.269, respectively. These period ratios place the planet pairs close to the 4:3 and 5:4 first-order mean-motion resonances (MMRs), respectively, motivating a search for transit timing variations (\S\ref{sec:TTVs}) and indications of resonant librations (\S\ref{sec:resonance}).

\subsection{Transit Timing Variations}\label{sec:TTVs}

Proximity to resonance enables the detection of planets perturbing each other, seen through transit timing variations (TTVs). TTVs can be used to infer planetary masses and orbital eccentricities, as well as being used to detect the presence of additional, non-transiting planets \cite[e.g.,][]{Agol2018}. If the planets are in low-order MMRs, then the 4:3 and 5:4 MMR pairs orbiting HD~101581 may have TTV super-periods of 38 and 105 days, and amplitudes on the order of a few minutes based on formulae described by \citet{Lithwick2012}. 

\begin{figure}
    \includegraphics[width=\linewidth]{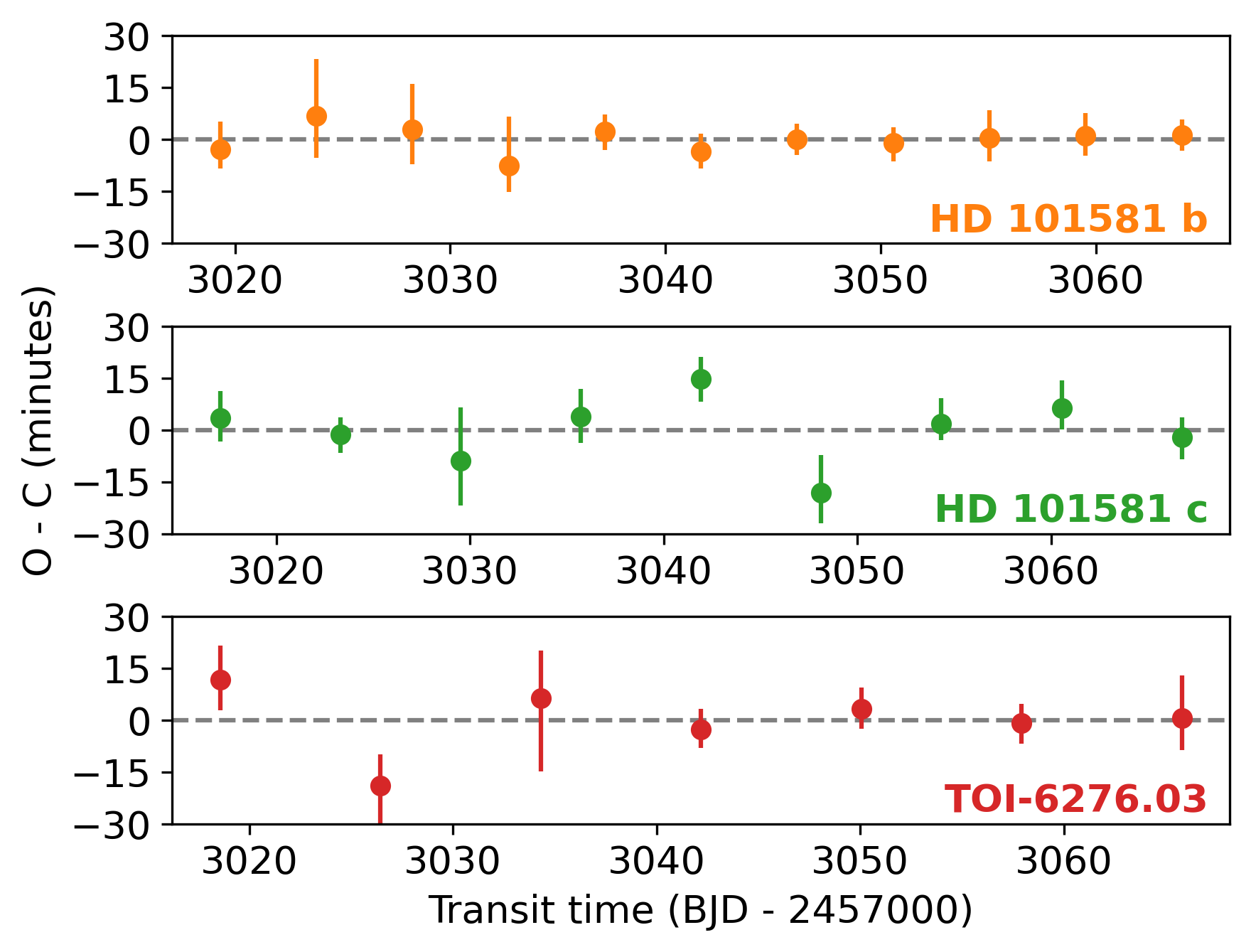}
    \caption{Differences between observed and calculated transit times (assuming linear ephemerides) for each planet in the TOI-6276 system. Each datapoint corresponds to the median with 16th to 84th percentile error bars. The transits of HD~101581 b are consistent with no variation, though we potentially see variations with a $\sim15$-minute amplitude for HD~101581 c. We also see $> 1\sigma$ variations among the first two transits of TOI-6276.03, but it is less clear if these are reliable given the low SNR of the transits.}\label{fig:ttvs}
\end{figure}

We searched for evidence of these TTVs by fitting the transit model described in \S\ref{sec:planets} to each individual transit. A normal prior was used for each transit time, with the mean set to the global best-fit transit time assuming linear ephemerides and standard deviation set to 30 minutes. The rest of the parameters were initialized following the global fit model, and sampling parameters similarly followed the global fit model. The resulting TTVs are shown in Figure \ref{fig:ttvs}.

The measured TTVs have average $1\sigma$ uncertainties of 8 minutes, which is larger than the expected TTV amplitudes. It is therefore no surprise that we do not see TTVs among the transits of HD~101581 b, and while we potentially see $> 1\sigma$ variations among the first two transits of TOI-6276.03, it is unclear if these are reliable given the low SNR of the transit signal. However, HD~101581 c potentially features variations with a $\sim15$-minute amplitude on a $\sim20$-day period. Additional photometry will be needed to better constrain the transit ephemerides and confirm the TTV periods and amplitudes, especially over longer timescales. 

\subsection{Resonance analysis}\label{sec:resonance}

Given the apparent proximity to the 4:3 and 5:4 resonances, we used $N$-body integrations to check whether the planets are indeed undergoing resonant librations. We drew 500 samples randomly from the two posterior distributions described in Section \ref{sec:planets}, one assuming circular orbits and the other allowing the eccentricities to float. We used the Wisdom-Holman \texttt{WHFast} integrator \citep{WisdomHolman1991, Rein2015} in REBOUND \citep{Rein2012} to evolve the system for 100 years. We then tracked the critical resonant angles associated with the 4:3 MMR for planets b and c ($\phi_{12,1} = 4\lambda_2 - 3\lambda_1 - \varpi_1$ and
$\phi_{12,2} = 4\lambda_2 - 3\lambda_1 - \varpi_2$) and with the 5:4 MMR for planet c and the planet candidate TOI-6276.03 ($\phi_{23,2} = 5\lambda_3 - 4\lambda_2 - \varpi_2$ and
$\phi_{23,3} = 5\lambda_3 - 4\lambda_2 - \varpi_3$). Here, $\lambda_i$ and $\varpi_i$ are the mean longitude and longitude of periapse of planet $i$. Physically, the critical resonant angles measure the evolution of the planetary conjunctions with respect to the pericenters of the two orbits.

For planet pairs actively participating in a mean-motion resonance, the critical resonant angles undergo bounded oscillations (or ``librations'') about their equilibria, such that the resonant libration amplitude is less than $180^{\circ}$  \citep{JensenMillholland2022}. We used two different approaches to numerically estimate the amplitude: $A_{\mathrm{lib}} = 0.5(\max \phi - \min \phi)$ and $A_{\mathrm{lib}} = \sqrt{\frac{2}{N}\sum_i{(\phi - \bar{\phi}})^2}$, where $N$ is the number of simulation data points and $\bar{\phi}$ is the average value of $\phi$. The latter definition is appropriate for approximately sinusoidal oscillations \citep{Millholland2018}.

Across all of our simulations, we did not find any indication of bounded librations of the critical resonant angles in either the 4:3 MMR for planets b and c or the 5:4 MMR for planet c and planet candidate TOI-6276.03. The first measure of the libration amplitude was $A_{\mathrm{lib}} \approx 180^{\circ}$ for all resonant angles and all simulations. However, we found that the second measure yielded $A_{\mathrm{lib}} < 180^{\circ}$ in some cases, which indicates that the critical resonant angles are often concentrated near $0^{\circ}$ but not librating. This is expected for planet pairs that are near but not in resonance. In summary our simulations suggest that the orbits of the planets (Table \ref{tab:parameters}) are not currently trapped in mutual mean motion resonance. The best-fit orbital configuration of the system is not consistent with a resonant chain. However, this analysis used estimated planet masses (from the \citet{ChenKipping2017} mass-radius relations), and it is possible that future analyses of the system using precise mass measurements may revisit this finding.

\subsection{Dynamical Stability}

We used the SPOCK stability classifier \citep{Tamayo2020} to assess the likely long-term stability of the system. SPOCK is a machine learning classifier trained with 100,000 compact 3-planet systems to predict the probability that a system will remain dynamically stable for $10^{9}$ orbits of the innermost planet. We ran the SPOCK classifier on 1000 random draws from the fit posteriors assuming circular orbits. All draws resulted in a stability probability of at least 50\%, and 95\% of draws resulted in a stability probability of more than 90\%, allowing us to conclude that this system is likely stable.

\subsection{Potential for Follow-Up}\label{sec:followup}

The apparent brightness of the host star ($V = 7.77$) and multiplicity of the transiting exoplanet system make HD~101581 an intriguing target for further characterization and comparative planetology.

\subsubsection{Radial Velocity Observations}

RV follow-up would be able to confirm all three planets through measurements of their masses, as well as place significantly stronger constraints on their orbital eccentricities. These observations would require sensitivity to semi-amplitudes of $K \approx (43, 43, 39)$ cm s$^{-1}$ at the predicted planet masses of $M_{p} \approx (0.83, 0.94, 0.91)~M_{\oplus}$, which is possibly within the reach of extreme precision RV (EPRV) instruments in the Southern Hemisphere such as VLT/ESPRESSO \citep{ESPRESSO}. ESPRESSO observations of K dwarf HD~23472 ($V = 9.73$) reached a median uncertainty of only 38 cm s$^{-1}$ and were capable of distinguishing the RV signals of two super-Earths and three Earth-sized planets orbiting the star \citep{Barros2022}. HD 101581 and HD 23472 are K5V and K4V dwarfs, respectively, are both single stars, metal-poor ([Fe/H] $= -0.34$ and $-0.20$), relatively inactive based on $\log{R^{\prime}_{HK}}$ ($-4.77$ and $-5.00$), have low $v\sin{i}$ ($2.5$ and $1.5$ km s$^{-1}$), and have similar rotation periods ($\sim30$ and $\sim44$ days) which are several times longer than the periods of their Earth-size planets. Exposures of 600s for K dwarf HD~85512 ($V = 7.65$) were also able to reach a photon-noise induced RV error of only $18$ cm s$^{-1}$ with ESPRESSO \citep{ESPRESSO}. Actual RV uncertainties for HD~101581 will be affected by stellar jitter.

\subsubsection{Atmospheric Characterization}

\begin{figure}
    \includegraphics[width=\linewidth]{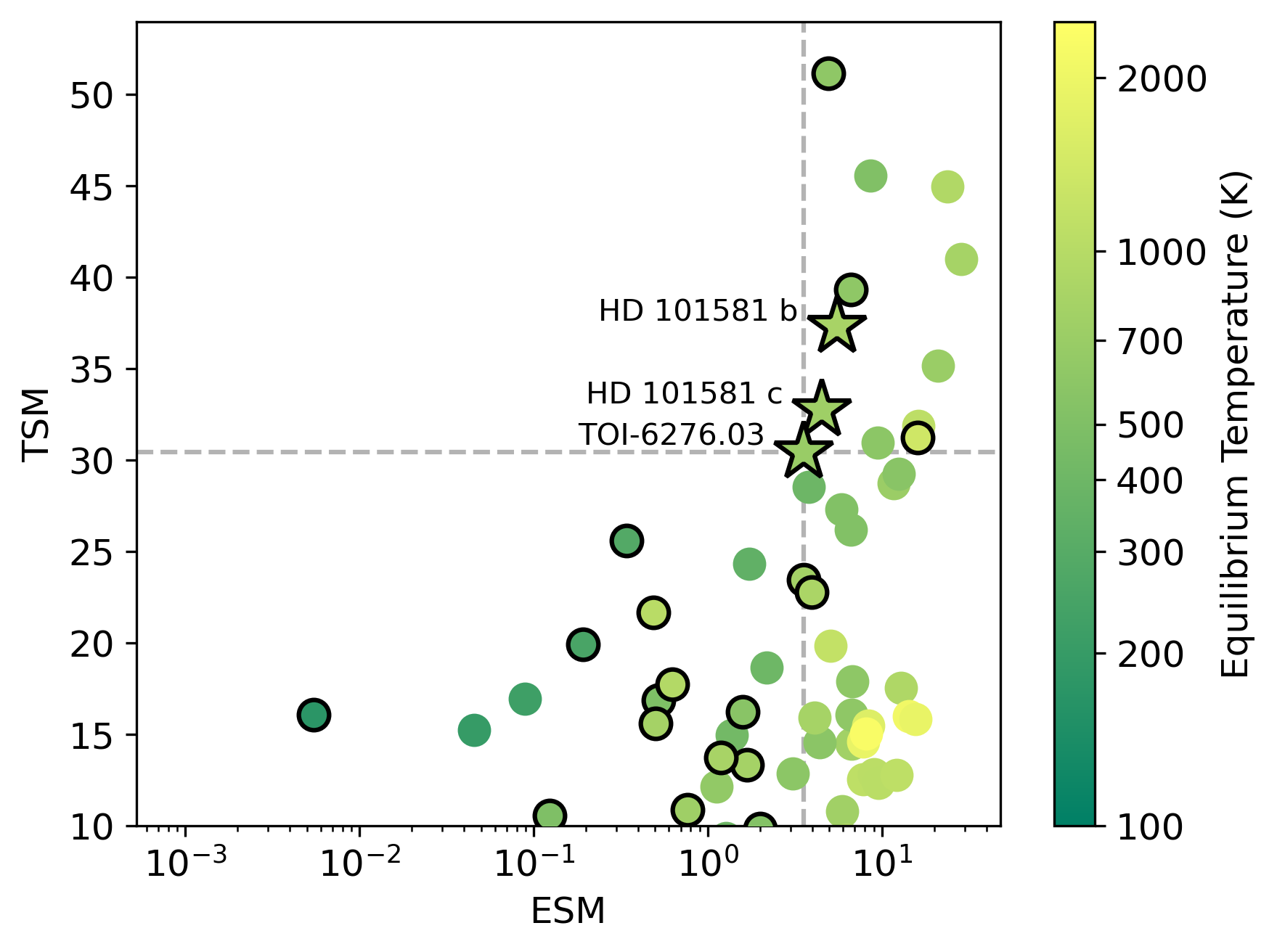}
    \caption{Estimated TSM and ESM values \citep{Kempton2018} for known terrestrial planets ($R_{p} < 1.5~R_{\oplus}$) with TSM $>$ 10, based on their properties given in the NASA Exoplanet Archive Planetary Systems Composite Data Table \citep{NEA}. Planets are colored by equilibrium temperature assuming zero albedo. Sub-Earths ($R_{p} < 1~R_{\oplus}$) are plotted with a black outline. The top-right box indicates the TSM/ESM parameter space in which the planets lie. The planets orbiting HD~101581 are among the best sub-Earths for characterization with both transmission and emission spectroscopy.}\label{fig:tsm}
\end{figure}

We computed the transmission and emission spectroscopy metrics \cite[TSM and ESM;][]{Kempton2018} for each transiting planet to measure their suitability for atmospheric characterization studies. We find TSM $= 37.3$, $32.8$, and $30.4$ for HD~101581 b and c and TOI-6276.03, respectively, well above the TSM = 10 threshold recommended by \citet{Kempton2018} to identify promising terrestrial planets ($R_{p} < 1.5~R_{\oplus}$) for atmospheric characterization. All three planets are among the top 15 terrestrial planets for atmospheric characterization with transmission spectroscopy (Figure \ref{fig:tsm}). The planets also have ESM $=5.5$, $4.5$, and $3.6$, placing them among the top 10 sub-Earths ($R_{p} < 1~R_{\oplus}$) for characterization with emission spectroscopy. Among other sub-Earths, HD~101581 b has the third highest TSM after L 98-59 b \citep{Kostov2019} and TOI-540 b \citep{Ment2021}, and the third highest ESM after TOI-540 b and GJ 367 b \citep{Lam2021}. 

\begin{figure}
    \includegraphics[width=\linewidth]{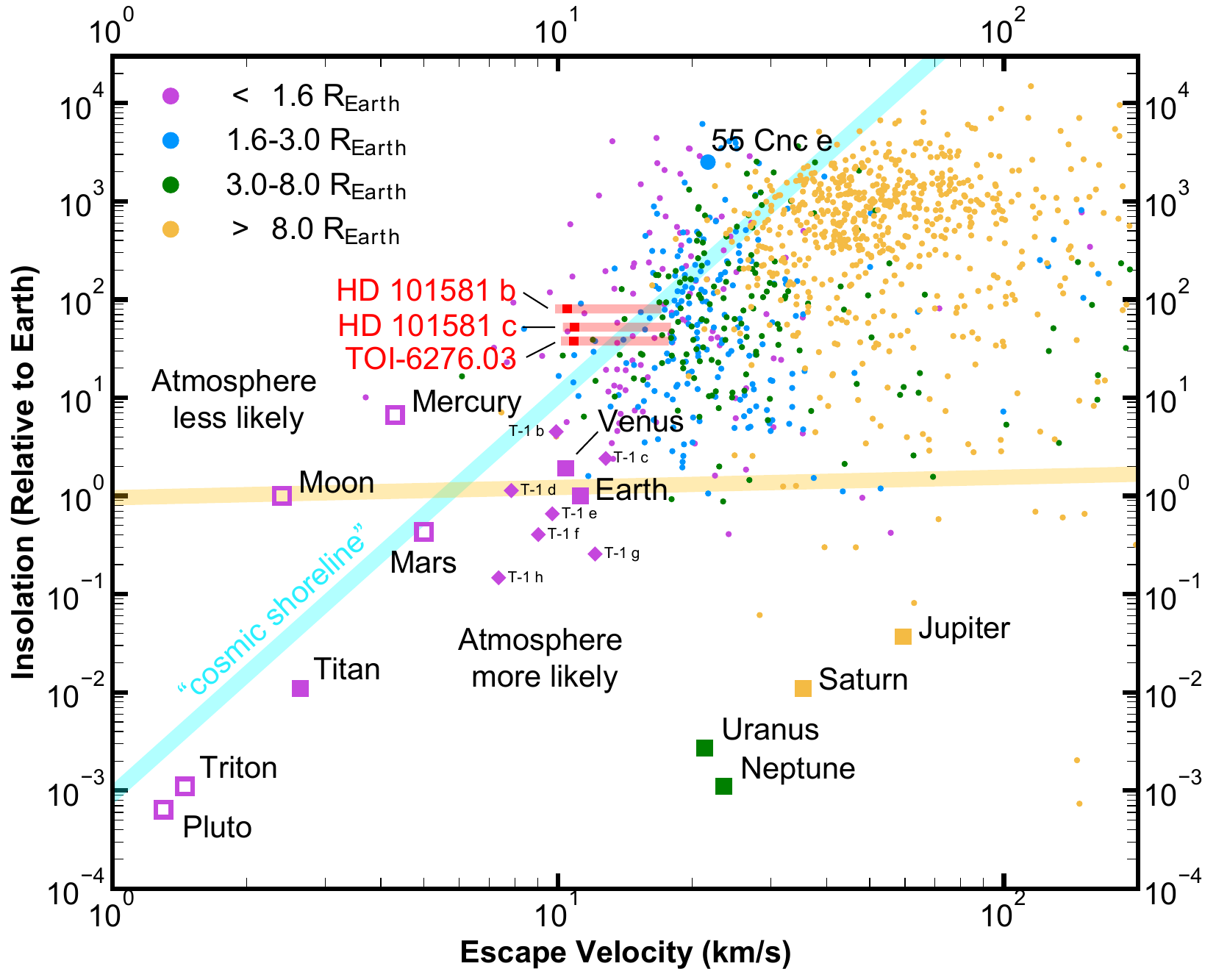}
    \caption{Instellation flux as a function of planet escape velocity, in log-log scale. Based on data from the \cite{NEA} downloaded on June 30, 2024. The empirical ``cosmic shoreline'' and the water vapor greenhouse runaway threshold \citep{Zahnle_cosmic_2017} are shown as cyan and yellow shaded regions, respectively. Planets are categorized into terrestrial planets (magenta), sub-Neptunes (blue), Neptune-like planets (green), and gas giants (yellow) based on radius, see legend. The red rectangles represent escape velocities calculated based on HD~101581 planets' mean estimated masses assuming Hypatia Catalog metallicity. Because only radius, but not mass, is known, we further plot light red regions covering all possible interior compositions, ranging from pure iron to pure silicates. Height of the rectangles represent uncertainties in insolation.}\label{fig:cosmic_shoreline}
\end{figure}

While the planets' small radii suggest that they are unlikely to host light hydrogen-rich atmospheres \citep{Rogers2015}, they may host Venus-like or even hotter atmospheres dominated by gases with high mean molecular weights (MMWs). To assess the likelihood of atmospheric retention, we compare the planets against the empirical $I \propto v_{\rm esc}^4$ ``cosmic shoreline,'' where $I$ is instellation flux and $v_{\rm esc}$ is escape velocity, that divides between planets likely and unlikely to sustain an atmosphere \citep{Zahnle_cosmic_2017}. All three planets are on the airless side of the empirical cosmic shoreline assuming star-like Fe/Si ratio (Figure \ref{fig:cosmic_shoreline}), but may cross the shoreline if their interior compositions are more iron-rich. Note that, however, planets above the shoreline may still possess atmospheres, as evidenced by the recent atmospheric detection on 55 Cnc e \citep{hu_secondary_2024}.

To evaluate the prospect of atmospheric characterization, we used \texttt{petitRADTRANS} \citep{Molliere2019} to model two possible atmospheric cases, namely a CO$_{2}$-dominated, Venus-like atmosphere (96.5\% CO$_{2}$ and 3.5\% N$_{2}$) and an O$_{2}$-dominated atmosphere (95\% O$_{2}$ and 5\% CO$_{2}$). We adopted a 10 bar surface pressure and temperature structure calculated using the \citet{Guillot2010} analytical pressure-temperature profile. We then used \texttt{PandExo}\footnote{\url{https://github.com/natashabatalha/PandExo}} and the JWST Exposure Time Calculator\footnote{\url{https://jwst.etc.stsci.edu/}} to assess the observability of these spectra with various spectroscopic modes. The star is too bright ($J = 5.792$, $K = 5.101$) to be observed with NIRISS-SOSS ($0.6 - 2.8~\mu$m), and only narrow wavelength ranges would be accessible using the SUB512S subarray with NIRSpec-G395H/F290LP ($3.0 - 3.4~\mu$m and $4.2 - 4.5~\mu$m), NIRSpec-G235H/F170LP ($1.8 - 2~\mu$m and $2.5 - 2.7~\mu$m), and NIRSpec-G140H/F100LP ($1.5 - 1.6~\mu$m) disperser-filter combinations. However, NIRCam and MIRI-LRS are suitable for spectroscopic follow-up of very bright targets \cite[e.g., 55 Cancri at $K = 4.015$ mag;][]{hu_secondary_2024}.

\begin{figure}
    \centering
    \includegraphics[width=0.8\linewidth]{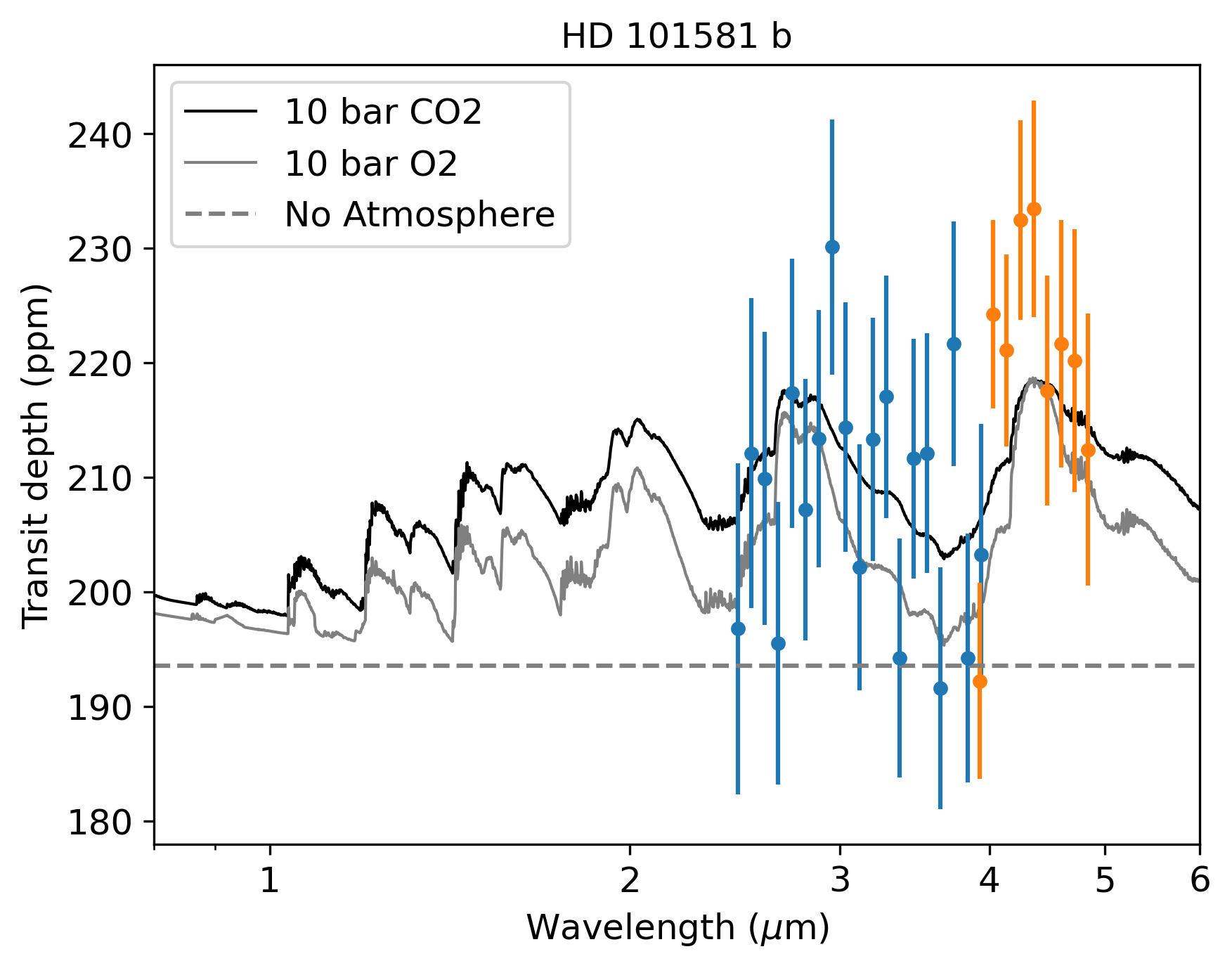}
    \includegraphics[width=0.8\linewidth]{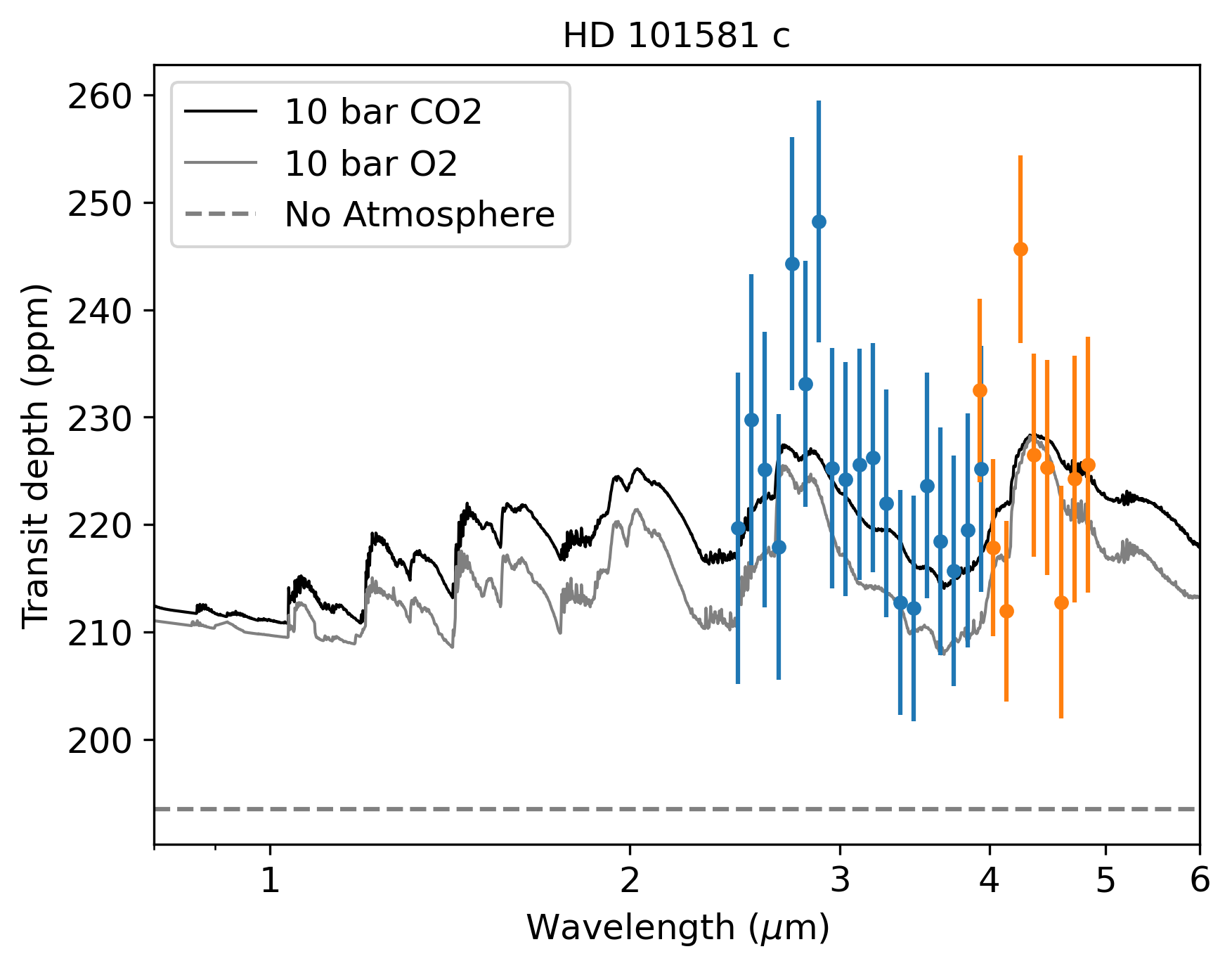}
    \includegraphics[width=0.8\linewidth]{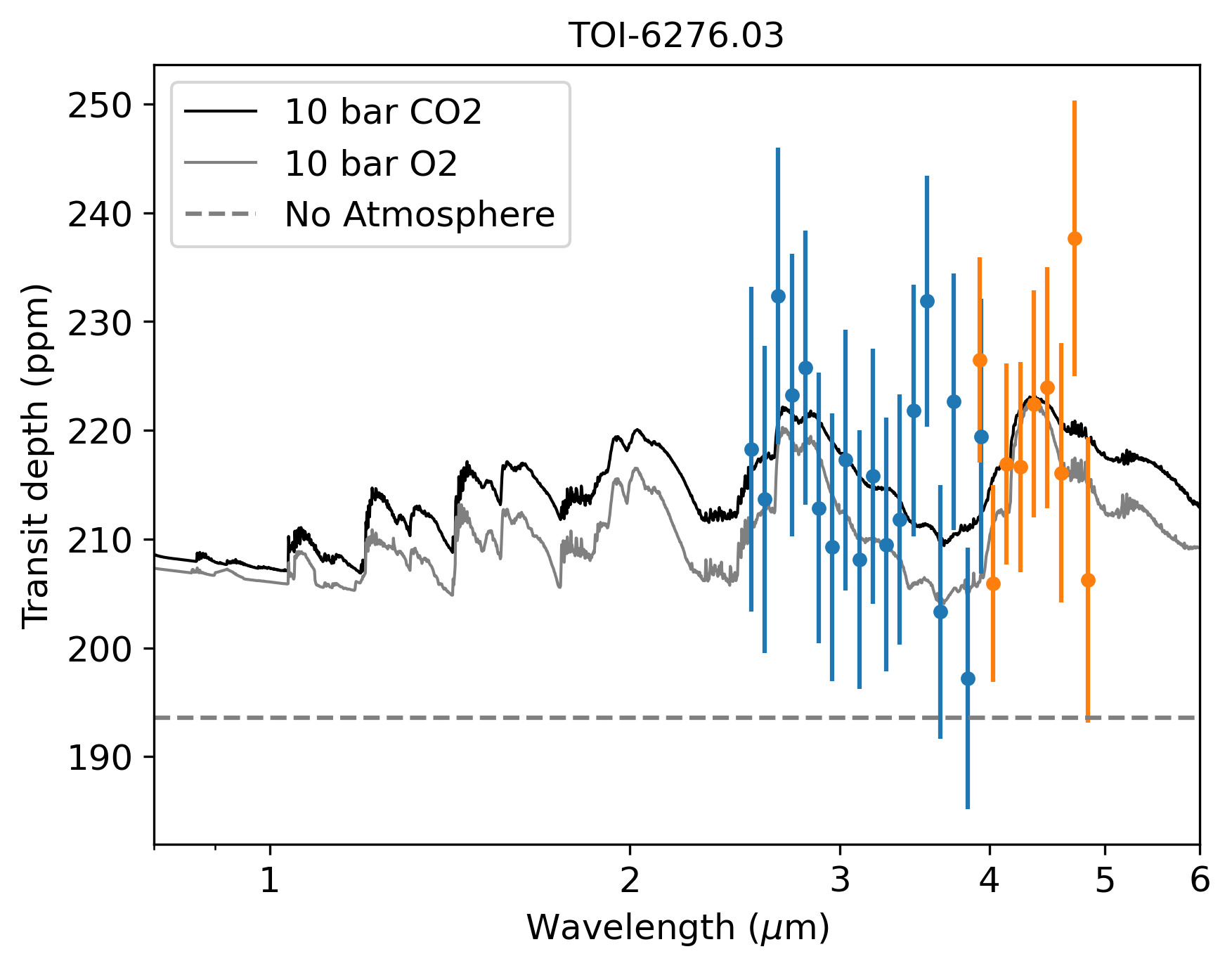}
    \caption{\label{fig:spectra} Expected transmission spectra for all three planets, assuming a CO$_{2}$-dominated, Venus-like atmosphere (black) or O$_{2}$-dominated atmosphere (grey) at 10 bar surface pressure. Simulated JWST/NIRCam long wavelength grism observations of the CO$_{2}$-dominated atmosphere using the SUBGRISM64 array (useful for extremely bright hosts) is shown for the F322W2 filter (blue) and F444W filter (orange). The JWST data have been re-binned to $R = 20$ and assume only 1 transit per planet. At TSM $= 37.3$, $32.8$, and $30.4$, all three planets are among the top six sub-Earth-size planets for transmission spectroscopy.}
\end{figure}

For example, NIRCam observations using the SUBGRISM64 subarray with two groups would avoid saturation and allow for observations with either the F322W2 filter ($2.5 - 4.2~\mu$m) or F444W filter ($3.8 - 5.0~\mu$m). Only one transit observation in either filter would be sufficient to detect the existence of an atmosphere for all three planets (Figure \ref{fig:spectra}). NIRCam will also support short wavelength grism time series observations starting in Cycle 4, which can provide a coverage of $0.6 - 5~\mu$m when combined with the long wavelength grism.

\section{Summary}\label{sec:conclusion}

We have statistically validated two transiting exoplanets orbiting HD~101581 and strengthened the candidacy of a third planet in the same system, all of which are remarkably uniform in size, with $R_{p} = (0.956_{-0.063}^{+0.061}, 0.990_{-0.070}^{+0.070}$,$0.982_{-0.098}^{+0.114})~R_{\oplus}$ for HD~101581 b, HD~101581 c, and TOI-6276.03, respectively. RV observations from AAT, PFS, and HARPS allow us to place $3\sigma$ upper limits on the planet masses of $M_{p} < (3.6, 4.2, 3.6)~M_{\oplus}$.

Their orbital periods, $P = (4.466, 6.204, 7.871)$ days place the planets near a 4:3 MMR for planets b and c, and 5:4 MMR for planet c and planet candidate TOI-6276.03. However, resonance analysis based on mass estimates from a mass-radius relation \citep{ChenKipping2017} did not reveal evidence that the orbital configuration is consistent with resonant libration.

At $V = 7.77$, HD~101581 is the brightest known star that hosts multiple transiting planets with $R_{p} < 1.5~R_{\oplus}$. Given that all three planets are among the top six sub-Earth-size planets by both transmission and emission spectroscopy metrics, HD~101581 is one of the best multi-planet systems for atmospheric characterization and comparative planetology of the small planets. Future analysis using precise mass measurements from extreme precision RV instruments may revisit the possible resonance of the system as well as the full confirmation of the planets and planet candidate.

\section{Acknowledgements}

We thank the referee for carefully reading our manuscript and for giving constructive comments that improved the quality of the manuscript.

Based on observations collected at the European Southern Observatory under ESO programmes 072.C-0488(E) and 075.C-0069(A).

The research shown here acknowledges the use of the Hypatia Catalog Database, an online compilation of stellar abundance data as described in \citet{Hinkel14}, which was supported by NASA's Nexus for Exoplanet System Science (NExSS) research coordination network and the Vanderbilt Initiative in Data-Intensive Astrophysics (VIDA).

Some of the observations in this paper made use of the High-Resolution Imaging instrument Zorro and were obtained under Gemini LLP Proposal Number: GN/S-2021A-LP-105. Zorro was funded by the NASA Exoplanet Exploration Program and built at the NASA Ames Research Center by Steve B. Howell, Nic Scott, Elliott P. Horch, and Emmett Quigley. Zorro was mounted on the Gemini South telescope of the international Gemini Observatory, a program of NSF’s OIR Lab, which is managed by the Association of Universities for Research in Astronomy (AURA) under a cooperative agreement with the National Science Foundation. on behalf of the Gemini partnership: the National Science Foundation (United States), National Research Council (Canada), Agencia Nacional de Investigación y Desarrollo (Chile), Ministerio de Ciencia, Tecnología e Innovación (Argentina), Ministério da Ciência, Tecnologia, Inovações e Comunicações (Brazil), and Korea Astronomy and Space Science Institute (Republic of Korea).

This work was funded by the Data Observatory Foundation. This work was funded by ANID Vinculación Internacional FOVI220091. L.V. acknowledges ANID projects Fondecyt n. 1211162, QUIMAL ASTRO20-0025, and BASAL FB210003. This publication was produced within the framework of institutional support for the development of the research organization of Masaryk University.

This work makes use of observations from the LCOGT network. Part of the LCOGT telescope time was granted by NOIRLab through the Mid-Scale Innovations Program (MSIP). MSIP is funded by NSF.

This research has made use of the Exoplanet Follow-up Observation Program (ExoFOP; DOI: 10.26134/ExoFOP5) website, which is operated by the California Institute of Technology, under contract with the National Aeronautics and Space Administration under the Exoplanet Exploration Program.

Funding for the TESS mission is provided by NASA's Science Mission Directorate. We acknowledge the use of public TESS data from pipelines at the TESS Science Office and at the TESS Science Processing Operations Center. This paper includes data collected by the TESS mission that are publicly available from the Mikulski Archive for Space Telescopes (MAST).

Resources supporting this work were provided by the NASA High-End Computing (HEC) Program through the NASA Advanced Supercomputing (NAS) Division at Ames Research Center for the production of the SPOC data products.

Part of this research was carried out at the Jet Propulsion Laboratory, California Institute of Technology, under a contract with the National Aeronautics and Space Administration (NASA).

\textsc{Minerva}-Australis is supported by Australian Research Council LIEF Grant LE160100001, Discovery Grants DP180100972 and DP220100365, Mount Cuba Astronomical Foundation, and institutional partners University of Southern Queensland, UNSW Sydney, MIT, Nanjing University, George Mason University, University of Louisville, University of California Riverside, University of Florida, and The University of Texas at Austin. We respectfully acknowledge the traditional custodians of all lands throughout Australia, and recognise their continued cultural and spiritual connection to the land, waterways, cosmos, and community. We pay our deepest respects to all Elders, ancestors and descendants of the Giabal, Jarowair, and Kambuwal nations, upon whose lands the \textsc{Minerva}-Australis facility at Mt Kent is situated.

MK acknowledges support by the Juan Carlos Torres postdoctoral fellowship from the MIT Kavli Institute for Astrophysics and Space Research. KAC acknowledges support from the TESS mission via subaward s3449 from MIT. ZL acknowledges funding from the Center for Matter at Atomic Pressures (CMAP), a NSF Physics Frontiers Center, under Award PHY-2020249. PK acknowledges the funding from grant LTT-20015. RB acknowledges support from FONDECYT Project 1241963 and from ANID -- Millennium  Science  Initiative -- ICN12\_009. The work of HPO has been carried out within the framework of the NCCR PlanetS supported by the Swiss National Science Foundation under grants 51NF40\_182901 and 51NF40\_205606. This research was partially supported by the project Partnership for Excellence in Superprecise Optics, reg. no. CZ.02.1.01/0.0/0.0/16\_026/0008390.

\facilities{AAT, ESO:1.52m, ESO:3.6m, Exoplanet Archive, Gaia, Gemini:South, LCOGT, Magellan:Clay, TESS, VLT:Yepun}

\software{\texttt{AstroImageJ} \citep{AstroImageJ}, \texttt{emcee} \citep{ForemanMackey2013}, \texttt{exoplanet} \citep{exoplanet:joss,
exoplanet:zenodo, exoplanet:agol20,
exoplanet:arviz, exoplanet:astropy13, exoplanet:astropy18, exoplanet:kipping13,
exoplanet:luger18, exoplanet:pymc3, exoplanet:theano}, \texttt{isochrones} \citep{Morton2015}, \texttt{matplotlib} \citep{Hunter2007}, \texttt{numpy} \citep{Harris2020}, \texttt{pandas} \citep{reback2020pandas, mckinney-proc-scipy-2010}, \texttt{PYMC3} \citep{exoplanet:pymc3}, \texttt{scipy} \citep{Virtanen2020}, \texttt{TAPIR} \citep{Jensen2013}, \texttt{tess-point} \citep{tesspoint}}

\bibliography{sample631}{}
\bibliographystyle{aasjournal}

\appendix

\renewcommand{\thefigure}{A\arabic{figure}}
\renewcommand{\thetable}{A\arabic{table}}

\setcounter{table}{0}
\setcounter{figure}{0}

\begin{figure*}
\centering
\includegraphics[width=0.7\linewidth]{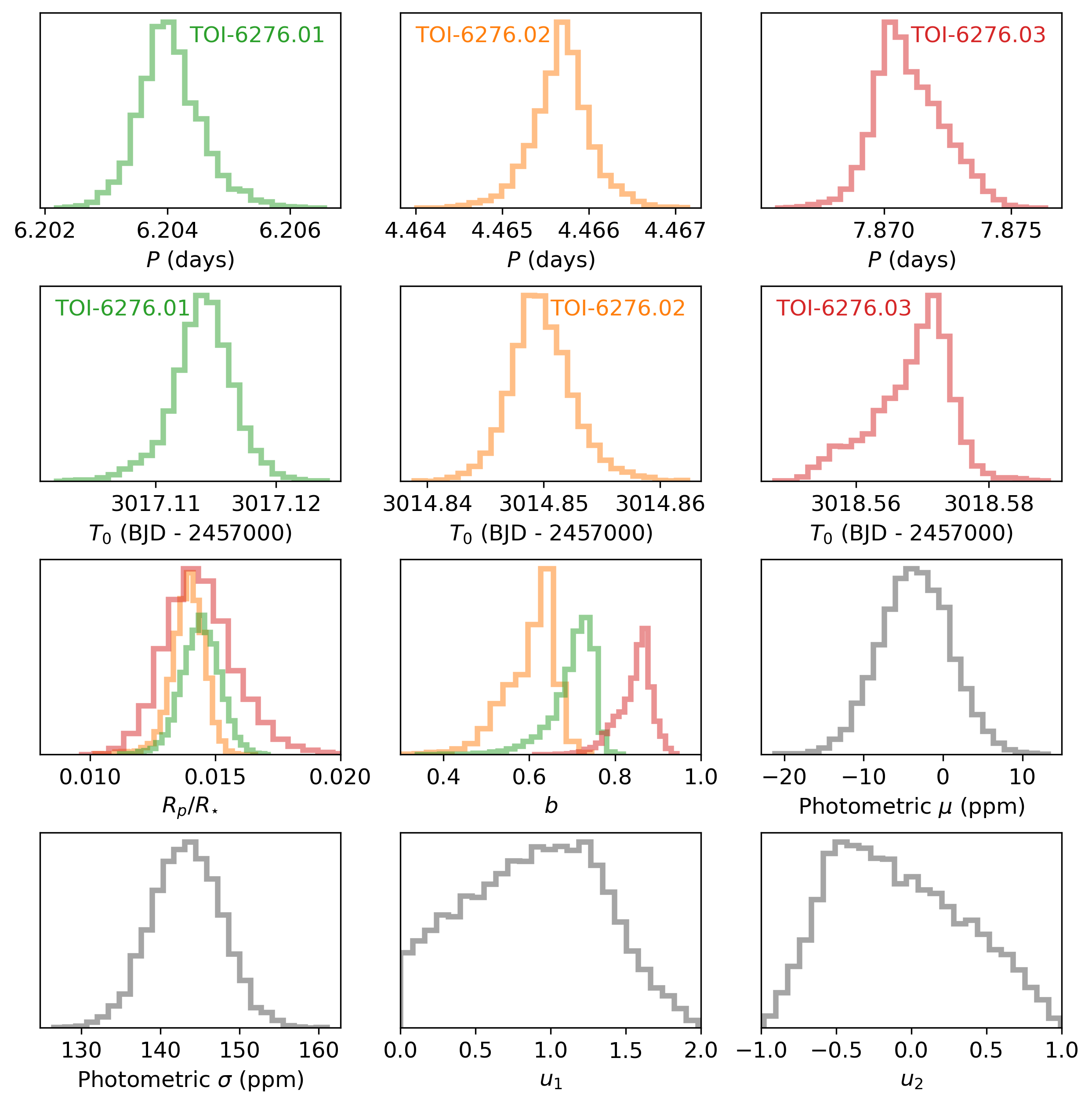}
\caption{Histograms of the distributions of all fit parameters from the three-planet transit model fit. The parameters fitted for TOI-6276.01 (HD~101581 c), TOI-6276.02 (HD~101581 b), and TOI-6276.03 are shown in green, orange, and red, respectively}\label{fig:posteriors}
\end{figure*}

Priors used for our multi-planet transit model fits (\S\ref{sec:planets}) are listed in Table \ref{tab:priors}, while the final distributions of all fit parameters are shown in Figure \ref{fig:posteriors}.

\begin{table*}[h!]
    \centering
    \begin{tabular}{l|lll|l}
    \hline\hline
    Parameter & Prior (TOI-6276.01) & Prior (TOI-6276.02) & Prior (TOI-6276.03) & Description\\
    \hline\hline
    \multicolumn{5}{c}{\textit{Planet Parameters}} \\
    \hline
    $P$ & $\mathcal{N}(6.21,0.621)$\footnote{$\mathcal{N}(\mu,\sigma)$: normal prior with mean $\mu$ and standard deviation $\sigma$} & $\mathcal{N}(4.47,0.447)$ & $\mathcal{N}(7.87,0.787)$ & Orbital period (days) \\
    $T_{0}$ & $\mathcal{N}(3017.103,0.621)$ & $\mathcal{N}(3014.859,0.447)$ & $\mathcal{N}(3018.560,0.787)$ & Transit epoch (BJD $-$ 2457000) \\
    $R_p/R_\star$ & $\mathcal{U}(0,0.5)$\footnote{$\mathcal{U}(a,b)$: uniform prior between $a$ and $b$} & $\mathcal{U}(0,0.5)$ & $\mathcal{U}(0,0.5)$ & Planet-to-star radius ratio \\
    $b$ & $\mathcal{U}(0,1)$ & $\mathcal{U}(0,1)$ & $\mathcal{U}(0,1)$ & Impact parameter \\
    \hline
    \multicolumn{5}{c}{\textit{Photometric Parameters}} \\
    \hline
    $\mu$ & & $\mathcal{U}(-1000,1000)$ & & FFI flux offset (ppm) \\
    $\sigma$ & & $\mathcal{U}(0,1000)$ & & Flux jitter (ppm) \\
    $q_{1}$ & & $\mathcal{U}(-1,1)$ & & Limb-darkening coefficient 1 \\
    $q_{2}$ & & $\mathcal{U}(-1,1)$ & & Limb-darkening coefficient 2\\
    \hline
    \end{tabular}
    \caption{Priors placed on all transit model fit parameters.}
    \label{tab:priors}
\end{table*}

\end{document}

%% file: macros.tex
\usepackage{xcolor}

\newcommand{\gaia}{{Gaia}}
\newcommand{\wise}{{WISE}}
\newcommand{\twomass}{{2MASS}}

\newcommand{\C}{\ensuremath{^{\circ}C\;}}


%% file: sample631.bbl
\begin{thebibliography}{}
\expandafter\ifx\csname natexlab\endcsname\relax\def\natexlab#1{#1}\fi
\providecommand{\url}[1]{\href{#1}{#1}}
\providecommand{\dodoi}[1]{doi:~\href{http://doi.org/#1}{\nolinkurl{#1}}}
\providecommand{\doeprint}[1]{\href{http://ascl.net/#1}{\nolinkurl{http://ascl.net/#1}}}
\providecommand{\doarXiv}[1]{\href{https://arxiv.org/abs/#1}{\nolinkurl{https://arxiv.org/abs/#1}}}

\bibitem[{Adibekyan {et~al.}(2021)Adibekyan, Dorn, Sousa, Santos, Bitsch, Israelian, Mordasini, Barros, Delgado~Mena, Demangeon, Faria, Figueira, Hakobyan, Oshagh, Soares, Kunitomo, Takeda, Jofré, Petrucci, \& Martioli}]{adibekyan_compositional_2021}
Adibekyan, V., Dorn, C., Sousa, S.~G., {et~al.} 2021, Science, 374, 330, \dodoi{10.1126/science.abg8794}

\bibitem[{{Agol} \& {Fabrycky}(2018)}]{Agol2018}
{Agol}, E., \& {Fabrycky}, D.~C. 2018, in Handbook of Exoplanets, ed. H.~J. {Deeg} \& J.~A. {Belmonte}, 7, \dodoi{10.1007/978-3-319-55333-7_7}

\bibitem[{{Agol} {et~al.}(2020){Agol}, {Luger}, \& {Foreman-Mackey}}]{exoplanet:agol20}
{Agol}, E., {Luger}, R., \& {Foreman-Mackey}, D. 2020, \aj, 159, 123, \dodoi{10.3847/1538-3881/ab4fee}

\bibitem[{{Astropy Collaboration} {et~al.}(2013){Astropy Collaboration}, {Robitaille}, {Tollerud}, {Greenfield}, {Droettboom}, {Bray}, {Aldcroft}, {Davis}, {Ginsburg}, {Price-Whelan}, {Kerzendorf}, {Conley}, {Crighton}, {Barbary}, {Muna}, {Ferguson}, {Grollier}, {Parikh}, {Nair}, {Unther}, {Deil}, {Woillez}, {Conseil}, {Kramer}, {Turner}, {Singer}, {Fox}, {Weaver}, {Zabalza}, {Edwards}, {Azalee Bostroem}, {Burke}, {Casey}, {Crawford}, {Dencheva}, {Ely}, {Jenness}, {Labrie}, {Lim}, {Pierfederici}, {Pontzen}, {Ptak}, {Refsdal}, {Servillat}, \& {Streicher}}]{exoplanet:astropy13}
{Astropy Collaboration}, {Robitaille}, T.~P., {Tollerud}, E.~J., {et~al.} 2013, \aap, 558, A33, \dodoi{10.1051/0004-6361/201322068}

\bibitem[{{Astropy Collaboration} {et~al.}(2018){Astropy Collaboration}, {Price-Whelan}, {Sip{\H o}cz}, {G{\"u}nther}, {Lim}, {Crawford}, {Conseil}, {Shupe}, {Craig}, {Dencheva}, {Ginsburg}, {VanderPlas}, {Bradley}, {P{\'e}rez-Su{\'a}rez}, {de Val-Borro}, {Aldcroft}, {Cruz}, {Robitaille}, {Tollerud}, {Ardelean}, {Babej}, {Bach}, {Bachetti}, {Bakanov}, {Bamford}, {Barentsen}, {Barmby}, {Baumbach}, {Berry}, {Biscani}, {Boquien}, {Bostroem}, {Bouma}, {Brammer}, {Bray}, {Breytenbach}, {Buddelmeijer}, {Burke}, {Calderone}, {Cano Rodr{\'{\i}}guez}, {Cara}, {Cardoso}, {Cheedella}, {Copin}, {Corrales}, {Crichton}, {D'Avella}, {Deil}, {Depagne}, {Dietrich}, {Donath}, {Droettboom}, {Earl}, {Erben}, {Fabbro}, {Ferreira}, {Finethy}, {Fox}, {Garrison}, {Gibbons}, {Goldstein}, {Gommers}, {Greco}, {Greenfield}, {Groener}, {Grollier}, {Hagen}, {Hirst}, {Homeier}, {Horton}, {Hosseinzadeh}, {Hu}, {Hunkeler}, {Ivezi{\'c}}, {Jain}, {Jenness}, {Kanarek}, {Kendrew}, {Kern}, {Kerzendorf}, {Khvalko}, {King}, {Kirkby}, {Kulkarni},
  {Kumar}, {Lee}, {Lenz}, {Littlefair}, {Ma}, {Macleod}, {Mastropietro}, {McCully}, {Montagnac}, {Morris}, {Mueller}, {Mumford}, {Muna}, {Murphy}, {Nelson}, {Nguyen}, {Ninan}, {N{\"o}the}, {Ogaz}, {Oh}, {Parejko}, {Parley}, {Pascual}, {Patil}, {Patil}, {Plunkett}, {Prochaska}, {Rastogi}, {Reddy Janga}, {Sabater}, {Sakurikar}, {Seifert}, {Sherbert}, {Sherwood-Taylor}, {Shih}, {Sick}, {Silbiger}, {Singanamalla}, {Singer}, {Sladen}, {Sooley}, {Sornarajah}, {Streicher}, {Teuben}, {Thomas}, {Tremblay}, {Turner}, {Terr{\'o}n}, {van Kerkwijk}, {de la Vega}, {Watkins}, {Weaver}, {Whitmore}, {Woillez}, {Zabalza}, \& {Astropy Contributors}}]{exoplanet:astropy18}
{Astropy Collaboration}, {Price-Whelan}, A.~M., {Sip{\H o}cz}, B.~M., {et~al.} 2018, \aj, 156, 123, \dodoi{10.3847/1538-3881/aabc4f}

\bibitem[{{Baranne} {et~al.}(1996){Baranne}, {Queloz}, {Mayor}, {Adrianzyk}, {Knispel}, {Kohler}, {Lacroix}, {Meunier}, {Rimbaud}, \& {Vin}}]{Baranne1996}
{Baranne}, A., {Queloz}, D., {Mayor}, M., {et~al.} 1996, \aaps, 119, 373

\bibitem[{{Barros} {et~al.}(2022){Barros}, {Demangeon}, {Alibert}, {Leleu}, {Adibekyan}, {Lovis}, {Bossini}, {Sousa}, {Hara}, {Bouchy}, {Lavie}, {Rodrigues}, {Gomes da Silva}, {Lillo-Box}, {Pepe}, {Tabernero}, {Zapatero Osorio}, {Sozzetti}, {Su{\'a}rez Mascare{\~n}o}, {Micela}, {Allende Prieto}, {Cristiani}, {Damasso}, {Di Marcantonio}, {Ehrenreich}, {Faria}, {Figueira}, {Gonz{\'a}lez Hern{\'a}ndez}, {Jenkins}, {Lo Curto}, {Martins}, {Micela}, {Nunes}, {Pall{\'e}}, {Santos}, {Rebolo}, {Seager}, {Twicken}, {Udry}, {Vanderspek}, \& {Winn}}]{Barros2022}
{Barros}, S.~C.~C., {Demangeon}, O.~D.~S., {Alibert}, Y., {et~al.} 2022, \aap, 665, A154, \dodoi{10.1051/0004-6361/202244293}

\bibitem[{{Borucki} {et~al.}(2010){Borucki}, {Koch}, {Basri}, {Batalha}, {Brown}, {Caldwell}, {Caldwell}, {Christensen-Dalsgaard}, {Cochran}, {DeVore}, {Dunham}, {Dupree}, {Gautier}, {Geary}, {Gilliland}, {Gould}, {Howell}, {Jenkins}, {Kondo}, {Latham}, {Marcy}, {Meibom}, {Kjeldsen}, {Lissauer}, {Monet}, {Morrison}, {Sasselov}, {Tarter}, {Boss}, {Brownlee}, {Owen}, {Buzasi}, {Charbonneau}, {Doyle}, {Fortney}, {Ford}, {Holman}, {Seager}, {Steffen}, {Welsh}, {Rowe}, {Anderson}, {Buchhave}, {Ciardi}, {Walkowicz}, {Sherry}, {Horch}, {Isaacson}, {Everett}, {Fischer}, {Torres}, {Johnson}, {Endl}, {MacQueen}, {Bryson}, {Dotson}, {Haas}, {Kolodziejczak}, {Van Cleve}, {Chandrasekaran}, {Twicken}, {Quintana}, {Clarke}, {Allen}, {Li}, {Wu}, {Tenenbaum}, {Verner}, {Bruhweiler}, {Barnes}, \& {Prsa}}]{Borucki2010}
{Borucki}, W.~J., {Koch}, D., {Basri}, G., {et~al.} 2010, Science, 327, 977, \dodoi{10.1126/science.1185402}

\bibitem[{{Bowler} {et~al.}(2021){Bowler}, {Cochran}, {Endl}, {Franson}, {Brandt}, {Dupuy}, {MacQueen}, {Kratter}, {Mawet}, \& {Ruane}}]{Bowler2021}
{Bowler}, B.~P., {Cochran}, W.~D., {Endl}, M., {et~al.} 2021, \aj, 161, 106, \dodoi{10.3847/1538-3881/abd243}

\bibitem[{{Brahm} {et~al.}(2017{\natexlab{a}}){Brahm}, {Jord{\'a}n}, \& {Espinoza}}]{CERES}
{Brahm}, R., {Jord{\'a}n}, A., \& {Espinoza}, N. 2017{\natexlab{a}}, \pasp, 129, 034002, \dodoi{10.1088/1538-3873/aa5455}

\bibitem[{{Brahm} {et~al.}(2017{\natexlab{b}}){Brahm}, {Jord{\'a}n}, {Hartman}, \& {Bakos}}]{ZASPE}
{Brahm}, R., {Jord{\'a}n}, A., {Hartman}, J., \& {Bakos}, G. 2017{\natexlab{b}}, \mnras, 467, 971, \dodoi{10.1093/mnras/stx144}

\bibitem[{{Brandt}(2018)}]{Brandt2018}
{Brandt}, T.~D. 2018, \apjs, 239, 31, \dodoi{10.3847/1538-4365/aaec06}

\bibitem[{{Brandt}(2021)}]{Brandt2021}
---. 2021, \apjs, 254, 42, \dodoi{10.3847/1538-4365/abf93c}

\bibitem[{{Brown} {et~al.}(2013){Brown}, {Baliber}, {Bianco}, {Bowman}, {Burleson}, {Conway}, {Crellin}, {Depagne}, {De Vera}, {Dilday}, {Dragomir}, {Dubberley}, {Eastman}, {Elphick}, {Falarski}, {Foale}, {Ford}, {Fulton}, {Garza}, {Gomez}, {Graham}, {Greene}, {Haldeman}, {Hawkins}, {Haworth}, {Haynes}, {Hidas}, {Hjelstrom}, {Howell}, {Hygelund}, {Lister}, {Lobdill}, {Martinez}, {Mullins}, {Norbury}, {Parrent}, {Paulson}, {Petry}, {Pickles}, {Posner}, {Rosing}, {Ross}, {Sand}, {Saunders}, {Shobbrook}, {Shporer}, {Street}, {Thomas}, {Tsapras}, {Tufts}, {Valenti}, {Vander Horst}, {Walker}, {White}, \& {Willis}}]{LCO}
{Brown}, T.~M., {Baliber}, N., {Bianco}, F.~B., {et~al.} 2013, \pasp, 125, 1031, \dodoi{10.1086/673168}

\bibitem[{{Burke} {et~al.}(2020){Burke}, {Levine}, {Fausnaugh}, {Vanderspek}, {Barclay}, {Libby-Roberts}, {Morris}, {Sipocz}, {Owens}, {Feinstein}, \& {Camacho}}]{tesspoint}
{Burke}, C.~J., {Levine}, A., {Fausnaugh}, M., {et~al.} 2020, {TESS-Point: High precision TESS pointing tool}, Astrophysics Source Code Library.
\newblock \doeprint{2003.001}

\bibitem[{{Butler} {et~al.}(1996){Butler}, {Marcy}, {Williams}, {McCarthy}, {Dosanjh}, \& {Vogt}}]{Butler1996}
{Butler}, R.~P., {Marcy}, G.~W., {Williams}, E., {et~al.} 1996, \pasp, 108, 500, \dodoi{10.1086/133755}

\bibitem[{{Castelli} \& {Kurucz}(2003)}]{ATLAS9}
{Castelli}, F., \& {Kurucz}, R.~L. 2003, in Modelling of Stellar Atmospheres, ed. N.~{Piskunov}, W.~W. {Weiss}, \& D.~F. {Gray}, Vol. 210, A20, \dodoi{10.48550/arXiv.astro-ph/0405087}

\bibitem[{{Castro-Ginard} {et~al.}(2024){Castro-Ginard}, {Penoyre}, {Casey}, {Brown}, {Belokurov}, {Cantat-Gaudin}, {Drimmel}, {Fouesneau}, {Khanna}, {Kurbatov}, {Price-Whelan}, {Rix}, \& {Smart}}]{CastroGinard2024}
{Castro-Ginard}, A., {Penoyre}, Z., {Casey}, A.~R., {et~al.} 2024, arXiv e-prints, arXiv:2404.14127, \dodoi{10.48550/arXiv.2404.14127}

\bibitem[{{Chen} \& {Kipping}(2017)}]{ChenKipping2017}
{Chen}, J., \& {Kipping}, D. 2017, \apj, 834, 17, \dodoi{10.3847/1538-4357/834/1/17}

\bibitem[{{Choi} {et~al.}(2016){Choi}, {Dotter}, {Conroy}, {Cantiello}, {Paxton}, \& {Johnson}}]{Choi2016}
{Choi}, J., {Dotter}, A., {Conroy}, C., {et~al.} 2016, \apj, 823, 102, \dodoi{10.3847/0004-637X/823/2/102}

\bibitem[{{Claytor} {et~al.}(2022){Claytor}, {van Saders}, {Llama}, {Sadowski}, {Quach}, \& {Avallone}}]{Claytor2022}
{Claytor}, Z.~R., {van Saders}, J.~L., {Llama}, J., {et~al.} 2022, \apj, 927, 219, \dodoi{10.3847/1538-4357/ac498f}

\bibitem[{{Claytor} {et~al.}(2020){Claytor}, {van Saders}, {Santos}, {Garc{\'\i}a}, {Mathur}, {Tayar}, {Pinsonneault}, \& {Shetrone}}]{Claytor2020}
{Claytor}, Z.~R., {van Saders}, J.~L., {Santos}, {\^A}. R.~G., {et~al.} 2020, \apj, 888, 43, \dodoi{10.3847/1538-4357/ab5c24}

\bibitem[{{Cloutier} {et~al.}(2019){Cloutier}, {Astudillo-Defru}, {Bonfils}, {Jenkins}, {Berdi{\~n}as}, {Ricker}, {Vanderspek}, {Latham}, {Seager}, {Winn}, {Jenkins}, {Almenara}, {Bouchy}, {Delfosse}, {D{\'\i}az}, {D{\'\i}az}, {Doyon}, {Figueira}, {Forveille}, {Kurtovic}, {Lovis}, {Mayor}, {Menou}, {Morgan}, {Morris}, {Muirhead}, {Murgas}, {Pepe}, {Santos}, {S{\'e}gransan}, {Smith}, {Tenenbaum}, {Torres}, {Udry}, {Vezie}, \& {Villasenor}}]{Cloutier2019}
{Cloutier}, R., {Astudillo-Defru}, N., {Bonfils}, X., {et~al.} 2019, \aap, 629, A111, \dodoi{10.1051/0004-6361/201935957}

\bibitem[{{Coelho} {et~al.}(2005){Coelho}, {Barbuy}, {Mel{\'e}ndez}, {Schiavon}, \& {Castilho}}]{Coelho2005}
{Coelho}, P., {Barbuy}, B., {Mel{\'e}ndez}, J., {Schiavon}, R.~P., \& {Castilho}, B.~V. 2005, \aap, 443, 735, \dodoi{10.1051/0004-6361:20053511}

\bibitem[{{Collins} {et~al.}(2017){Collins}, {Kielkopf}, {Stassun}, \& {Hessman}}]{AstroImageJ}
{Collins}, K.~A., {Kielkopf}, J.~F., {Stassun}, K.~G., \& {Hessman}, F.~V. 2017, \aj, 153, 77, \dodoi{10.3847/1538-3881/153/2/77}

\bibitem[{{Crane} {et~al.}(2006){Crane}, {Shectman}, \& {Butler}}]{Crane2006}
{Crane}, J.~D., {Shectman}, S.~A., \& {Butler}, R.~P. 2006, in Society of Photo-Optical Instrumentation Engineers (SPIE) Conference Series, Vol. 6269, Ground-based and Airborne Instrumentation for Astronomy, ed. I.~S. {McLean} \& M.~{Iye}, 626931, \dodoi{10.1117/12.672339}

\bibitem[{{Crane} {et~al.}(2010){Crane}, {Shectman}, {Butler}, {Thompson}, {Birk}, {Jones}, \& {Burley}}]{Crane2010}
{Crane}, J.~D., {Shectman}, S.~A., {Butler}, R.~P., {et~al.} 2010, in Society of Photo-Optical Instrumentation Engineers (SPIE) Conference Series, Vol. 7735, Ground-based and Airborne Instrumentation for Astronomy III, ed. I.~S. {McLean}, S.~K. {Ramsay}, \& H.~{Takami}, 773553, \dodoi{10.1117/12.857792}

\bibitem[{{Crane} {et~al.}(2008){Crane}, {Shectman}, {Butler}, {Thompson}, \& {Burley}}]{Crane2008}
{Crane}, J.~D., {Shectman}, S.~A., {Butler}, R.~P., {Thompson}, I.~B., \& {Burley}, G.~S. 2008, in Society of Photo-Optical Instrumentation Engineers (SPIE) Conference Series, Vol. 7014, Ground-based and Airborne Instrumentation for Astronomy II, ed. I.~S. {McLean} \& M.~M. {Casali}, 701479, \dodoi{10.1117/12.789637}

\bibitem[{{Delisle}(2017)}]{Delisle2017}
{Delisle}, J.~B. 2017, \aap, 605, A96, \dodoi{10.1051/0004-6361/201730857}

\bibitem[{{Demarque} {et~al.}(2008){Demarque}, {Guenther}, {Li}, {Mazumdar}, \& {Straka}}]{Demarque2008}
{Demarque}, P., {Guenther}, D.~B., {Li}, L.~H., {Mazumdar}, A., \& {Straka}, C.~W. 2008, \apss, 316, 31, \dodoi{10.1007/s10509-007-9698-y}

\bibitem[{{Diego} {et~al.}(1990){Diego}, {Charalambous}, {Fish}, \& {Walker}}]{Diego1990}
{Diego}, F., {Charalambous}, A., {Fish}, A.~C., \& {Walker}, D.~D. 1990, in Society of Photo-Optical Instrumentation Engineers (SPIE) Conference Series, Vol. 1235, Instrumentation in Astronomy VII, ed. D.~L. {Crawford}, 562--576, \dodoi{10.1117/12.19119}

\bibitem[{{Dotter}(2016)}]{Dotter2016}
{Dotter}, A. 2016, \apjs, 222, 8, \dodoi{10.3847/0067-0049/222/1/8}

\bibitem[{{Dotter} {et~al.}(2008){Dotter}, {Chaboyer}, {Jevremovi{\'c}}, {Kostov}, {Baron}, \& {Ferguson}}]{Dotter2008}
{Dotter}, A., {Chaboyer}, B., {Jevremovi{\'c}}, D., {et~al.} 2008, \apjs, 178, 89, \dodoi{10.1086/589654}

\bibitem[{{Dragomir} {et~al.}(2019){Dragomir}, {Teske}, {G{\"u}nther}, {S{\'e}gransan}, {Burt}, {Huang}, {Vanderburg}, {Matthews}, {Dumusque}, {Stassun}, {Pepper}, {Ricker}, {Vanderspek}, {Latham}, {Seager}, {Winn}, {Jenkins}, {Beatty}, {Bouchy}, {Brown}, {Butler}, {Ciardi}, {Crane}, {Eastman}, {Fossati}, {Francis}, {Fulton}, {Gaudi}, {Goeke}, {James}, {Klaus}, {Kuhn}, {Lovis}, {Lund}, {McDermott}, {Paegert}, {Pepe}, {Rodriguez}, {Sha}, {Shectman}, {Shporer}, {Siverd}, {Garcia Soto}, {Stevens}, {Twicken}, {Udry}, {Villanueva}, {Wang}, {Wohler}, {Yao}, \& {Zhan}}]{Dragomir2019}
{Dragomir}, D., {Teske}, J., {G{\"u}nther}, M.~N., {et~al.} 2019, \apjl, 875, L7, \dodoi{10.3847/2041-8213/ab12ed}

\bibitem[{{Eastman} {et~al.}(2019){Eastman}, {Rodriguez}, {Agol}, {Stassun}, {Beatty}, {Vanderburg}, {Gaudi}, {Collins}, \& {Luger}}]{Eastman2019}
{Eastman}, J.~D., {Rodriguez}, J.~E., {Agol}, E., {et~al.} 2019, arXiv e-prints, arXiv:1907.09480, \dodoi{10.48550/arXiv.1907.09480}

\bibitem[{{Eggenberger} {et~al.}(2007){Eggenberger}, {Udry}, {Chauvin}, {Beuzit}, {Lagrange}, {S{\'e}gransan}, \& {Mayor}}]{NACO}
{Eggenberger}, A., {Udry}, S., {Chauvin}, G., {et~al.} 2007, \aap, 474, 273, \dodoi{10.1051/0004-6361:20077447}

\bibitem[{{Fellgett}(1955)}]{Fellgett1955}
{Fellgett}, P. 1955, Optica Acta, 2, 9, \dodoi{10.1080/713820996}

\bibitem[{{Foreman-Mackey} {et~al.}(2013){Foreman-Mackey}, {Hogg}, {Lang}, \& {Goodman}}]{ForemanMackey2013}
{Foreman-Mackey}, D., {Hogg}, D.~W., {Lang}, D., \& {Goodman}, J. 2013, \pasp, 125, 306, \dodoi{10.1086/670067}

\bibitem[{Foreman-Mackey {et~al.}(2021)Foreman-Mackey, Savel, Luger, Agol, Czekala, Price-Whelan, Hedges, Gilbert, Bouma, Brandt, \& Barclay}]{exoplanet:zenodo}
Foreman-Mackey, D., Savel, A., Luger, R., {et~al.} 2021, exoplanet-dev/exoplanet v0.5.1, \dodoi{10.5281/zenodo.1998447}

\bibitem[{{Foreman-Mackey} {et~al.}(2021){Foreman-Mackey}, {Luger}, {Agol}, {Barclay}, {Bouma}, {Brandt}, {Czekala}, {David}, {Dong}, {Gilbert}, {Gordon}, {Hedges}, {Hey}, {Morris}, {Price-Whelan}, \& {Savel}}]{exoplanet:joss}
{Foreman-Mackey}, D., {Luger}, R., {Agol}, E., {et~al.} 2021, arXiv e-prints, arXiv:2105.01994.
\newblock \doarXiv{2105.01994}

\bibitem[{{Gaia Collaboration} {et~al.}(2016){Gaia Collaboration}, {Prusti}, {de Bruijne}, {Brown}, {Vallenari}, {Babusiaux}, {Bailer-Jones}, {Bastian}, {Biermann}, {Evans}, {Eyer}, {Jansen}, {Jordi}, {Klioner}, {Lammers}, {Lindegren}, {Luri}, {Mignard}, {Milligan}, {Panem}, {Poinsignon}, {Pourbaix}, {Randich}, {Sarri}, {Sartoretti}, {Siddiqui}, {Soubiran}, {Valette}, {van Leeuwen}, {Walton}, {Aerts}, {Arenou}, {Cropper}, {Drimmel}, {H{\o}g}, {Katz}, {Lattanzi}, {O'Mullane}, {Grebel}, {Holland}, {Huc}, {Passot}, {Bramante}, {Cacciari}, {Casta{\~n}eda}, {Chaoul}, {Cheek}, {De Angeli}, {Fabricius}, {Guerra}, {Hern{\'a}ndez}, {Jean-Antoine-Piccolo}, {Masana}, {Messineo}, {Mowlavi}, {Nienartowicz}, {Ord{\'o}{\~n}ez-Blanco}, {Panuzzo}, {Portell}, {Richards}, {Riello}, {Seabroke}, {Tanga}, {Th{\'e}venin}, {Torra}, {Els}, {Gracia-Abril}, {Comoretto}, {Garcia-Reinaldos}, {Lock}, {Mercier}, {Altmann}, {Andrae}, {Astraatmadja}, {Bellas-Velidis}, {Benson}, {Berthier}, {Blomme}, {Busso}, {Carry}, {Cellino}, {Clementini},
  {Cowell}, {Creevey}, {Cuypers}, {Davidson}, {De Ridder}, {de Torres}, {Delchambre}, {Dell'Oro}, {Ducourant}, {Fr{\'e}mat}, {Garc{\'\i}a-Torres}, {Gosset}, {Halbwachs}, {Hambly}, {Harrison}, {Hauser}, {Hestroffer}, {Hodgkin}, {Huckle}, {Hutton}, {Jasniewicz}, {Jordan}, {Kontizas}, {Korn}, {Lanzafame}, {Manteiga}, {Moitinho}, {Muinonen}, {Osinde}, {Pancino}, {Pauwels}, {Petit}, {Recio-Blanco}, {Robin}, {Sarro}, {Siopis}, {Smith}, {Smith}, {Sozzetti}, {Thuillot}, {van Reeven}, {Viala}, {Abbas}, {Abreu Aramburu}, {Accart}, {Aguado}, {Allan}, {Allasia}, {Altavilla}, {{\'A}lvarez}, {Alves}, {Anderson}, {Andrei}, {Anglada Varela}, {Antiche}, {Antoja}, {Ant{\'o}n}, {Arcay}, {Atzei}, {Ayache}, {Bach}, {Baker}, {Balaguer-N{\'u}{\~n}ez}, {Barache}, {Barata}, {Barbier}, {Barblan}, {Baroni}, {Barrado y Navascu{\'e}s}, {Barros}, {Barstow}, {Becciani}, {Bellazzini}, {Bellei}, {Bello Garc{\'\i}a}, {Belokurov}, {Bendjoya}, {Berihuete}, {Bianchi}, {Bienaym{\'e}}, {Billebaud}, {Blagorodnova}, {Blanco-Cuaresma}, {Boch},
  {Bombrun}, {Borrachero}, {Bouquillon}, {Bourda}, {Bouy}, {Bragaglia}, {Breddels}, {Brouillet}, {Br{\"u}semeister}, {Bucciarelli}, {Budnik}, {Burgess}, {Burgon}, {Burlacu}, {Busonero}, {Buzzi}, {Caffau}, {Cambras}, {Campbell}, {Cancelliere}, {Cantat-Gaudin}, {Carlucci}, {Carrasco}, {Castellani}, {Charlot}, {Charnas}, {Charvet}, {Chassat}, {Chiavassa}, {Clotet}, {Cocozza}, {Collins}, {Collins}, {Costigan}, {Crifo}, {Cross}, {Crosta}, {Crowley}, {Dafonte}, {Damerdji}, {Dapergolas}, {David}, {David}, {De Cat}, {de Felice}, {de Laverny}, {De Luise}, {De March}, {de Martino}, {de Souza}, {Debosscher}, {del Pozo}, {Delbo}, {Delgado}, {Delgado}, {di Marco}, {Di Matteo}, {Diakite}, {Distefano}, {Dolding}, {Dos Anjos}, {Drazinos}, {Dur{\'a}n}, {Dzigan}, {Ecale}, {Edvardsson}, {Enke}, {Erdmann}, {Escolar}, {Espina}, {Evans}, {Eynard Bontemps}, {Fabre}, {Fabrizio}, {Faigler}, {Falc{\~a}o}, {Farr{\`a}s Casas}, {Faye}, {Federici}, {Fedorets}, {Fern{\'a}ndez-Hern{\'a}ndez}, {Fernique}, {Fienga}, {Figueras}, {Filippi},
  {Findeisen}, {Fonti}, {Fouesneau}, {Fraile}, {Fraser}, {Fuchs}, {Furnell}, {Gai}, {Galleti}, {Galluccio}, {Garabato}, {Garc{\'\i}a-Sedano}, {Gar{\'e}}, {Garofalo}, {Garralda}, {Gavras}, {Gerssen}, {Geyer}, {Gilmore}, {Girona}, {Giuffrida}, {Gomes}, {Gonz{\'a}lez-Marcos}, {Gonz{\'a}lez-N{\'u}{\~n}ez}, {Gonz{\'a}lez-Vidal}, {Granvik}, {Guerrier}, {Guillout}, {Guiraud}, {G{\'u}rpide}, {Guti{\'e}rrez-S{\'a}nchez}, {Guy}, {Haigron}, {Hatzidimitriou}, {Haywood}, {Heiter}, {Helmi}, {Hobbs}, {Hofmann}, {Holl}, {Holland}, {Hunt}, {Hypki}, {Icardi}, {Irwin}, {Jevardat de Fombelle}, {Jofr{\'e}}, {Jonker}, {Jorissen}, {Julbe}, {Karampelas}, {Kochoska}, {Kohley}, {Kolenberg}, {Kontizas}, {Koposov}, {Kordopatis}, {Koubsky}, {Kowalczyk}, {Krone-Martins}, {Kudryashova}, {Kull}, {Bachchan}, {Lacoste-Seris}, {Lanza}, {Lavigne}, {Le Poncin-Lafitte}, {Lebreton}, {Lebzelter}, {Leccia}, {Leclerc}, {Lecoeur-Taibi}, {Lemaitre}, {Lenhardt}, {Leroux}, {Liao}, {Licata}, {Lindstr{\o}m}, {Lister}, {Livanou}, {Lobel}, {L{\"o}ffler},
  {L{\'o}pez}, {Lopez-Lozano}, {Lorenz}, {Loureiro}, {MacDonald}, {Magalh{\~a}es Fernandes}, {Managau}, {Mann}, {Mantelet}, {Marchal}, {Marchant}, {Marconi}, {Marie}, {Marinoni}, {Marrese}, {Marschalk{\'o}}, {Marshall}, {Mart{\'\i}n-Fleitas}, {Martino}, {Mary}, {Matijevi{\v{c}}}, {Mazeh}, {McMillan}, {Messina}, {Mestre}, {Michalik}, {Millar}, {Miranda}, {Molina}, {Molinaro}, {Molinaro}, {Moln{\'a}r}, {Moniez}, {Montegriffo}, {Monteiro}, {Mor}, {Mora}, {Morbidelli}, {Morel}, {Morgenthaler}, {Morley}, {Morris}, {Mulone}, {Muraveva}, {Musella}, {Narbonne}, {Nelemans}, {Nicastro}, {Noval}, {Ord{\'e}novic}, {Ordieres-Mer{\'e}}, {Osborne}, {Pagani}, {Pagano}, {Pailler}, {Palacin}, {Palaversa}, {Parsons}, {Paulsen}, {Pecoraro}, {Pedrosa}, {Pentik{\"a}inen}, {Pereira}, {Pichon}, {Piersimoni}, {Pineau}, {Plachy}, {Plum}, {Poujoulet}, {Pr{\v{s}}a}, {Pulone}, {Ragaini}, {Rago}, {Rambaux}, {Ramos-Lerate}, {Ranalli}, {Rauw}, {Read}, {Regibo}, {Renk}, {Reyl{\'e}}, {Ribeiro}, {Rimoldini}, {Ripepi}, {Riva}, {Rixon},
  {Roelens}, {Romero-G{\'o}mez}, {Rowell}, {Royer}, {Rudolph}, {Ruiz-Dern}, {Sadowski}, {Sagrist{\`a} Sell{\'e}s}, {Sahlmann}, {Salgado}, {Salguero}, {Sarasso}, {Savietto}, {Schnorhk}, {Schultheis}, {Sciacca}, {Segol}, {Segovia}, {Segransan}, {Serpell}, {Shih}, {Smareglia}, {Smart}, {Smith}, {Solano}, {Solitro}, {Sordo}, {Soria Nieto}, {Souchay}, {Spagna}, {Spoto}, {Stampa}, {Steele}, {Steidelm{\"u}ller}, {Stephenson}, {Stoev}, {Suess}, {S{\"u}veges}, {Surdej}, {Szabados}, {Szegedi-Elek}, {Tapiador}, {Taris}, {Tauran}, {Taylor}, {Teixeira}, {Terrett}, {Tingley}, {Trager}, {Turon}, {Ulla}, {Utrilla}, {Valentini}, {van Elteren}, {Van Hemelryck}, {van Leeuwen}, {Varadi}, {Vecchiato}, {Veljanoski}, {Via}, {Vicente}, {Vogt}, {Voss}, {Votruba}, {Voutsinas}, {Walmsley}, {Weiler}, {Weingrill}, {Werner}, {Wevers}, {Whitehead}, {Wyrzykowski}, {Yoldas}, {{\v{Z}}erjal}, {Zucker}, {Zurbach}, {Zwitter}, {Alecu}, {Allen}, {Allende Prieto}, {Amorim}, {Anglada-Escud{\'e}}, {Arsenijevic}, {Azaz}, {Balm}, {Beck}, {Bernstein},
  {Bigot}, {Bijaoui}, {Blasco}, {Bonfigli}, {Bono}, {Boudreault}, {Bressan}, {Brown}, {Brunet}, {Bunclark}, {Buonanno}, {Butkevich}, {Carret}, {Carrion}, {Chemin}, {Ch{\'e}reau}, {Corcione}, {Darmigny}, {de Boer}, {de Teodoro}, {de Zeeuw}, {Delle Luche}, {Domingues}, {Dubath}, {Fodor}, {Fr{\'e}zouls}, {Fries}, {Fustes}, {Fyfe}, {Gallardo}, {Gallegos}, {Gardiol}, {Gebran}, {Gomboc}, {G{\'o}mez}, {Grux}, {Gueguen}, {Heyrovsky}, {Hoar}, {Iannicola}, {Isasi Parache}, {Janotto}, {Joliet}, {Jonckheere}, {Keil}, {Kim}, {Klagyivik}, {Klar}, {Knude}, {Kochukhov}, {Kolka}, {Kos}, {Kutka}, {Lainey}, {LeBouquin}, {Liu}, {Loreggia}, {Makarov}, {Marseille}, {Martayan}, {Martinez-Rubi}, {Massart}, {Meynadier}, {Mignot}, {Munari}, {Nguyen}, {Nordlander}, {Ocvirk}, {O'Flaherty}, {Olias Sanz}, {Ortiz}, {Osorio}, {Oszkiewicz}, {Ouzounis}, {Palmer}, {Park}, {Pasquato}, {Peltzer}, {Peralta}, {P{\'e}turaud}, {Pieniluoma}, {Pigozzi}, {Poels}, {Prat}, {Prod'homme}, {Raison}, {Rebordao}, {Risquez}, {Rocca-Volmerange}, {Rosen},
  {Ruiz-Fuertes}, {Russo}, {Sembay}, {Serraller Vizcaino}, {Short}, {Siebert}, {Silva}, {Sinachopoulos}, {Slezak}, {Soffel}, {Sosnowska}, {Strai{\v{z}}ys}, {ter Linden}, {Terrell}, {Theil}, {Tiede}, {Troisi}, {Tsalmantza}, {Tur}, {Vaccari}, {Vachier}, {Valles}, {Van Hamme}, {Veltz}, {Virtanen}, {Wallut}, {Wichmann}, {Wilkinson}, {Ziaeepour}, \& {Zschocke}}]{Gaia}
{Gaia Collaboration}, {Prusti}, T., {de Bruijne}, J.~H.~J., {et~al.} 2016, \aap, 595, A1, \dodoi{10.1051/0004-6361/201629272}

\bibitem[{{Gaia Collaboration} {et~al.}(2021){Gaia Collaboration}, {Brown}, {Vallenari}, {Prusti}, {de Bruijne}, {Babusiaux}, {Biermann}, {Creevey}, {Evans}, {Eyer}, {Hutton}, {Jansen}, {Jordi}, {Klioner}, {Lammers}, {Lindegren}, {Luri}, {Mignard}, {Panem}, {Pourbaix}, {Randich}, {Sartoretti}, {Soubiran}, {Walton}, {Arenou}, {Bailer-Jones}, {Bastian}, {Cropper}, {Drimmel}, {Katz}, {Lattanzi}, {van Leeuwen}, {Bakker}, {Cacciari}, {Casta{\~n}eda}, {De Angeli}, {Ducourant}, {Fabricius}, {Fouesneau}, {Fr{\'e}mat}, {Guerra}, {Guerrier}, {Guiraud}, {Jean-Antoine Piccolo}, {Masana}, {Messineo}, {Mowlavi}, {Nicolas}, {Nienartowicz}, {Pailler}, {Panuzzo}, {Riclet}, {Roux}, {Seabroke}, {Sordo}, {Tanga}, {Th{\'e}venin}, {Gracia-Abril}, {Portell}, {Teyssier}, {Altmann}, {Andrae}, {Bellas-Velidis}, {Benson}, {Berthier}, {Blomme}, {Brugaletta}, {Burgess}, {Busso}, {Carry}, {Cellino}, {Cheek}, {Clementini}, {Damerdji}, {Davidson}, {Delchambre}, {Dell'Oro}, {Fern{\'a}ndez-Hern{\'a}ndez}, {Galluccio}, {Garc{\'\i}a-Lario},
  {Garcia-Reinaldos}, {Gonz{\'a}lez-N{\'u}{\~n}ez}, {Gosset}, {Haigron}, {Halbwachs}, {Hambly}, {Harrison}, {Hatzidimitriou}, {Heiter}, {Hern{\'a}ndez}, {Hestroffer}, {Hodgkin}, {Holl}, {Jan{\ss}en}, {Jevardat de Fombelle}, {Jordan}, {Krone-Martins}, {Lanzafame}, {L{\"o}ffler}, {Lorca}, {Manteiga}, {Marchal}, {Marrese}, {Moitinho}, {Mora}, {Muinonen}, {Osborne}, {Pancino}, {Pauwels}, {Petit}, {Recio-Blanco}, {Richards}, {Riello}, {Rimoldini}, {Robin}, {Roegiers}, {Rybizki}, {Sarro}, {Siopis}, {Smith}, {Sozzetti}, {Ulla}, {Utrilla}, {van Leeuwen}, {van Reeven}, {Abbas}, {Abreu Aramburu}, {Accart}, {Aerts}, {Aguado}, {Ajaj}, {Altavilla}, {{\'A}lvarez}, {{\'A}lvarez Cid-Fuentes}, {Alves}, {Anderson}, {Anglada Varela}, {Antoja}, {Audard}, {Baines}, {Baker}, {Balaguer-N{\'u}{\~n}ez}, {Balbinot}, {Balog}, {Barache}, {Barbato}, {Barros}, {Barstow}, {Bartolom{\'e}}, {Bassilana}, {Bauchet}, {Baudesson-Stella}, {Becciani}, {Bellazzini}, {Bernet}, {Bertone}, {Bianchi}, {Blanco-Cuaresma}, {Boch}, {Bombrun}, {Bossini},
  {Bouquillon}, {Bragaglia}, {Bramante}, {Breedt}, {Bressan}, {Brouillet}, {Bucciarelli}, {Burlacu}, {Busonero}, {Butkevich}, {Buzzi}, {Caffau}, {Cancelliere}, {C{\'a}novas}, {Cantat-Gaudin}, {Carballo}, {Carlucci}, {Carnerero}, {Carrasco}, {Casamiquela}, {Castellani}, {Castro-Ginard}, {Castro Sampol}, {Chaoul}, {Charlot}, {Chemin}, {Chiavassa}, {Cioni}, {Comoretto}, {Cooper}, {Cornez}, {Cowell}, {Crifo}, {Crosta}, {Crowley}, {Dafonte}, {Dapergolas}, {David}, {David}, {de Laverny}, {De Luise}, {De March}, {De Ridder}, {de Souza}, {de Teodoro}, {de Torres}, {del Peloso}, {del Pozo}, {Delbo}, {Delgado}, {Delgado}, {Delisle}, {Di Matteo}, {Diakite}, {Diener}, {Distefano}, {Dolding}, {Eappachen}, {Edvardsson}, {Enke}, {Esquej}, {Fabre}, {Fabrizio}, {Faigler}, {Fedorets}, {Fernique}, {Fienga}, {Figueras}, {Fouron}, {Fragkoudi}, {Fraile}, {Franke}, {Gai}, {Garabato}, {Garcia-Gutierrez}, {Garc{\'\i}a-Torres}, {Garofalo}, {Gavras}, {Gerlach}, {Geyer}, {Giacobbe}, {Gilmore}, {Girona}, {Giuffrida}, {Gomel}, {Gomez},
  {Gonzalez-Santamaria}, {Gonz{\'a}lez-Vidal}, {Granvik}, {Guti{\'e}rrez-S{\'a}nchez}, {Guy}, {Hauser}, {Haywood}, {Helmi}, {Hidalgo}, {Hilger}, {H{\l}adczuk}, {Hobbs}, {Holland}, {Huckle}, {Jasniewicz}, {Jonker}, {Juaristi Campillo}, {Julbe}, {Karbevska}, {Kervella}, {Khanna}, {Kochoska}, {Kontizas}, {Kordopatis}, {Korn}, {Kostrzewa-Rutkowska}, {Kruszy{\'n}ska}, {Lambert}, {Lanza}, {Lasne}, {Le Campion}, {Le Fustec}, {Lebreton}, {Lebzelter}, {Leccia}, {Leclerc}, {Lecoeur-Taibi}, {Liao}, {Licata}, {Lindstr{\o}m}, {Lister}, {Livanou}, {Lobel}, {Madrero Pardo}, {Managau}, {Mann}, {Marchant}, {Marconi}, {Marcos Santos}, {Marinoni}, {Marocco}, {Marshall}, {Martin Polo}, {Mart{\'\i}n-Fleitas}, {Masip}, {Massari}, {Mastrobuono-Battisti}, {Mazeh}, {McMillan}, {Messina}, {Michalik}, {Millar}, {Mints}, {Molina}, {Molinaro}, {Moln{\'a}r}, {Montegriffo}, {Mor}, {Morbidelli}, {Morel}, {Morris}, {Mulone}, {Munoz}, {Muraveva}, {Murphy}, {Musella}, {Noval}, {Ord{\'e}novic}, {Orr{\`u}}, {Osinde}, {Pagani}, {Pagano},
  {Palaversa}, {Palicio}, {Panahi}, {Pawlak}, {Pe{\~n}alosa Esteller}, {Penttil{\"a}}, {Piersimoni}, {Pineau}, {Plachy}, {Plum}, {Poggio}, {Poretti}, {Poujoulet}, {Pr{\v{s}}a}, {Pulone}, {Racero}, {Ragaini}, {Rainer}, {Raiteri}, {Rambaux}, {Ramos}, {Ramos-Lerate}, {Re Fiorentin}, {Regibo}, {Reyl{\'e}}, {Ripepi}, {Riva}, {Rixon}, {Robichon}, {Robin}, {Roelens}, {Rohrbasser}, {Romero-G{\'o}mez}, {Rowell}, {Royer}, {Rybicki}, {Sadowski}, {Sagrist{\`a} Sell{\'e}s}, {Sahlmann}, {Salgado}, {Salguero}, {Samaras}, {Sanchez Gimenez}, {Sanna}, {Santove{\~n}a}, {Sarasso}, {Schultheis}, {Sciacca}, {Segol}, {Segovia}, {S{\'e}gransan}, {Semeux}, {Shahaf}, {Siddiqui}, {Siebert}, {Siltala}, {Slezak}, {Smart}, {Solano}, {Solitro}, {Souami}, {Souchay}, {Spagna}, {Spoto}, {Steele}, {Steidelm{\"u}ller}, {Stephenson}, {S{\"u}veges}, {Szabados}, {Szegedi-Elek}, {Taris}, {Tauran}, {Taylor}, {Teixeira}, {Thuillot}, {Tonello}, {Torra}, {Torra}, {Turon}, {Unger}, {Vaillant}, {van Dillen}, {Vanel}, {Vecchiato}, {Viala}, {Vicente},
  {Voutsinas}, {Weiler}, {Wevers}, {Wyrzykowski}, {Yoldas}, {Yvard}, {Zhao}, {Zorec}, {Zucker}, {Zurbach}, \& {Zwitter}}]{GaiaDR3}
{Gaia Collaboration}, {Brown}, A.~G.~A., {Vallenari}, A., {et~al.} 2021, \aap, 649, A1, \dodoi{10.1051/0004-6361/202039657}

\bibitem[{{Gandolfi} {et~al.}(2018){Gandolfi}, {Barrag{\'a}n}, {Livingston}, {Fridlund}, {Justesen}, {Redfield}, {Fossati}, {Mathur}, {Grziwa}, {Cabrera}, {Garc{\'\i}a}, {Persson}, {Van Eylen}, {Hatzes}, {Hidalgo}, {Albrecht}, {Bugnet}, {Cochran}, {Csizmadia}, {Deeg}, {Eigm{\"u}ller}, {Endl}, {Erikson}, {Esposito}, {Guenther}, {Korth}, {Luque}, {Monta{\~n}es Rodr{\'\i}guez}, {Nespral}, {Nowak}, {P{\"a}tzold}, \& {Prieto-Arranz}}]{Gandolfi2018}
{Gandolfi}, D., {Barrag{\'a}n}, O., {Livingston}, J.~H., {et~al.} 2018, \aap, 619, L10, \dodoi{10.1051/0004-6361/201834289}

\bibitem[{{Gelman} \& {Rubin}(1992)}]{Gelman1992}
{Gelman}, A., \& {Rubin}, D.~B. 1992, Statistical Science, 7, 457, \dodoi{10.1214/ss/1177011136}

\bibitem[{{Giacalone} {et~al.}(2021){Giacalone}, {Dressing}, {Jensen}, {Collins}, {Ricker}, {Vanderspek}, {Seager}, {Winn}, {Jenkins}, {Barclay}, {Barkaoui}, {Cadieux}, {Charbonneau}, {Collins}, {Conti}, {Doyon}, {Evans}, {Ghachoui}, {Gillon}, {Guerrero}, {Hart}, {Jehin}, {Kielkopf}, {McLean}, {Murgas}, {Palle}, {Parviainen}, {Pozuelos}, {Relles}, {Shporer}, {Socia}, {Stockdale}, {Tan}, {Torres}, {Twicken}, {Waalkes}, \& {Waite}}]{TRICERATOPS}
{Giacalone}, S., {Dressing}, C.~D., {Jensen}, E. L.~N., {et~al.} 2021, \aj, 161, 24, \dodoi{10.3847/1538-3881/abc6af}

\bibitem[{{Gillon} {et~al.}(2016){Gillon}, {Jehin}, {Lederer}, {Delrez}, {de Wit}, {Burdanov}, {Van Grootel}, {Burgasser}, {Triaud}, {Opitom}, {Demory}, {Sahu}, {Bardalez Gagliuffi}, {Magain}, \& {Queloz}}]{Gillon2016}
{Gillon}, M., {Jehin}, E., {Lederer}, S.~M., {et~al.} 2016, \nat, 533, 221, \dodoi{10.1038/nature17448}

\bibitem[{{Goodman} \& {Weare}(2010)}]{Goodman2010}
{Goodman}, J., \& {Weare}, J. 2010, Communications in Applied Mathematics and Computational Science, 5, 65, \dodoi{10.2140/camcos.2010.5.65}

\bibitem[{{Gray} {et~al.}(2006){Gray}, {Corbally}, {Garrison}, {McFadden}, {Bubar}, {McGahee}, {O'Donoghue}, \& {Knox}}]{Gray2006}
{Gray}, R.~O., {Corbally}, C.~J., {Garrison}, R.~F., {et~al.} 2006, \aj, 132, 161, \dodoi{10.1086/504637}

\bibitem[{{Griffin}(1967)}]{Griffin1967}
{Griffin}, R.~F. 1967, \apj, 148, 465, \dodoi{10.1086/149168}

\bibitem[{{Guerrero} {et~al.}(2021){Guerrero}, {Seager}, {Huang}, {Vanderburg}, {Garcia Soto}, {Mireles}, {Hesse}, {Fong}, {Glidden}, {Shporer}, {Latham}, {Collins}, {Quinn}, {Burt}, {Dragomir}, {Crossfield}, {Vanderspek}, {Fausnaugh}, {Burke}, {Ricker}, {Daylan}, {Essack}, {G{\"u}nther}, {Osborn}, {Pepper}, {Rowden}, {Sha}, {Villanueva}, {Yahalomi}, {Yu}, {Ballard}, {Batalha}, {Berardo}, {Chontos}, {Dittmann}, {Esquerdo}, {Mikal-Evans}, {Jayaraman}, {Krishnamurthy}, {Louie}, {Mehrle}, {Niraula}, {Rackham}, {Rodriguez}, {Rowden}, {Sousa-Silva}, {Watanabe}, {Wong}, {Zhan}, {Zivanovic}, {Christiansen}, {Ciardi}, {Swain}, {Lund}, {Mullally}, {Fleming}, {Rodriguez}, {Boyd}, {Quintana}, {Barclay}, {Col{\'o}n}, {Rinehart}, {Schlieder}, {Clampin}, {Jenkins}, {Twicken}, {Caldwell}, {Coughlin}, {Henze}, {Lissauer}, {Morris}, {Rose}, {Smith}, {Tenenbaum}, {Ting}, {Wohler}, {Bakos}, {Bean}, {Berta-Thompson}, {Bieryla}, {Bouma}, {Buchhave}, {Butler}, {Charbonneau}, {Doty}, {Ge}, {Holman}, {Howard}, {Kaltenegger}, {Kane},
  {Kjeldsen}, {Kreidberg}, {Lin}, {Minsky}, {Narita}, {Paegert}, {P{\'a}l}, {Palle}, {Sasselov}, {Spencer}, {Sozzetti}, {Stassun}, {Torres}, {Udry}, \& {Winn}}]{TOI}
{Guerrero}, N.~M., {Seager}, S., {Huang}, C.~X., {et~al.} 2021, \apjs, 254, 39, \dodoi{10.3847/1538-4365/abefe1}

\bibitem[{{Guillot}(2010)}]{Guillot2010}
{Guillot}, T. 2010, \aap, 520, A27, \dodoi{10.1051/0004-6361/200913396}

\bibitem[{Harris {et~al.}(2020)Harris, Millman, van~der Walt, Gommers, Virtanen, Cournapeau, Wieser, Taylor, Berg, Smith, Kern, Picus, Hoyer, van Kerkwijk, Brett, Haldane, del R{\'{i}}o, Wiebe, Peterson, G{\'{e}}rard-Marchant, Sheppard, Reddy, Weckesser, Abbasi, Gohlke, \& Oliphant}]{Harris2020}
Harris, C.~R., Millman, K.~J., van~der Walt, S.~J., {et~al.} 2020, Nature, 585, 357, \dodoi{10.1038/s41586-020-2649-2}

\bibitem[{{Hedges} {et~al.}(2020){Hedges}, {Angus}, {Barentsen}, {Saunders}, {Montet}, \& {Gully-Santiago}}]{Hedges2020}
{Hedges}, C., {Angus}, R., {Barentsen}, G., {et~al.} 2020, Research Notes of the American Astronomical Society, 4, 220, \dodoi{10.3847/2515-5172/abd106}

\bibitem[{{Hinkel} {et~al.}(2014){Hinkel}, {Timmes}, {Young}, {Pagano}, \& {Turnbull}}]{Hinkel14}
{Hinkel}, N.~R., {Timmes}, F.~X., {Young}, P.~A., {Pagano}, M.~D., \& {Turnbull}, M.~C. 2014, \aj, 148, 54, \dodoi{10.1088/0004-6256/148/3/54}

\bibitem[{{Hinkel} \& {Unterborn}(2018)}]{Hinkel18}
{Hinkel}, N.~R., \& {Unterborn}, C.~T. 2018, \apj, 853, 83, \dodoi{10.3847/1538-4357/aaa5b4}

\bibitem[{{Hippke} {et~al.}(2019){Hippke}, {David}, {Mulders}, \& {Heller}}]{Wotan}
{Hippke}, M., {David}, T.~J., {Mulders}, G.~D., \& {Heller}, R. 2019, \aj, 158, 143, \dodoi{10.3847/1538-3881/ab3984}

\bibitem[{{Hippke} \& {Heller}(2019)}]{TLS}
{Hippke}, M., \& {Heller}, R. 2019, \aap, 623, A39, \dodoi{10.1051/0004-6361/201834672}

\bibitem[{{Hord} {et~al.}(2024){Hord}, {Kempton}, {Evans-Soma}, {Latham}, {Ciardi}, {Dragomir}, {Col{\'o}n}, {Ross}, {Vanderburg}, {de Beurs}, {Collins}, {Watkins}, {Bean}, {Cowan}, {Daylan}, {Morley}, {Ih}, {Baker}, {Barkaoui}, {Batalha}, {Behmard}, {Belinski}, {Benkhaldoun}, {Benni}, {Bernacki}, {Bieryla}, {Binnenfeld}, {Bosch-Cabot}, {Bouchy}, {Bozza}, {Brahm}, {Buchhave}, {Calkins}, {Chontos}, {Clark}, {Cloutier}, {Cointepas}, {Collins}, {Conti}, {Crossfield}, {Dai}, {de Leon}, {Dransfield}, {Dressing}, {Dustor}, {Esquerdo}, {Evans}, {Fajardo-Acosta}, {Fio{\l}ka}, {For{\'e}s-Toribio}, {Frasca}, {Fukui}, {Fulton}, {Furlan}, {Gan}, {Gandolfi}, {Ghachoui}, {Giacalone}, {Gilbert}, {Gillon}, {Girardin}, {Gonzales}, {Grau Horta}, {Gregorio}, {Greklek-McKeon}, {Guerra}, {Hartman}, {Hellier}, {Helm}, {He{\l}miniak}, {Henning}, {Hill}, {Horne}, {Howard}, {Howell}, {Huber}, {Isopi}, {Jehin}, {Jenkins}, {Jensen}, {Johnson}, {Jord{\'a}n}, {Kane}, {Kielkopf}, {Krushinsky}, {Lasota}, {Lee}, {Lewin}, {Livingston},
  {Lubin}, {Lund}, {Mallia}, {Mann}, {Marino}, {Maslennikova}, {Massey}, {Matson}, {Matthews}, {Mayo}, {Mazeh}, {McLeod}, {Michaels}, {Mo{\v{c}}nik}, {Mori}, {Mraz}, {Mu{\~n}oz}, {Narita}, {Natarajan}, {Dyregaard Nielsen}, {Osborn}, {Palle}, {Panahi}, {Papini}, {Plavchan}, {Polanski}, {Popowicz}, {Pozuelos}, {Quinn}, {Radford}, {Reed}, {Relles}, {Rice}, {Robertson}, {Rodriguez}, {Rosenthal}, {Rubenzahl}, {Schanche}, {Schlieder}, {Schwarz}, {Sefako}, {Shporer}, {Sozzetti}, {Srdoc}, {Stockdale}, {Tarasenkov}, {Tan}, {Timmermans}, {Ting}, {Van Zandt}, {Vignes}, {Waite}, {Watanabe}, {Weiss}, {Wittrock}, {Zhou}, {Ziegler}, \& {Zucker}}]{Hord2024}
{Hord}, B.~J., {Kempton}, E. M.~R., {Evans-Soma}, T.~M., {et~al.} 2024, \aj, 167, 233, \dodoi{10.3847/1538-3881/ad3068}

\bibitem[{{Howell} {et~al.}(2011){Howell}, {Everett}, {Sherry}, {Horch}, \& {Ciardi}}]{Howell2011}
{Howell}, S.~B., {Everett}, M.~E., {Sherry}, W., {Horch}, E., \& {Ciardi}, D.~R. 2011, \aj, 142, 19, \dodoi{10.1088/0004-6256/142/1/19}

\bibitem[{{Hu} {et~al.}(2024){Hu}, {Bello-Arufe}, {Zhang}, {Paragas}, {Zilinskas}, {van Buchem}, {Bess}, {Patel}, {Ito}, {Damiano}, {Scheucher}, {Oza}, {Knutson}, {Miguel}, {Dragomir}, {Brandeker}, \& {Demory}}]{hu_secondary_2024}
{Hu}, R., {Bello-Arufe}, A., {Zhang}, M., {et~al.} 2024, \nat, 630, 609, \dodoi{10.1038/s41586-024-07432-x}

\bibitem[{{Huang} \& {Ormel}(2022)}]{Huang2022}
{Huang}, S., \& {Ormel}, C.~W. 2022, \mnras, 511, 3814, \dodoi{10.1093/mnras/stac288}

\bibitem[{Hunter(2007)}]{Hunter2007}
Hunter, J.~D. 2007, Computing in Science \& Engineering, 9, 90, \dodoi{10.1109/MCSE.2007.55}

\bibitem[{{Husser} {et~al.}(2013){Husser}, {Wende-von Berg}, {Dreizler}, {Homeier}, {Reiners}, {Barman}, \& {Hauschildt}}]{Husser:2013}
{Husser}, T.~O., {Wende-von Berg}, S., {Dreizler}, S., {et~al.} 2013, \aap, 553, A6, \dodoi{10.1051/0004-6361/201219058}

\bibitem[{{Jenkins}(2002)}]{Jenkins2002}
{Jenkins}, J.~M. 2002, \apj, 575, 493, \dodoi{10.1086/341136}

\bibitem[{{Jenkins} {et~al.}(2020){Jenkins}, {Tenenbaum}, {Seader}, {Burke}, {McCauliff}, {Smith}, {Twicken}, \& {Chandrasekaran}}]{Jenkins2020}
{Jenkins}, J.~M., {Tenenbaum}, P., {Seader}, S., {et~al.} 2020, {Kepler Data Processing Handbook: Transiting Planet Search}, Kepler Science Document KSCI-19081-003, id. 9. Edited by Jon M. Jenkins.

\bibitem[{{Jenkins} {et~al.}(2010){Jenkins}, {Chandrasekaran}, {McCauliff}, {Caldwell}, {Tenenbaum}, {Li}, {Klaus}, {Cote}, \& {Middour}}]{Jenkins2010}
{Jenkins}, J.~M., {Chandrasekaran}, H., {McCauliff}, S.~D., {et~al.} 2010, in Society of Photo-Optical Instrumentation Engineers (SPIE) Conference Series, Vol. 7740, Software and Cyberinfrastructure for Astronomy, ed. N.~M. {Radziwill} \& A.~{Bridger}, 77400D, \dodoi{10.1117/12.856764}

\bibitem[{{Jenkins} {et~al.}(2016){Jenkins}, {Twicken}, {McCauliff}, {Campbell}, {Sanderfer}, {Lung}, {Mansouri-Samani}, {Girouard}, {Tenenbaum}, {Klaus}, {Smith}, {Caldwell}, {Chacon}, {Henze}, {Heiges}, {Latham}, {Morgan}, {Swade}, {Rinehart}, \& {Vanderspek}}]{SPOC}
{Jenkins}, J.~M., {Twicken}, J.~D., {McCauliff}, S., {et~al.} 2016, in \procspie, Vol. 9913, Software and Cyberinfrastructure for Astronomy IV, 99133E, \dodoi{10.1117/12.2233418}

\bibitem[{{Jenkins} {et~al.}(2006){Jenkins}, {Jones}, {Tinney}, {Butler}, {McCarthy}, {Marcy}, {Pinfield}, {Carter}, \& {Penny}}]{Jenkins2006}
{Jenkins}, J.~S., {Jones}, H.~R.~A., {Tinney}, C.~G., {et~al.} 2006, \mnras, 372, 163, \dodoi{10.1111/j.1365-2966.2006.10811.x}

\bibitem[{{Jensen} \& {Millholland}(2022)}]{JensenMillholland2022}
{Jensen}, D., \& {Millholland}, S.~C. 2022, \aj, 164, 144, \dodoi{10.3847/1538-3881/ac86c5}

\bibitem[{{Jensen}(2013)}]{Jensen2013}
{Jensen}, E. 2013, {Tapir: A web interface for transit/eclipse observability}, Astrophysics Source Code Library, record ascl:1306.007

\bibitem[{{Kempton} {et~al.}(2018){Kempton}, {Bean}, {Louie}, {Deming}, {Koll}, {Mansfield}, {Christiansen}, {L{\'o}pez-Morales}, {Swain}, {Zellem}, {Ballard}, {Barclay}, {Barstow}, {Batalha}, {Beatty}, {Berta-Thompson}, {Birkby}, {Buchhave}, {Charbonneau}, {Cowan}, {Crossfield}, {de Val-Borro}, {Doyon}, {Dragomir}, {Gaidos}, {Heng}, {Hu}, {Kane}, {Kreidberg}, {Mallonn}, {Morley}, {Narita}, {Nascimbeni}, {Pall{\'e}}, {Quintana}, {Rauscher}, {Seager}, {Shkolnik}, {Sing}, {Sozzetti}, {Stassun}, {Valenti}, \& {von Essen}}]{Kempton2018}
{Kempton}, E. M.~R., {Bean}, J.~L., {Louie}, D.~R., {et~al.} 2018, \pasp, 130, 114401, \dodoi{10.1088/1538-3873/aadf6f}

\bibitem[{{Kipping}(2013)}]{exoplanet:kipping13}
{Kipping}, D.~M. 2013, \mnras, 435, 2152, \dodoi{10.1093/mnras/stt1435}

\bibitem[{{Kostov} {et~al.}(2019){Kostov}, {Schlieder}, {Barclay}, {Quintana}, {Col{\'o}n}, {Brande}, {Collins}, {Feinstein}, {Hadden}, {Kane}, {Kreidberg}, {Kruse}, {Lam}, {Matthews}, {Montet}, {Pozuelos}, {Stassun}, {Winters}, {Ricker}, {Vanderspek}, {Latham}, {Seager}, {Winn}, {Jenkins}, {Afanasev}, {Armstrong}, {Arney}, {Boyd}, {Barentsen}, {Barkaoui}, {Batalha}, {Beichman}, {Bayliss}, {Burke}, {Burdanov}, {Cacciapuoti}, {Carson}, {Charbonneau}, {Christiansen}, {Ciardi}, {Clampin}, {Collins}, {Conti}, {Coughlin}, {Covone}, {Crossfield}, {Delrez}, {Domagal-Goldman}, {Dressing}, {Ducrot}, {Essack}, {Everett}, {Fauchez}, {Foreman-Mackey}, {Gan}, {Gilbert}, {Gillon}, {Gonzales}, {Hamann}, {Hedges}, {Hocutt}, {Hoffman}, {Horch}, {Horne}, {Howell}, {Hynes}, {Ireland}, {Irwin}, {Isopi}, {Jensen}, {Jehin}, {Kaltenegger}, {Kielkopf}, {Kopparapu}, {Lewis}, {Lopez}, {Lissauer}, {Mann}, {Mallia}, {Mandell}, {Matson}, {Mazeh}, {Monsue}, {Moran}, {Moran}, {Morley}, {Morris}, {Muirhead}, {Mukai}, {Mullally}, {Mullally},
  {Murray}, {Narita}, {Palle}, {Pidhorodetska}, {Quinn}, {Relles}, {Rinehart}, {Ritsko}, {Rodriguez}, {Rowden}, {Rowe}, {Sebastian}, {Sefako}, {Shahaf}, {Shporer}, {Ta{\~n}{\'o}n Reyes}, {Tenenbaum}, {Ting}, {Twicken}, {van Belle}, {Vega}, {Volosin}, {Walkowicz}, \& {Youngblood}}]{Kostov2019}
{Kostov}, V.~B., {Schlieder}, J.~E., {Barclay}, T., {et~al.} 2019, \aj, 158, 32, \dodoi{10.3847/1538-3881/ab2459}

\bibitem[{Kumar {et~al.}(2019)Kumar, Carroll, Hartikainen, \& Martin}]{exoplanet:arviz}
Kumar, R., Carroll, C., Hartikainen, A., \& Martin, O.~A. 2019, The Journal of Open Source Software, \dodoi{10.21105/joss.01143}

\bibitem[{{K{\"u}rster} {et~al.}(2003){K{\"u}rster}, {Endl}, {Rouesnel}, {Els}, {Kaufer}, {Brillant}, {Hatzes}, {Saar}, \& {Cochran}}]{Kurster2003}
{K{\"u}rster}, M., {Endl}, M., {Rouesnel}, F., {et~al.} 2003, \aap, 403, 1077, \dodoi{10.1051/0004-6361:20030396}

\bibitem[{{Lam} {et~al.}(2021){Lam}, {Csizmadia}, {Astudillo-Defru}, {Bonfils}, {Gandolfi}, {Padovan}, {Esposito}, {Hellier}, {Hirano}, {Livingston}, {Murgas}, {Smith}, {Collins}, {Mathur}, {Garcia}, {Howell}, {Santos}, {Dai}, {Ricker}, {Vanderspek}, {Latham}, {Seager}, {Winn}, {Jenkins}, {Albrecht}, {Almenara}, {Artigau}, {Barrag{\'a}n}, {Bouchy}, {Cabrera}, {Charbonneau}, {Chaturvedi}, {Chaushev}, {Christiansen}, {Cochran}, {De Meideiros}, {Delfosse}, {D{\'\i}az}, {Doyon}, {Eigm{\"u}ller}, {Figueira}, {Forveille}, {Fridlund}, {Gaisn{\'e}}, {Goffo}, {Georgieva}, {Grziwa}, {Guenther}, {Hatzes}, {Johnson}, {Kab{\'a}th}, {Knudstrup}, {Korth}, {Lewin}, {Lissauer}, {Lovis}, {Luque}, {Melo}, {Morgan}, {Morris}, {Mayor}, {Narita}, {Osborne}, {Palle}, {Pepe}, {Persson}, {Quinn}, {Rauer}, {Redfield}, {Schlieder}, {S{\'e}gransan}, {Serrano}, {Smith}, {{\v{S}}ubjak}, {Twicken}, {Udry}, {Van Eylen}, \& {Vezie}}]{Lam2021}
{Lam}, K. W.~F., {Csizmadia}, S., {Astudillo-Defru}, N., {et~al.} 2021, Science, 374, 1271, \dodoi{10.1126/science.aay3253}

\bibitem[{{Lavie} {et~al.}(2023){Lavie}, {Bouchy}, {Lovis}, {Zapatero Osorio}, {Deline}, {Barros}, {Figueira}, {Sozzetti}, {Gonz{\'a}lez Hern{\'a}ndez}, {Lillo-Box}, {Rodrigues}, {Mehner}, {Damasso}, {Adibekyan}, {Alibert}, {Allende Prieto}, {Cristiani}, {D'Odorico}, {Di Marcantonio}, {Ehrenreich}, {G{\'e}nova Santos}, {Lo Curto}, {Martins}, {Micela}, {Molaro}, {Nunes}, {Palle}, {Pepe}, {Poretti}, {Rebolo}, {Santos}, {Sousa}, {Su{\'a}rez Mascare{\~n}o}, {Tabrenero}, \& {Udry}}]{Lavie2023}
{Lavie}, B., {Bouchy}, F., {Lovis}, C., {et~al.} 2023, \aap, 673, A69, \dodoi{10.1051/0004-6361/202143007}

\bibitem[{{Li} {et~al.}(2019){Li}, {Tenenbaum}, {Twicken}, {Burke}, {Jenkins}, {Quintana}, {Rowe}, \& {Seader}}]{Li2019}
{Li}, J., {Tenenbaum}, P., {Twicken}, J.~D., {et~al.} 2019, \pasp, 131, 024506, \dodoi{10.1088/1538-3873/aaf44d}

\bibitem[{{Lissauer} {et~al.}(2011){Lissauer}, {Ragozzine}, {Fabrycky}, {Steffen}, {Ford}, {Jenkins}, {Shporer}, {Holman}, {Rowe}, {Quintana}, {Batalha}, {Borucki}, {Bryson}, {Caldwell}, {Carter}, {Ciardi}, {Dunham}, {Fortney}, {Gautier}, {Howell}, {Koch}, {Latham}, {Marcy}, {Morehead}, \& {Sasselov}}]{Lissauer2011}
{Lissauer}, J.~J., {Ragozzine}, D., {Fabrycky}, D.~C., {et~al.} 2011, \apjs, 197, 8, \dodoi{10.1088/0067-0049/197/1/8}

\bibitem[{{Lissauer} {et~al.}(2012){Lissauer}, {Marcy}, {Rowe}, {Bryson}, {Adams}, {Buchhave}, {Ciardi}, {Cochran}, {Fabrycky}, {Ford}, {Fressin}, {Geary}, {Gilliland}, {Holman}, {Howell}, {Jenkins}, {Kinemuchi}, {Koch}, {Morehead}, {Ragozzine}, {Seader}, {Tanenbaum}, {Torres}, \& {Twicken}}]{Lissauer2012}
{Lissauer}, J.~J., {Marcy}, G.~W., {Rowe}, J.~F., {et~al.} 2012, \apj, 750, 112, \dodoi{10.1088/0004-637X/750/2/112}

\bibitem[{{Lithwick} {et~al.}(2012){Lithwick}, {Xie}, \& {Wu}}]{Lithwick2012}
{Lithwick}, Y., {Xie}, J., \& {Wu}, Y. 2012, \apj, 761, 122, \dodoi{10.1088/0004-637X/761/2/122}

\bibitem[{{Luger} {et~al.}(2019){Luger}, {Agol}, {Foreman-Mackey}, {Fleming}, {Lustig-Yaeger}, \& {Deitrick}}]{exoplanet:luger18}
{Luger}, R., {Agol}, E., {Foreman-Mackey}, D., {et~al.} 2019, \aj, 157, 64, \dodoi{10.3847/1538-3881/aae8e5}

\bibitem[{{Luque} {et~al.}(2023){Luque}, {Osborn}, {Leleu}, {Pall{\'e}}, {Bonfanti}, {Barrag{\'a}n}, {Wilson}, {Broeg}, {Cameron}, {Lendl}, {Maxted}, {Alibert}, {Gandolfi}, {Delisle}, {Hooton}, {Egger}, {Nowak}, {Lafarga}, {Rapetti}, {Twicken}, {Morales}, {Carleo}, {Orell-Miquel}, {Adibekyan}, {Alonso}, {Alqasim}, {Amado}, {Anderson}, {Anglada-Escud{\'e}}, {Bandy}, {B{\'a}rczy}, {Barrado Navascues}, {Barros}, {Baumjohann}, {Bayliss}, {Bean}, {Beck}, {Beck}, {Benz}, {Billot}, {Bonfils}, {Borsato}, {Boyle}, {Brandeker}, {Bryant}, {Cabrera}, {Carrazco-Gaxiola}, {Charbonneau}, {Charnoz}, {Ciardi}, {Cochran}, {Collins}, {Crossfield}, {Csizmadia}, {Cubillos}, {Dai}, {Davies}, {Deeg}, {Deleuil}, {Deline}, {Delrez}, {Demangeon}, {Demory}, {Ehrenreich}, {Erikson}, {Esparza-Borges}, {Falk}, {Fortier}, {Fossati}, {Fridlund}, {Fukui}, {Garcia-Mejia}, {Gill}, {Gillon}, {Goffo}, {G{\'o}mez Maqueo Chew}, {G{\"u}del}, {Guenther}, {G{\"u}nther}, {Hatzes}, {Helling}, {Hesse}, {Howell}, {Hoyer}, {Ikuta}, {Isaak}, {Jenkins},
  {Kagetani}, {Kiss}, {Kodama}, {Korth}, {Lam}, {Laskar}, {Latham}, {Lecavelier des Etangs}, {Leon}, {Livingston}, {Magrin}, {Matson}, {Matthews}, {Mordasini}, {Mori}, {Moyano}, {Munari}, {Murgas}, {Narita}, {Nascimbeni}, {Olofsson}, {Osborne}, {Ottensamer}, {Pagano}, {Parviainen}, {Peter}, {Piotto}, {Pollacco}, {Queloz}, {Quinn}, {Quirrenbach}, {Ragazzoni}, {Rando}, {Ratti}, {Rauer}, {Redfield}, {Ribas}, {Ricker}, {Rudat}, {Sabin}, {Salmon}, {Santos}, {Scandariato}, {Schanche}, {Schlieder}, {Seager}, {S{\'e}gransan}, {Shporer}, {Simon}, {Smith}, {Sousa}, {Stalport}, {Szab{\'o}}, {Thomas}, {Tuson}, {Udry}, {Vanderburg}, {Van Eylen}, {Van Grootel}, {Venturini}, {Walter}, {Walton}, {Watanabe}, {Winn}, \& {Zingales}}]{Luque2023}
{Luque}, R., {Osborn}, H.~P., {Leleu}, A., {et~al.} 2023, \nat, 623, 932, \dodoi{10.1038/s41586-023-06692-3}

\bibitem[{{Mayor} {et~al.}(2003){Mayor}, {Pepe}, {Queloz}, {Bouchy}, {Rupprecht}, {Lo Curto}, {Avila}, {Benz}, {Bertaux}, {Bonfils}, {Dall}, {Dekker}, {Delabre}, {Eckert}, {Fleury}, {Gilliotte}, {Gojak}, {Guzman}, {Kohler}, {Lizon}, {Longinotti}, {Lovis}, {Megevand}, {Pasquini}, {Reyes}, {Sivan}, {Sosnowska}, {Soto}, {Udry}, {van Kesteren}, {Weber}, \& {Weilenmann}}]{Mayor2003}
{Mayor}, M., {Pepe}, F., {Queloz}, D., {et~al.} 2003, The Messenger, 114, 20

\bibitem[{{McCully} {et~al.}(2018){McCully}, {Volgenau}, {Harbeck}, {Lister}, {Saunders}, {Turner}, {Siiverd}, \& {Bowman}}]{BANZAI}
{McCully}, C., {Volgenau}, N.~H., {Harbeck}, D.-R., {et~al.} 2018, in Society of Photo-Optical Instrumentation Engineers (SPIE) Conference Series, Vol. 10707, Software and Cyberinfrastructure for Astronomy V, ed. J.~C. {Guzman} \& J.~{Ibsen}, 107070K, \dodoi{10.1117/12.2314340}

\bibitem[{{Ment} {et~al.}(2021){Ment}, {Irwin}, {Charbonneau}, {Winters}, {Medina}, {Cloutier}, {D{\'\i}az}, {Jenkins}, {Ziegler}, {Law}, {Mann}, {Ricker}, {Vanderspek}, {Latham}, {Seager}, {Winn}, {Jenkins}, {Goeke}, {Levine}, {Rojas-Ayala}, {Rowden}, {Ting}, \& {Twicken}}]{Ment2021}
{Ment}, K., {Irwin}, J., {Charbonneau}, D., {et~al.} 2021, \aj, 161, 23, \dodoi{10.3847/1538-3881/abbd91}

\bibitem[{{Millholland} {et~al.}(2017){Millholland}, {Wang}, \& {Laughlin}}]{Millholland2017}
{Millholland}, S., {Wang}, S., \& {Laughlin}, G. 2017, \apjl, 849, L33, \dodoi{10.3847/2041-8213/aa9714}

\bibitem[{{Millholland} {et~al.}(2018){Millholland}, {Laughlin}, {Teske}, {Butler}, {Burt}, {Holden}, {Vogt}, {Crane}, {Shectman}, \& {Thompson}}]{Millholland2018}
{Millholland}, S., {Laughlin}, G., {Teske}, J., {et~al.} 2018, \aj, 155, 106, \dodoi{10.3847/1538-3881/aaa894}

\bibitem[{{Molli{\`e}re} {et~al.}(2019){Molli{\`e}re}, {Wardenier}, {van Boekel}, {Henning}, {Molaverdikhani}, \& {Snellen}}]{Molliere2019}
{Molli{\`e}re}, P., {Wardenier}, J.~P., {van Boekel}, R., {et~al.} 2019, \aap, 627, A67, \dodoi{10.1051/0004-6361/201935470}

\bibitem[{{Morton}(2015)}]{Morton2015}
{Morton}, T.~D. 2015, {isochrones: Stellar model grid package}, Astrophysics Source Code Library, record ascl:1503.010.
\newblock \doeprint{1503.010}

\bibitem[{{NASA Exoplanet Archive}(2024)}]{NEA}
{NASA Exoplanet Archive}. 2024, Planetary Systems Composite Parameters, Version: 2024-04-16 01:06,  NExScI-Caltech/IPAC, \dodoi{10.26133/NEA13}

\bibitem[{NExScI(2023)}]{ExoFOP}
NExScI. 2023, Exoplanet Follow-up Observing Program Web Service,  IPAC, \dodoi{10.26134/ExoFOP5}

\bibitem[{{Noyes} {et~al.}(1984){Noyes}, {Hartmann}, {Baliunas}, {Duncan}, \& {Vaughan}}]{Noyes1984}
{Noyes}, R.~W., {Hartmann}, L.~W., {Baliunas}, S.~L., {Duncan}, D.~K., \& {Vaughan}, A.~H. 1984, \apj, 279, 763, \dodoi{10.1086/161945}

\bibitem[{{Otegi} {et~al.}(2022){Otegi}, {Helled}, \& {Bouchy}}]{Otegi2022}
{Otegi}, J.~F., {Helled}, R., \& {Bouchy}, F. 2022, \aap, 658, A107, \dodoi{10.1051/0004-6361/202142110}

\bibitem[{{Paegert} {et~al.}(2021){Paegert}, {Stassun}, {Collins}, {Pepper}, {Torres}, {Jenkins}, {Twicken}, \& {Latham}}]{Paegert2021}
{Paegert}, M., {Stassun}, K.~G., {Collins}, K.~A., {et~al.} 2021, arXiv e-prints, arXiv:2108.04778.
\newblock \doarXiv{2108.04778}

\bibitem[{{Pepe} {et~al.}(2021){Pepe}, {Cristiani}, {Rebolo}, {Santos}, {Dekker}, {Cabral}, {Di Marcantonio}, {Figueira}, {Lo Curto}, {Lovis}, {Mayor}, {M{\'e}gevand}, {Molaro}, {Riva}, {Zapatero Osorio}, {Amate}, {Manescau}, {Pasquini}, {Zerbi}, {Adibekyan}, {Abreu}, {Affolter}, {Alibert}, {Aliverti}, {Allart}, {Allende Prieto}, {{\'A}lvarez}, {Alves}, {Avila}, {Baldini}, {Bandy}, {Barros}, {Benz}, {Bianco}, {Borsa}, {Bourrier}, {Bouchy}, {Broeg}, {Calderone}, {Cirami}, {Coelho}, {Conconi}, {Coretti}, {Cumani}, {Cupani}, {D'Odorico}, {Damasso}, {Deiries}, {Delabre}, {Demangeon}, {Dumusque}, {Ehrenreich}, {Faria}, {Fragoso}, {Genolet}, {Genoni}, {G{\'e}nova Santos}, {Gonz{\'a}lez Hern{\'a}ndez}, {Hughes}, {Iwert}, {Kerber}, {Knudstrup}, {Landoni}, {Lavie}, {Lillo-Box}, {Lizon}, {Maire}, {Martins}, {Mehner}, {Micela}, {Modigliani}, {Monteiro}, {Monteiro}, {Moschetti}, {Murphy}, {Nunes}, {Oggioni}, {Oliveira}, {Oshagh}, {Pall{\'e}}, {Pariani}, {Poretti}, {Rasilla}, {Rebord{\~a}o}, {Redaelli}, {Santana Tschudi},
  {Santin}, {Santos}, {S{\'e}gransan}, {Schmidt}, {Segovia}, {Sosnowska}, {Sozzetti}, {Sousa}, {Span{\`o}}, {Su{\'a}rez Mascare{\~n}o}, {Tabernero}, {Tenegi}, {Udry}, \& {Zanutta}}]{ESPRESSO}
{Pepe}, F., {Cristiani}, S., {Rebolo}, R., {et~al.} 2021, \aap, 645, A96, \dodoi{10.1051/0004-6361/202038306}

\bibitem[{{Perdelwitz} {et~al.}(2024){Perdelwitz}, {Trifonov}, {Teklu}, {Sreenivas}, \& {Tal-Or}}]{Perdelwitz2024}
{Perdelwitz}, V., {Trifonov}, T., {Teklu}, J.~T., {Sreenivas}, K.~R., \& {Tal-Or}, L. 2024, \aap, 683, A125, \dodoi{10.1051/0004-6361/202348263}

\bibitem[{{Perryman} {et~al.}(1997){Perryman}, {Lindegren}, {Kovalevsky}, {Hog}, {Bastian}, {Bernacca}, {Creze}, {Donati}, {Grenon}, {Grewing}, {van Leeuwen}, {van der Marel}, {Mignard}, {Murray}, {Le Poole}, {Schrijver}, {Turon}, {Arenou}, {Froeschle}, \& {Petersen}}]{Hipparcos}
{Perryman}, M.~A.~C., {Lindegren}, L., {Kovalevsky}, J., {et~al.} 1997, \aap, 500, 501

\bibitem[{{Petit} {et~al.}(2020){Petit}, {Pichierri}, {Davies}, \& {Johansen}}]{Petit2020}
{Petit}, A.~C., {Pichierri}, G., {Davies}, M.~B., \& {Johansen}, A. 2020, \aap, 641, A176, \dodoi{10.1051/0004-6361/202038764}

\bibitem[{{Pu} \& {Wu}(2015)}]{PuWu2015}
{Pu}, B., \& {Wu}, Y. 2015, \apj, 807, 44, \dodoi{10.1088/0004-637X/807/1/44}

\bibitem[{{Rein} \& {Liu}(2012)}]{Rein2012}
{Rein}, H., \& {Liu}, S.~F. 2012, \aap, 537, A128, \dodoi{10.1051/0004-6361/201118085}

\bibitem[{{Rein} \& {Tamayo}(2015)}]{Rein2015}
{Rein}, H., \& {Tamayo}, D. 2015, \mnras, 452, 376, \dodoi{10.1093/mnras/stv1257}

\bibitem[{{Ricker} {et~al.}(2015){Ricker}, {Winn}, {Vanderspek}, {Latham}, {Bakos}, {Bean}, {Berta-Thompson}, {Brown}, {Buchhave}, {Butler}, {Butler}, {Chaplin}, {Charbonneau}, {Christensen-Dalsgaard}, {Clampin}, {Deming}, {Doty}, {De Lee}, {Dressing}, {Dunham}, {Endl}, {Fressin}, {Ge}, {Henning}, {Holman}, {Howard}, {Ida}, {Jenkins}, {Jernigan}, {Johnson}, {Kaltenegger}, {Kawai}, {Kjeldsen}, {Laughlin}, {Levine}, {Lin}, {Lissauer}, {MacQueen}, {Marcy}, {McCullough}, {Morton}, {Narita}, {Paegert}, {Palle}, {Pepe}, {Pepper}, {Quirrenbach}, {Rinehart}, {Sasselov}, {Sato}, {Seager}, {Sozzetti}, {Stassun}, {Sullivan}, {Szentgyorgyi}, {Torres}, {Udry}, \& {Villasenor}}]{Ricker2015}
{Ricker}, G.~R., {Winn}, J.~N., {Vanderspek}, R., {et~al.} 2015, Journal of Astronomical Telescopes, Instruments, and Systems, 1, 014003, \dodoi{10.1117/1.JATIS.1.1.014003}

\bibitem[{{Rogers}(2015)}]{Rogers2015}
{Rogers}, L.~A. 2015, \apj, 801, 41, \dodoi{10.1088/0004-637X/801/1/41}

\bibitem[{Salvatier {et~al.}(2016)Salvatier, Wiecki, \& Fonnesbeck}]{exoplanet:pymc3}
Salvatier, J., Wiecki, T.~V., \& Fonnesbeck, C. 2016, PeerJ Computer Science, 2, e55

\bibitem[{{Santos} {et~al.}(2000){Santos}, {Mayor}, {Naef}, {Pepe}, {Queloz}, {Udry}, \& {Blecha}}]{Santos2000}
{Santos}, N.~C., {Mayor}, M., {Naef}, D., {et~al.} 2000, \aap, 361, 265

\bibitem[{{Schlegel} {et~al.}(1998){Schlegel}, {Finkbeiner}, \& {Davis}}]{Schlegel:1998}
{Schlegel}, D.~J., {Finkbeiner}, D.~P., \& {Davis}, M. 1998, \apj, 500, 525, \dodoi{10.1086/305772}

\bibitem[{{Scott} {et~al.}(2021){Scott}, {Howell}, {Gnilka}, {Stephens}, {Salinas}, {Matson}, {Furlan}, {Horch}, {Everett}, {Ciardi}, {Mills}, \& {Quigley}}]{Scott2021}
{Scott}, N.~J., {Howell}, S.~B., {Gnilka}, C.~L., {et~al.} 2021, Frontiers in Astronomy and Space Sciences, 8, 138, \dodoi{10.3389/fspas.2021.716560}

\bibitem[{{Seager} \& {Mall{\'e}n-Ornelas}(2003)}]{Seager2003}
{Seager}, S., \& {Mall{\'e}n-Ornelas}, G. 2003, \apj, 585, 1038, \dodoi{10.1086/346105}

\bibitem[{{Skrutskie} {et~al.}(2006){Skrutskie}, {Cutri}, {Stiening}, {Weinberg}, {Schneider}, {Carpenter}, {Beichman}, {Capps}, {Chester}, {Elias}, {Huchra}, {Liebert}, {Lonsdale}, {Monet}, {Price}, {Seitzer}, {Jarrett}, {Kirkpatrick}, {Gizis}, {Howard}, {Evans}, {Fowler}, {Fullmer}, {Hurt}, {Light}, {Kopan}, {Marsh}, {McCallon}, {Tam}, {Van Dyk}, \& {Wheelock}}]{2MASS}
{Skrutskie}, M.~F., {Cutri}, R.~M., {Stiening}, R., {et~al.} 2006, \aj, 131, 1163, \dodoi{10.1086/498708}

\bibitem[{{Smith} {et~al.}(2012){Smith}, {Stumpe}, {Van Cleve}, {Jenkins}, {Barclay}, {Fanelli}, {Girouard}, {Kolodziejczak}, {McCauliff}, {Morris}, \& {Twicken}}]{Smith2012}
{Smith}, J.~C., {Stumpe}, M.~C., {Van Cleve}, J.~E., {et~al.} 2012, \pasp, 124, 1000, \dodoi{10.1086/667697}

\bibitem[{{Soto} \& {Jenkins}(2018)}]{Soto2018}
{Soto}, M.~G., \& {Jenkins}, J.~S. 2018, \aap, 615, A76, \dodoi{10.1051/0004-6361/201731533}

\bibitem[{{Sousa} {et~al.}(2008){Sousa}, {Santos}, {Mayor}, {Udry}, {Casagrande}, {Israelian}, {Pepe}, {Queloz}, \& {Monteiro}}]{Sousa2008}
{Sousa}, S.~G., {Santos}, N.~C., {Mayor}, M., {et~al.} 2008, \aap, 487, 373, \dodoi{10.1051/0004-6361:200809698}

\bibitem[{{Stassun} {et~al.}(2017){Stassun}, {Collins}, \& {Gaudi}}]{Stassun:2017}
{Stassun}, K.~G., {Collins}, K.~A., \& {Gaudi}, B.~S. 2017, \aj, 153, 136, \dodoi{10.3847/1538-3881/aa5df3}

\bibitem[{{Stassun} {et~al.}(2018){Stassun}, {Corsaro}, {Pepper}, \& {Gaudi}}]{Stassun:2018}
{Stassun}, K.~G., {Corsaro}, E., {Pepper}, J.~A., \& {Gaudi}, B.~S. 2018, \aj, 155, 22, \dodoi{10.3847/1538-3881/aa998a}

\bibitem[{{Stassun} \& {Torres}(2016)}]{Stassun:2016}
{Stassun}, K.~G., \& {Torres}, G. 2016, \aj, 152, 180, \dodoi{10.3847/0004-6256/152/6/180}

\bibitem[{{Stumpe} {et~al.}(2014){Stumpe}, {Smith}, {Catanzarite}, {Van Cleve}, {Jenkins}, {Twicken}, \& {Girouard}}]{Stumpe2014}
{Stumpe}, M.~C., {Smith}, J.~C., {Catanzarite}, J.~H., {et~al.} 2014, \pasp, 126, 100, \dodoi{10.1086/674989}

\bibitem[{{Stumpe} {et~al.}(2012){Stumpe}, {Smith}, {Van Cleve}, {Twicken}, {Barclay}, {Fanelli}, {Girouard}, {Jenkins}, {Kolodziejczak}, {McCauliff}, \& {Morris}}]{Stumpe2012}
{Stumpe}, M.~C., {Smith}, J.~C., {Van Cleve}, J.~E., {et~al.} 2012, \pasp, 124, 985, \dodoi{10.1086/667698}

\bibitem[{{Su{\'a}rez Mascare{\~n}o} {et~al.}(2016){Su{\'a}rez Mascare{\~n}o}, {Rebolo}, \& {Gonz{\'a}lez Hern{\'a}ndez}}]{Mascareno2016}
{Su{\'a}rez Mascare{\~n}o}, A., {Rebolo}, R., \& {Gonz{\'a}lez Hern{\'a}ndez}, J.~I. 2016, \aap, 595, A12, \dodoi{10.1051/0004-6361/201628586}

\bibitem[{Tamayo {et~al.}(2020)Tamayo, Cranmer, Hadden, Rein, Battaglia, Obertas, Armitage, Ho, Spergel, Gilbertson, {et~al.}}]{Tamayo2020}
Tamayo, D., Cranmer, M., Hadden, S., {et~al.} 2020, Proceedings of the National Academy of Sciences, 117, 18194

\bibitem[{Team(2020)}]{reback2020pandas}
Team, T. P.~D. 2020, pandas-dev/pandas: Pandas, latest,  Zenodo, \dodoi{10.5281/zenodo.3509134}

\bibitem[{{Theano Development Team}(2016)}]{exoplanet:theano}
{Theano Development Team}. 2016, arXiv e-prints, abs/1605.02688.
\newblock \url{http://arxiv.org/abs/1605.02688}

\bibitem[{{Thompson} {et~al.}(2018){Thompson}, {Coughlin}, {Hoffman}, {Mullally}, {Christiansen}, {Burke}, {Bryson}, {Batalha}, {Haas}, {Catanzarite}, {Rowe}, {Barentsen}, {Caldwell}, {Clarke}, {Jenkins}, {Li}, {Latham}, {Lissauer}, {Mathur}, {Morris}, {Seader}, {Smith}, {Klaus}, {Twicken}, {Van Cleve}, {Wohler}, {Akeson}, {Ciardi}, {Cochran}, {Henze}, {Howell}, {Huber}, {Pr{\v{s}}a}, {Ram{\'\i}rez}, {Morton}, {Barclay}, {Campbell}, {Chaplin}, {Charbonneau}, {Christensen-Dalsgaard}, {Dotson}, {Doyle}, {Dunham}, {Dupree}, {Ford}, {Geary}, {Girouard}, {Isaacson}, {Kjeldsen}, {Quintana}, {Ragozzine}, {Shabram}, {Shporer}, {Silva Aguirre}, {Steffen}, {Still}, {Tenenbaum}, {Welsh}, {Wolfgang}, {Zamudio}, {Koch}, \& {Borucki}}]{Thompson2018}
{Thompson}, S.~E., {Coughlin}, J.~L., {Hoffman}, K., {et~al.} 2018, \apjs, 235, 38, \dodoi{10.3847/1538-4365/aab4f9}

\bibitem[{{Tinney} {et~al.}(2001){Tinney}, {Butler}, {Marcy}, {Jones}, {Penny}, {Vogt}, {Apps}, \& {Henry}}]{Tinney2001}
{Tinney}, C.~G., {Butler}, R.~P., {Marcy}, G.~W., {et~al.} 2001, \apj, 551, 507, \dodoi{10.1086/320097}

\bibitem[{{Torres} {et~al.}(2010){Torres}, {Andersen}, \& {Gim{\'e}nez}}]{Torres:2010}
{Torres}, G., {Andersen}, J., \& {Gim{\'e}nez}, A. 2010, \aapr, 18, 67, \dodoi{10.1007/s00159-009-0025-1}

\bibitem[{{Trifonov} {et~al.}(2019){Trifonov}, {Rybizki}, \& {K{\"u}rster}}]{Trifonov2019}
{Trifonov}, T., {Rybizki}, J., \& {K{\"u}rster}, M. 2019, \aap, 622, L7, \dodoi{10.1051/0004-6361/201834817}

\bibitem[{{Trifonov} {et~al.}(2020){Trifonov}, {Tal-Or}, {Zechmeister}, {Kaminski}, {Zucker}, \& {Mazeh}}]{Trifonov2020}
{Trifonov}, T., {Tal-Or}, L., {Zechmeister}, M., {et~al.} 2020, \aap, 636, A74, \dodoi{10.1051/0004-6361/201936686}

\bibitem[{{Twicken} {et~al.}(2018){Twicken}, {Catanzarite}, {Clarke}, {Girouard}, {Jenkins}, {Klaus}, {Li}, {McCauliff}, {Seader}, {Tenenbaum}, {Wohler}, {Bryson}, {Burke}, {Caldwell}, {Haas}, {Henze}, \& {Sanderfer}}]{Twicken2018}
{Twicken}, J.~D., {Catanzarite}, J.~H., {Clarke}, B.~D., {et~al.} 2018, \pasp, 130, 064502, \dodoi{10.1088/1538-3873/aab694}

\bibitem[{{Van Eylen} {et~al.}(2019){Van Eylen}, {Albrecht}, {Huang}, {MacDonald}, {Dawson}, {Cai}, {Foreman-Mackey}, {Lundkvist}, {Silva Aguirre}, {Snellen}, \& {Winn}}]{VanEylen2019}
{Van Eylen}, V., {Albrecht}, S., {Huang}, X., {et~al.} 2019, \aj, 157, 61, \dodoi{10.3847/1538-3881/aaf22f}

\bibitem[{{van Leeuwen}(2007)}]{HipparcosNew}
{van Leeuwen}, F. 2007, \aap, 474, 653, \dodoi{10.1051/0004-6361:20078357}

\bibitem[{{Vanzi} {et~al.}(2012){Vanzi}, {Chacon}, {Helminiak}, {Baffico}, {Rivinius}, {{\v{S}}tefl}, {Baade}, {Avila}, \& {Guirao}}]{Vanzi2012}
{Vanzi}, L., {Chacon}, J., {Helminiak}, K.~G., {et~al.} 2012, \mnras, 424, 2770, \dodoi{10.1111/j.1365-2966.2012.21382.x}

\bibitem[{Virtanen {et~al.}(2020)Virtanen, Gommers, Oliphant, Haberland, Reddy, Cournapeau, Burovski, Peterson, Weckesser, Bright, {van der Walt}, Brett, Wilson, Millman, Mayorov, Nelson, Jones, Kern, Larson, Carey, Polat, Feng, Moore, {VanderPlas}, Laxalde, Perktold, Cimrman, Henriksen, Quintero, Harris, Archibald, Ribeiro, Pedregosa, {van Mulbregt}, \& {SciPy 1.0 Contributors}}]{Virtanen2020}
Virtanen, P., Gommers, R., Oliphant, T.~E., {et~al.} 2020, Nature Methods, 17, 261, \dodoi{10.1038/s41592-019-0686-2}

\bibitem[{{Weiss} \& {Schlattl}(2008)}]{WeissSchlattl2008}
{Weiss}, A., \& {Schlattl}, H. 2008, \apss, 316, 99, \dodoi{10.1007/s10509-007-9606-5}

\bibitem[{{Weiss} {et~al.}(2023){Weiss}, {Millholland}, {Petigura}, {Adams}, {Batygin}, {Block}, \& {Mordasini}}]{Weiss2023}
{Weiss}, L.~M., {Millholland}, S.~C., {Petigura}, E.~A., {et~al.} 2023, in Astronomical Society of the Pacific Conference Series, Vol. 534, Protostars and Planets VII, ed. S.~{Inutsuka}, Y.~{Aikawa}, T.~{Muto}, K.~{Tomida}, \& M.~{Tamura}, 863

\bibitem[{{Weiss} {et~al.}(2018){Weiss}, {Marcy}, {Petigura}, {Fulton}, {Howard}, {Winn}, {Isaacson}, {Morton}, {Hirsch}, {Sinukoff}, {Cumming}, {Hebb}, \& {Cargile}}]{Weiss2018}
{Weiss}, L.~M., {Marcy}, G.~W., {Petigura}, E.~A., {et~al.} 2018, \aj, 155, 48, \dodoi{10.3847/1538-3881/aa9ff6}

\bibitem[{{W}es {M}c{K}inney(2010)}]{mckinney-proc-scipy-2010}
{W}es {M}c{K}inney. 2010, in {P}roceedings of the 9th {P}ython in {S}cience {C}onference, ed. {S}t\'efan van~der {W}alt \& {J}arrod {M}illman, 56 -- 61, \dodoi{10.25080/Majora-92bf1922-00a}

\bibitem[{{Wilson}(1978)}]{Wilson1978}
{Wilson}, O.~C. 1978, \apj, 226, 379, \dodoi{10.1086/156618}

\bibitem[{{Wisdom} \& {Holman}(1991)}]{WisdomHolman1991}
{Wisdom}, J., \& {Holman}, M. 1991, \aj, 102, 1528, \dodoi{10.1086/115978}

\bibitem[{{Wood} {et~al.}(2021){Wood}, {Mann}, \& {Kraus}}]{MOLUSC}
{Wood}, M.~L., {Mann}, A.~W., \& {Kraus}, A.~L. 2021, \aj, 162, 128, \dodoi{10.3847/1538-3881/ac0ae9}

\bibitem[{{Wright} {et~al.}(2010){Wright}, {Eisenhardt}, {Mainzer}, {Ressler}, {Cutri}, {Jarrett}, {Kirkpatrick}, {Padgett}, {McMillan}, {Skrutskie}, {Stanford}, {Cohen}, {Walker}, {Mather}, {Leisawitz}, {Gautier}, {McLean}, {Benford}, {Lonsdale}, {Blain}, {Mendez}, {Irace}, {Duval}, {Liu}, {Royer}, {Heinrichsen}, {Howard}, {Shannon}, {Kendall}, {Walsh}, {Larsen}, {Cardon}, {Schick}, {Schwalm}, {Abid}, {Fabinsky}, {Naes}, \& {Tsai}}]{WISE}
{Wright}, E.~L., {Eisenhardt}, P. R.~M., {Mainzer}, A.~K., {et~al.} 2010, \aj, 140, 1868, \dodoi{10.1088/0004-6256/140/6/1868}

\bibitem[{{Zacharias} {et~al.}(2013){Zacharias}, {Finch}, {Girard}, {Henden}, {Bartlett}, {Monet}, \& {Zacharias}}]{UCAC4}
{Zacharias}, N., {Finch}, C.~T., {Girard}, T.~M., {et~al.} 2013, \aj, 145, 44, \dodoi{10.1088/0004-6256/145/2/44}

\bibitem[{{Zahnle} \& {Catling}(2017)}]{Zahnle_cosmic_2017}
{Zahnle}, K.~J., \& {Catling}, D.~C. 2017, \apj, 843, 122, \dodoi{10.3847/1538-4357/aa7846}

\bibitem[{{Zechmeister} \& {K{\"u}rster}(2009)}]{Zechmeister2009}
{Zechmeister}, M., \& {K{\"u}rster}, M. 2009, \aap, 496, 577, \dodoi{10.1051/0004-6361:200811296}

\bibitem[{{Zechmeister} {et~al.}(2018){Zechmeister}, {Reiners}, {Amado}, {Azzaro}, {Bauer}, {B{\'e}jar}, {Caballero}, {Guenther}, {Hagen}, {Jeffers}, {Kaminski}, {K{\"u}rster}, {Launhardt}, {Montes}, {Morales}, {Quirrenbach}, {Reffert}, {Ribas}, {Seifert}, {Tal-Or}, \& {Wolthoff}}]{Zechmeister2018}
{Zechmeister}, M., {Reiners}, A., {Amado}, P.~J., {et~al.} 2018, \aap, 609, A12, \dodoi{10.1051/0004-6361/201731483}

\end{thebibliography}
